\documentclass[structabstract]{aa}

\usepackage{graphicx}
\usepackage{txfonts}
\usepackage{natbib}
\usepackage{subfig}

\usepackage{supertabular}

\usepackage{sidecap}
\usepackage{wrapfig}
\usepackage{amssymb}

\def\asec{\ifmmode ^{\prime\prime}\else$^{\prime\prime}$\fi}

\def\degs{\ifmmode ^{\circ}\else$^{\circ}$\fi}
\def\amin{\ifmmode ^{\prime}\else$^{\prime}$\fi}
\def\asec{\ifmmode ^{\prime\prime}\else$^{\prime\prime}$\fi}

\def\degs{\ifmmode ^{\circ}\else$^{\circ}$\fi}
\def\amin{\ifmmode ^{\prime}\else$^{\prime}$\fi}
\def\cm{\mbox{\,cm}}

\def\cm3{\mbox{\,cm$^{-3}$}}

\def\lsim{\!\!\!\phantom{\le}\smash{\buildrel{}\over
 {\lower2.5dd\hbox{$\buildrel{\lower2dd\hbox{$\displaystyle<$}}\over
                                 \sim$}}}\,\,}
\def\gsim{\!\!\!\phantom{\ge}\smash{\buildrel{}\over
{\lower2.5dd\hbox{$\buildrel{\lower2dd\hbox{$\displaystyle>$}}\over
                               \sim$}}}\,\,}

\begin{document}

 \title{Metallicity determination of M dwarfs \thanks{Based on data obtained at ESO-VLT, Paranal Observatory, Chile, Program ID 082.D-0838(A) and 084.D-1042(A).}  \\ {\Large High-resolution IR spectroscopy} } 
   
   \author{Sara Lindgren\inst{1} 
   \and Ulrike Heiter\inst{1}
   \and Andreas Seifahrt\inst{2}
               }
          
\institute{$^1$ Uppsala University, Department of Physics and Astronomy, Division of Astronomy and Space Physics, Box 516, 751 20 Uppsala, Sweden\\
{\it \email{sara.lindgren@physics.uu.se}}\\          
$^2$ University of Chicago, The Department of Astronomy and Astrophysics, 5640 S. Ellis Ave. Chicago, IL 60637, USA.
           }
\date{}

\abstract
{Several new techniques to determine the metallicity of M dwarfs with better precision have been developed over the last decades. However, most of these studies were based on empirical methods. In order to enable detailed abundance analysis,  standard methods established for warmer solar-like stars, i.e. model-dependent methods using fitting of synthetic spectra,   still  need to be used.}
{In this work we continue the reliability confirmation and development of metallicity determinations of M dwarfs using high-resolution infrared spectra. The reliability was confirmed though analysis of M dwarfs in four binary systems with FGK dwarf companions and by comparison with previous optical studies of the FGK dwarfs.}
{The metallicity determination was based on spectra taken in the $J$ band (1.1-1.4 $\mu$m) with the CRIRES spectrograph. In this part of the infrared, the density of stellar molecular lines is limited, reducing the amount of blends with atomic lines enabling an accurate continuum placement. Lines of several atomic species were used to determine the stellar metallicity.}
{All binaries show excellent agreement between the derived metallicity of the M dwarf and its binary companion. Our results are also in good agreement with values found in the literature. Furthermore, we propose an alternative way to determine the effective temperature of M dwarfs of spectral types later than M2 through synthetic spectral fitting of the FeH lines in our observed spectra.}
{We have confirmed that a reliable metallicity determination of M dwarfs can be achieved using high-resolution infrared spectroscopy. We also note that metallicites obtained with photometric metallicity calibrations available for M dwarfs only partly agree with the results we obtain from high-resolution spectroscopy. }


\authorrunning{S. Lindgren et al.} 
\titlerunning{Metallicity of M dwarfs}

   \maketitle
%

\section{Introduction}

\noindent
M dwarfs are the most numerous type of main-sequence stars; they make up about 70\% of all stars in our Galaxy \citep{Covey2008, Bochanski2010} and about 50\% of its baryonic matter \citep{Chabrier2003}. Accurate knowledge of their composition is therefore essential in order to  advance the understanding of large-scale processes such as the galactic chemical evolution and the initial/present-day mass function. Another area where M dwarfs are of increasing interest is for planet formation theory. Thanks to large surveys like Kepler\footnote{http://kepler.nasa.gov/} and CoRoT\footnote{http://sci.esa.int/corot/} there has been a tremendous increase in discovered exoplanets. Many previous surveys focused mainly on solar-like hosts, while some current and future projects like MEarth \citep{Irwin2009}, CARMENES \citep{Quirrenbach2014}, and K2 \citep{Howell2014} will shift focus towards cooler K and M dwarfs. The vast majority of all exoplanets have been discovered either with the radial velocity technique \citep{Struve1952} or the transit technique \citep{Rosenblatt1971}. However, both the radial velocity and transit techniques are biased towards finding large, close-in planets (e.g. RV: \citealt{Cumming1999}; Transit: \citealt{Seager2003}). One of the goals of today's exoplanet research is to find Earth-like planets. The first step is to find a sample of small planets located within the habitable zone of the host star where liquid water can exist. The next step is to characterise the atmospheres of these planets to find traces of life (e.g. \citealt{Hu2012}; \citealt{Tian2014}). When performing abundance analysis of exoplanet atmospheres, observations from multiple transits is highly advantageous. Around solar-like dwarfs the habitable zone is located around 1~AU, giving year-long periods that make the acquisition of spectra from multiple transits very time consuming. Around M dwarfs the habitable zone is much tighter owing to their lower luminosity, and transit spectra of these planets can be observed multiple times per year. Another advantage of observing M dwarfs is that it is easier to discover Earth-sized planets around these smaller stars. The smaller mass and radius difference between an M dwarf and an Earth-sized planet will give larger velocity shifts and deeper transit depths compared to solar-like hosts. Previous studies have shown that the probability of finding small planets and planets within the habitable zone around M dwarfs is relatively high. For example, \citet{Dressing2013} used a sample of 3897 dwarf stars with temperatures below 4000~K from the Kepler Input Catalog and found an occurrence rate of 0.5-4~R$_E$ sized planets to be 0.90$^{+0.04}_{-0.03}$ per star. For Earth-size planets (0.5-1.4~R$_E$) they found an occurrence rate of 0.51$^{+0.06}_{-0.05}$ per star, and 0.15$^{+0.13}_{-0.06}$ per star for Earth-size planets within the habitable zone.

Shortly after the first exoplanet was discovered \citep{Mayor1995} it was found that solar-like stars hosting giant planets showed a strong trend towards being more metal-rich than   the Sun \citep{Gonzalez1997}. This result was later  confirmed by several other studies (e.g. \citealt{Santos2004, Valenti2005, Laws2003, Gonzalez2001, Gonzalez2007}). For lower mass planets (Neptunian or super-Earths) this metallicity trend seems to vanish (e.g. \citealt{Sousa2008, Sousa2011, Ghezzi2010}). Simulations based on core accretion models reproduce similar trends with metallicity (e.g. \citealt{Ida2004a, Ida2004b, Kornet2005, Mordasini2008}), leading to the conclusion that metallicity plays a crucial role for planet formation around solar-like stars.

The first study that investigated a possible metallicity correlation for M dwarfs was done by \citet[hereafter Bo05]{Bonfils2005a}. Their sample contained only two known planet hosts (GJ~876 and GJ~436). Both were found to have solar metallicity, but because of the limited sample the authors made no conclusion regarding a planet-metallicty correlation. Later \citet[hereafter JA09]{Johnson2009} applied the photometric calibration by Bo05 to seven M dwarf hosts, which gave a subsolar mean metallicity. This seemed to indicate the opposite metallicity trend to FGK hosts. However, both JA09 and later \citet[hereafter Ne12]{Neves2012} found that the calibration by Bo05 tends to underestimate the metallicity. This will be discussed  in section \ref{prev_meh}. As more planets have been discovered around M dwarfs and abundance determinations are becoming more precise, the majority of recent studies point toward a metal enhancement also for M dwarfs hosting giant planets \citep{Bonfils2007, JA09, Schlaufman2010, Terrien2012, Neves2013}. However, there is still a large spread in metallicity of M dwarfs derived by different methods, and a more accurate method is needed. With this work we aim to provide the  most accurate method to date for determining the metallicity of individual M dwarfs. If applied to a larger sample of M dwarfs, we hope this can provide further insight to the planet-metallicity correlation and planet formation theory.

\section{Sample selection and observations}

The sample was selected to include a number of binaries containing M dwarfs with an F, G, or K dwarf companion. In addition, a number of individual M dwarfs, mainly known planet hosts, were included in the observational programme. Target selection was based on information from the catalogue of nearby wide binary and multiple systems \citep{Poveda1994} and from the Interactive Catalog of the on-line Extrasolar Planets Encyclopaedia \citep{Schneider2011}, as well as from a programme searching for stellar companions of exoplanet host stars \citep{Mugrauer2004, Mugrauer2005, Mugrauer2007}.

\begin{table}
\caption{Observational information.} 
\label{observation}
\centering
\begin{tabular}{lcc}
\hline\hline
Target & Observational period & S/N ratio \\ \hline
HIP~12048 A             & 82,84                 & 450$^\dagger$         \\
HIP~12048 B             & 82,84                 & 30$^\dagger$  \\
GJ~527 A                & 82                    & 500   \\
GJ~527 B                & 82                    & 50    \\
GJ~250 A                & 82                    & 80            \\
GJ~250 B                & 82                    & 100   \\
HIP~57172 A             & 82                    & 120   \\
HIP~57172 B             & 82                    & 150   \\
GJ~176                  & 82                    & 70            \\
GJ~317                  & 84                    & 100   \\
GJ~436                  & 84                    & 140   \\
GJ~581                  & 84                    & 110   \\
GJ~628                  & 84                    & 130   \\
GJ~674                  & 84                    & 140   \\
GJ~849                  & 82,84                 & 90,120        \\
GJ~876                  & 84                    & 100   \\
\hline
\end{tabular}
\noindent
\begin{flushleft}
{\bf Notes.} $\dagger$ S/N was calculated from the combined spectra from observational period 82 and 84 for target HIP~12048.
 \end{flushleft}
\end{table}

\begin{table*}
\caption{Our sample, containing four FGK+M binaries and eight single M dwarfs.} 
\label{sampleinfo}
\centering
\begin{tabular}{lcccccccccc}
\hline\hline
Target & Sp. Type & $V$ & $K\rm_s$ & $\pi$ [mas] & Ang. sep. & $v$~sin$i$ [km/s] & log(L$_{\rm X}$/L$_{\rm bol}$)  & Planets & References \\ \hline
HIP~12048~A             & G5            & 6.816         & 5.266                         & 25.67           & 6.20''                & 1.9                           &                                       & yes             & 1, 10, 21, 31         \\
HIP~12048~B             & M2.5          &                       & 8.766                 & 25.67           & 6.20''                &                               &                                       &                         & 5                             \\
GJ~527~A                & F6                    & 4.480                 & 3.36$^\dagger$  & 64.03         & 1.80''                & 15                            & $-$6.14$^\ddagger$      & yes           & 2, 10, 19, 22, 34             \\
GJ~527~B                & M2            & 11.1          &                               & 64.03           & 1.80''                &                               &                                       &                         & 3, 12                 \\
GJ~250~A                & K3.5          & 6.581                 & 4.107                 & 114.81          & 58.70''               & 1.8                   &                                       & no                      & 6,10, 21                      \\
GJ~250~B                & M2.3          & 10.032                & 5.723                         & 114.81          & 58.70''               & $<$ 2.5                       &                                       &                         & 5, 10, 14             \\
HIP~57172~A             & K2            & 8.215                 & 6.147                 & 34.24           & 73.10''               & 0.7                   &                                       & yes             & 4, 10, 18, 29         \\
HIP~57172~B             & M0-M1                 & 10.604                & 7.107                   & 34.24         & 73.10''               &                               &                                       &                         & 6, 10                 \\
GJ~176                  & M2.2          & 9.951                 & 5.607                         & 107.83          &                       & $<$ 0.8                       & $-$4.46                         & yes           & 4, 11, 17, 27, 33             \\
GJ~317                  & M3.5          & 12.00                 & 7.028                         & 65.3            &                       & $<$ 2.5                       & $-$4.57                         & yes           & 7, 9, 14, 28, 35              \\
GJ~436                  & M2.8          & 10.663                & 6.073                         & 98.61           &                       & 1.0 $\pm$ 0.9         & $-$6.36                               & yes             & 4, 10, 19, 23, 35             \\
GJ~581                  & M3.2          & 10.567                & 5.837                         & 158.79          &                       & 0.4 $\pm$ 0.3         & $<-$4.57                      & yes             & 4, 11, 19, 25, 35             \\
GJ~628                  & M3.6          & 10.075                & 5.075                         & 232.98          &                       & 1.5                   & $-$5.16                               & no                      & 4, 11, 20, 32         \\
GJ~674                  & M2.5          & 9.407                 & 4.855                         & 220.24          &                       & $<$ 1                 & $-$3.80                               & yes             & 4, 11, 13, 26, 35             \\
GJ~849                  & M3.1          & 10.366                & 5.594                 & 116.05          &                       & $<$ 2.5                       & $-$6.73                         & yes           & 8, 11, 16, 24, 35             \\
GJ~876                  & M3.7          & 10.179                & 5.010                         & 213.28          &                       & 0.16                  & $-$5.11                               & yes             & 4, 11, 15, 30, 35             \\
\hline
\end{tabular}
\noindent
\begin{flushleft}
{\bf References.} \emph{Spectral types}: 1.~\citet{Eggen1960}, 2.~\citet{Gray2003}, 3.~\citet{Gray2006}, 4.~\citet{Mann2015}, 5.~\citet{Mason2001}, 6.~\citet{Mugrauer2007}, 7.~\citet{Reid1995}, 8.~\citet{Torres2006}. \emph{V-band photometry}: 9.~\citet{Bessel1990}, 10.~\citet{Kharchenko2001}, 11.~\citet{Koen2010}, 12.~\citet{Turon1993}. All K$_S$ values are from 2MASS \citet{Cutri2003}. \emph{Parallax}: All parallax values are from Hipparcos \citep{vanLeeuwe2007}, which exception of GJ~317 that is taken from \citet{Anglada-Escude2012}. \emph{Angular separation}: Values taken from the Washington Double Star Catalog \citep{Mason2001}. \emph{Rotation velocities:} 13.~\citet{Bonfils2007}, 14.~\citet{Browning2010}, 15.~\citet{Correia2010},  16.~\citet{Delfosse1998}, 17.~\citet{Forveille2009}, 18.~\citet{Lovis2005}, 19.~\citet{Marcy1992}, 20.~\citet{Reiners2007b}, 21.~\citet{Valenti2005}. \emph{Planet discoveries:} 22.~\citet{Butler1997}, 23.~\citet{Butler2004}, 24.~\citet{Butler2006}, 25. \citealt{Bonfils2005b}, 26.~\citet{Bonfils2007}, 27.~\citet{Endl2008}, 28.~\citet{Johnson2007}, 29.~\citet{Lovis2005}, 30.~\citet{Marcy1998}, 31.~\citet{Marcy2000}. \emph{Activity}: 32.~\citet{Delfosse1998}, 33.~\citet{Forveille2009}, 34.~\citet{Lovis2005}, 35.~\citet{Poppenhaeger2010}.\\
\noindent
\newline
{\bf Notes.} $\dagger$ No value for GJ~527~B was found in the 2MASS catalogue, and we believe that the value in 2MASS contains both components due to the small separation in this binary. $\ddagger$ The companions are unresolved in the XMM-Newton observations. Additionally, no V~band magnitude could be found in the literature for HIP~12048~B and no previous determination of $v$~sin$i$ was found for HIP~12048~B, GJ~527~B, and HIP~57172~B.
 \end{flushleft}
\end{table*}

Spectra of all targets were obtained in the $J$ band (1100-1400~nm) with the CRIRES spectrograph at ESO-VLT \citep{Kaeufl2004}. The observations were carried out in service mode during periods 82 (1 October 2008 to 31 March 2009) and 84 (1 October 2009 to 31 March 2010). For all targets a slit-width of 0.4'' was used, giving a resolving power of R~$\sim$~50,000. Each target was observed in four wavelength intervals centred at 1177, 1181, 1204, and 1258~nm in period 82, and 1177, 1205, 1258, and 1303~nm in period 84. Observational information is found in Table \ref{observation} and some of the properties of the stars in our sample are given in Table \ref{sampleinfo}. Some of the targets, including two wide K+M binaries GJ~250 \citep{Dommanget2002, Newton2014} and HIP~57172 \citep{Mugrauer2007}, were previously  analysed by \citet[hereafter On12]{Onehag2012}. An additional binary, GJ~105, was included in On12. However, our analysis of the M dwarf in this binary did not converge to a unique solution. As further investigation is needed, this binary is not included in our paper. In addition to the sample in On12, two close FG+M binaries (separation $\leq$20") observed in the same programme were added to the analysis in this paper: HIP~12048 \citep{Mugrauer2005, Mugrauer2007} and GJ~527 \citep{Patience2002}. These two close binaries were observed simultaneously by orienting the slit at the spectrograph to cover both components. The additional data reduction required for these two binaries is described in section \ref{reduction}.

The choice of $J$ band was motivated by the greatly reduced number of stellar molecular lines in this part of the infrared, and also by the presence of sufficient numbers of atomic lines to perform a reliable metallicity determination.\footnote{We used 15-20 lines to determine the metallicity (see Sect. 4.3.4). For comparison, \citet{Valenti2005} used around 50 lines in the optical region in the analysis of the F dwarfs in their sample to determine several stellar parameters simultaneously.} However, only data from chips 2 and 3 were used for the analysis since chips 1 and 4 on the CRIRES spectrograph were heavily vignetted and contaminated by overlapping orders. In addition to the science observations, a rapidly rotating early-type star was observed for each target in order to remove the telluric lines from the science spectra.

\section{Methodology}

Metallicity determination of M dwarfs is a spectroscopic challenge. These cool dwarfs are intrinsically faint and their low surface temperatures allows the existence of diatomic and triatomic molecules in the photospheric layers. Until recently, observations with high spectral resolution were limited to the optical. In this regime, these molecules give rise to many millions of weak lines, leaving almost no unblended atomic lines and making accurate continuum placement nearly impossible (e.g. \citealt{Gustafsson1989}). With new instruments operating in the infrared, for example  TripleSpec at Palomar \citep{Herter2008}, SpeX at IRTF \citep{Rayner2003}, and CRIRES at VLT \citep{Kaeufl2004}, the situation for M dwarfs  dramatically improved. In certain bands in the infrared, stellar molecular lines are few \citep{Onehag2012}, which greatly improves the reliability of abundance determinations based on atomic lines.

Fundamental parameters, such as [M/H], $T_{\rm eff}$, and log~$g$ for FGK dwarfs are today determined with high precision through synthetic spectra or equivalent width fitting, using software like MOOG \citep{Sneden1973} or SME \citep{Valenti1996}. The first attempt to determine the chemical abundances of M dwarfs was done by \citet{Mould1976a, Mould1978}. Since then there has been tremendous progress both  observationally and theoretically. Accurate abundance analysis requires detailed knowledge of a variety of physical data for line absorption and continuum formation, as well as a detailed physical description of the environment where the spectral lines and continuum is formed, i.e. a model atmosphere.

Early model atmospheres for M dwarfs were developed in the 1960-70s (e.g. \cite{Tsuji1966, Auman1969, Mould1976b}). Today, the most commonly used models are MARCS \citep{Gustafsson2008}, ATLAS \citep{Castelli2004}, and PHOENIX \citep{Hauschildt1999b}. In the range of effective temperatures typical for M dwarfs (2700 $<$ $T_{\rm eff}$ $<$ 4000~K), molecules are as important as atomic species. While the most relevant molecules were identified several decades ago (e.g. \citealt{Russell1934}; \citealt{deJager1957}) molecular data is still one of the limiting factors when working with cool dwarfs. Dominating molecules are TiO and H$_2$O, but also other oxides (e.g. VO and CO) and several hydrides (e.g. CaH, MgH, SiH, OH, CH, FeH, CrH) are important. Great improvements in the molecular line data have been made in the last decades; H$_2$O (e.g. \citealt{Miller1994, Barber2006}), TiO \citep{Plez1998}, CrH \citep{Burrows2002}, and FeH (e.g. \citealt{Dulick2003}; Plez, private communication, 2012) to mention some. Another complication when modelling cool stellar atmospheres is the occurrence of grains, especially important for $T_{\rm eff}$ $<$~3000~K, which is where the phase transition is believed to occur for M dwarfs. The first molecule to condensate is ZrO$_2$, followed by Al$_2$O$_3$ and CaTiO$_3$. However, for temperatures above 2600~K simulations indicate that the grain formation is not efficient enough to affect the spectral energy distribution, e.g. \citet{Allard2013}. Grain formation is therefore believed not to affect early and mid M-type dwarfs, which are analysed in this work.

The majority of today's abundance analyses, particularly for M dwarfs, are based on plane-parallel model atmospheres and line formation assuming local thermodynamic equilibrium (LTE). For solar-like dwarfs and giants it has been shown that these assumptions are not always valid and large differences in derived abundances can be found for certain species and stellar parameters (mainly metal-poor stars). A good review is given in \citet{Asplund2005}. Only a few studies have investigated non-LTE effects for M dwarfs, mainly since previous work show that deviations from LTE in general increase with higher effective temperature, lower surface gravity, and lower metallicity. This indicates that the non-LTE corrections needed for M dwarfs should be limited. On the other hand, M dwarfs have much lower electron temperature and very low electron density. In solar-like stars collisions with electrons is an efficient restoring force toward LTE, while collision with H$_2$ and He is less efficient. Therefore, the possibility exists that the lower electron collision rate in M dwarfs may increase the non-LTE effects despite the high surface gravity and low effective temperature (e.g. \citealt{Hauschildt1997}). However, owing to lack of determined non-LTE factors, we will not include any non-LTE corrections when deriving the metallicity in this work.

Previous studies have shown that most M dwarfs are magnetically active with non-negligible magnetic field strengths (e.g. \citealt{Johns-Krull1996, Berger2006b}), but in most  abundance determinations the effects of magnetic fields are yet to be included. Stars are predicted to spin down and become less magnetically active with time (e.g. \citealt{Noyes1984, Rutten1987}). \citet{Reiners2012} analysed a sample of 334 M dwarfs of spectral types M0-M4.5, a range that covers all subtypes included in our work. They found that only about 5\% of the M0-M2 dwarfs show H$\alpha$ emission. For subtype M3 about 25\% were found to be active, and the percentage of active stars was found to be rapidly increasing towards later subtypes. However, for the most active M dwarfs (M3-M4.5) almost 75\% have a $v$~sin$i$ greater than 3~km~s$^{-1}$.

To minimise the influence of magnetic fields in our analysis, targets with low $v$~sin$i$ were selected. As shown in  Table \ref{sampleinfo}, our sample mainly contains M0-M3.5 dwarfs with $v$~sin$i$ $<$ 3~km~s$^{-1}$. For all individual M dwarfs we also found estimates of their activity level and most have log(L$_{\rm X}$/L$_{\rm bol}$)~$<$~$-$4.5. An exception is GJ~674, which is the youngest star in our sample. GJ~674 has an estimated age of 0.1-1 Gyr \citep{Bonfils2007} and shows modest activity with a measured log(L$_{\rm X}$/L$_{\rm bol}$) of $-$3.80 \citep{Poppenhaeger2010}. The other M~dwarf that may exhibit higher magnetic activity is the relatively young GJ~527~B, with an estimated age of 1.3-3.1 Gyr \citep{Saffe2005}. Both components in this binary have a faster rotation than our remaining sample. GJ~527~A have a previously determined $v$~sin$i$ of 15~km~s$^{-1}$ \citep{Marcy1992}, and we estimate GJ~527~B to have a $v$~sin$i$ of 5~km~s$^{-1}$, see section 5.1.2. Unfortunatley, no activity estimate is  available for GJ~527 B since the binary components are unresolved by ROSAT \citep{Hunsch1999} and XMM-Newton \citep{Jansen2001}, and the log(L$_{\rm X}$/L$_{\rm bol}$) value of $-$6.11 is believed to mainly reflect the activity level of the much brighter GJ~527~A.

\subsection{Previous metallicity determinations}
\label{prev_meh}
During the last decades there has been progress to determine the metallicity of M dwarfs with better precision using different methods and observational datasets. Since M dwarfs have such complicated spectra, most studies concentrate on FGK+M binaries. Both components are assumed to have been formed from the same molecular cloud and therefore to exhibit very similar overall chemical composition. Since the warmer solar-like stars have less complex spectra, they provide a more reliable abundance determination. For cooler dwarfs, a common method has been to establish empirical calibration relations, which  can then be used to determine a star's metallicity. There are two approaches to establishing these relations, one photometric and one spectroscopic. The most used photometric calibration relates the star's position in a colour-magnitude diagram to its metallicity. One of the first relations was found by Bo05. They used the mass-luminosity relationships of $V$, $J$, $H,$ and $K$ bands found by \citet{Delfosse2000}, and argued that the flux distribution of the visible versus infrared is correlated with the metallicity. Metallicity has been found to have two effects on the spectra. The higher metallicity will decrease the overall bolometric luminosity and the increased line opacity of the TiO and VO bands will redistribute a part of the flux from the visible to the near-IR. These two effects will work together and lower the flux in the visible bands, while models predict that the redistributed flux into the infrared will counteract the effect of the lowered bolometric flux making the infrared bands largely insensitive to the metallicity of the star. \citet{Johnson2009} found two issues with the calibration by Bo05. First, when applying the photometric calibration to M dwarfs known to host planets, a mean metallicity of $-$0.11~dex was found, i.e. the opposite to hosts being metal enhanced as established for solar-like stars. Second, the volume-limited sample of 47 M dwarfs was found to have a mean metallicity of $-$0.17~dex, which is 0.09~dex lower than their sample of FGK dwarfs. \citet{Bonfils2005a} argued that the difference in mean metallicity between the samples may be due to the longer lifetime of the M dwarfs, while JA09 argued that it was due to the M dwarf sample itself. Analysing a larger volume-limited sample, which included six selected high-metallicity FGK+M binaries, they derived a new photometric relation. Using this calibration they found a better agreement with the mean metallicity of the FGK sample and derived a mean metallicity of +0.16~dex for the M dwarf hosts. In the work by \citet{Schlaufman2010} the kinematic properties of the stars were considered to account for the various populations of our Galaxy. They also used of stellar evolution models by \citet{Baraffe1998} to guide their parameterisation of the colour-magnitude space to further improve on the previous calibrations. \citet{Neves2012} compared the performance of those three photometric calibrations using a sample of 23 wide FGK + M dwarfs binaries with spectroscopically determined metallicity for all primaries. Compared to their calibration they found that the calibration by Bo05 tends to underestimate the metallicity, while JA09 overestimated it except for the most metal-rich M dwarfs. They also made a slight update of the coefficients in the calibration by \citet{Schlaufman2010}, which reduced the spread between the values determined with the photmometric calibration relative to the spectroscopically determined values.

For the spectroscopic approach, different molecular and/or atomic features were used as an indicator of the metallicity. One of the more precise calibrations was found by \citet{Rojas-Ayala2010, Rojas-Ayala2012} who used moderate-resolution $K$ band spectra. They showed that the equivalent width of the Ca~I lines (2.261, 2.263, and 2.265~$\mu$m) and Na~I lines (2.206 and 2.209~$\mu$m), together with their H$_2$O-K2 index, can be used to derive the metallicity of M dwarfs with a precision of 0.12~dex. Other spectroscopic calibrations were derived by e.g. \citet{Woolf2006, Woolf2009}, and  \citet{Dhital2012} in the visible, \citet{Terrien2012} and \citet{Newton2014} in the infrared and \citet{Mann2013a, Mann2014, Mann2015} using metallicity sensitive features in both the visible and the infrared.

The methods described above all use empirical relationships, while very few studies have explored model-dependent methods to derive the stellar parameters, e.g. effective temperature, surface gravity, and metallicity, as is standard for solar-like stars. \citet{Valenti1998} pioneered this technique for M dwarfs. They used spectral fitting to a combination of molecular (TiO) and atomic (Ti~I, Fe~I) features in an optical high-resolution spectrum, in order to determine $T_{\rm eff}$, log~$g$, and [M/H] for the M dwarf Gliese 725~B. Improvements and reliability testing of the method was done by \citet{Bean2006b}. They analysed five nearby, common proper motion, visual FGK+M binaries, where the warmer companion was used as a reference for the derived metallicity. The same wavelength region as in \citet{Valenti1998} was used, but the analysis included several improvements such as improved line/molecular data, and their fit also included more atomic lines. However, the analysis was still affected by the presence of molecular lines, especially by TiO, which limited the metallicity determination to strong Ti and Fe lines. \citet{Onehag2012} observed several M dwarfs and K+M binaries in the $J$ band and showed that the amount of blends with molecular lines is highly reduced in the infrared. This opens up the possibility of using lines from several atomic species and lines with different strengths. Our paper is a continuation of that study, and we refer to On12 for any details of the analysis not explicitly described in the following sections.

\section{Analysis}

\subsection{Data reduction}
\label{reduction}
We reduced the raw data using a custom IDL pipeline developed by \citet{Bean2010}. The pipeline is based on standard techniques and includes bias subtraction, non-linearity correction, and flat fielding. For each observation, several frames in two nod positions were taken. All frames from the same nod position were co-added before proceeding to the flux extraction. For consistency, the same procedure was applied to the telluric standard stars. Frames from different nod positions cannot  be combined as the slit curvature and chip rotation caused a notable wavelength offset between the nod positions. Hence, two 1D spectra per target and observation were extracted and later combined in 1D after proper wavelength calibration. 

The pipeline  uses an optimum extraction algorithm based on \citet{Horn1986}. Iterative masking is applied to remove pixels where the actual flux deviates more than 10 $\sigma$ from the expected flux, derived from a simple box extraction times the averaged spatial profile.

An optical ghost on chip two of the setting centred at 1177\,nm contaminates one of the two nod positions and renders a small part of the spectrum in this nod position useless. The affected region was excluded from the combined 1D spectrum and only flux from the unaffected nod position  is used for the final spectrum. 

One target, GJ~527, showed some blending between the two components of this close binary system, i.e. the extended halo of the primary does affect the secondary. A simple de-blending routine was implemented to clean up the spectra. The unaffected wing of the stellar PSF (the opposite of the respective companion) was used to estimate the flux contamination as a function of wavelength. At the cost of slightly lower S/N (due to the loss of information and, hence, poorer outlier rejection) the de-blending recovers the flux of each component without the need for multiple iterations of the procedure.

\subsection{Further data processing}
Accurate abundance determinations require a reliable continuum placement. Each spectrum was rectified individually by fitting a polynomial of second degree to the data points determined to belong to the continuum for that particular spectrum. In order to define which parts of the data belonged to the continuum, each spectrum was divided into several short wavelength intervals of $\sim$1~\AA, where each spectrum is approximately 60~\AA. For each interval the standard deviation was calculated. The mean of the three lowest standard deviation values was defined as the noise level of that particular spectrum and star. Subsequently, an iterative process was applied to the whole spectrum of each chip where data points with the lowest flux values were removed until the standard deviation of the remaining points was not larger than three times the noise level. 

The continuum placement is based on the assumption that no suppression of the continuum due to a haze of weak molecular lines exist. The main molecules identified as being present in the $J$~band are FeH, CrH, and H$_2$O. \citet{Onehag2012} used the line data from \citet{Burrows2002} to conclude that the observed spectral region only contains a few CrH lines and that no blends with the atomic lines were present. They also used the line data from \citet{Barber2006} to investigate the influence of water. By calculating several different atmospheric models with different effective temperatures, they concluded that for M dwarfs with effective temperatures of 3200~K the continuum is suppressed with only 2\% and at higher effective temperatures the water lines quickly disappear. Since all M dwarfs in our sample have an effective temperature above 3200~K, we concluded that the suppression of the continuum due to molecules is negligible. Several FeH lines are present in our spectra, but the majority are distinguishable from each other and do not form a pseudo-continuum.

The standard wavelength calibration procedure for CRIRES is based on thorium and argon lines, but in the infrared the number of Th or Ar lines is not sufficient to ensure the precise wavelength correction needed. An additional calibration was done using a polynomial fit to the telluric lines in a high-resolution solar spectrum by \citet{Livingston1991}. This extra calibration was made both for the spectra of the science and the telluric standard targets. After removal of the telluric features, all science spectra for a target taken at the same wavelength interval were co-added to increase the signal-to-noise (S/N) ratio. \citet{Onehag2012} already showed the consistency between period 82 and 84, hence overlapping data for HIP~12048 from both period 82 and 84 were co-added to increase the S/N of that M dwarf.

\subsection{Determination of stellar parameters}

Determination of the fundamental parameters was done with the software package Spectroscopy Made Easy (SME; \cite{Valenti1996}), version 423. Synthetic spectra are computed on the fly, based on a grid of model atmospheres and a list of atomic and molecular data. Stellar parameters that the user wishes to determine (in SME referred to as free parameters) are found by computing synthetic spectra with small offsets in different directions for this subset of parameters. The optimal fit is then determined through $\chi^2$ minimisation between the synthetic and observed science spectra. The synthetic spectra are calculated by interpolation from a grid of pre-calculated model atmospheres. In this work MARCS models were used, which are hydrostatic, plane-parallel LTE model atmospheres \citep{Gustafsson2008}. Additional information used by SME are line- and continuum masks. These are subsections of the spectrum defined by the user, and are used to control which part of the spectrum is to be used for the $\chi^2$ fit. In this work only a line mask was used, which we used to remove part of the spectra affected by bad pixels, line blends, or telluric features that had not been removed properly. Each observation with the CRIRES spectrograph covers a rather short wavelength range, and especially for the M dwarfs the spectra contain too few atomic lines to accurately determine the surface gravity, effective temperature, and metallicity simultaneously using spectral analysis. Examples of obtained spectra are shown in Appendix A, available in the online material. The surface gravity and the effective temperature were therefore determined prior to the metallicity determination (see Sections \ref{teff} and \ref{logg}).

\subsubsection{Line data}

Most of the atomic line data were taken from the VALD database \citep{Kupka2000, Heiter2008}, with some additional data from \citet{Melendez1999}. Line data for FeH were provided by Bertrand Plez (private communication, 2012)\footnote{The FeH line list will soon be available on the MARCS webpage (http://marcs.astro.uu.se)}, and were used to estimate the effective temperatures of the M dwarfs (see Section \ref{teff}), but not for the metallicity determination. The solar abundances by \citet{Grevesse2007} were used. 

Before the analysis, the oscillator strengths (log~$gf$) and van der Waals (vdW) broadening parameters in our line list were adjusted using a high-resolution solar spectrum from Kitt Peak \citep{Livingston1991}\footnote{ftp://nsokp.nso.edu/pub/Kurucz\_1984\_atlas/photatl/}. Adopted solar parameters were $T_{\rm eff}$ = 5777~K, log~$g$ (cm~s$^{-2}$) = 4.44, [M/H] = 0.00, and $v$~sin$i$ = 1.7~km~s$^{-1}$, and the solar chemical composition by \citet{Grevesse2007}. First the micro- ($\xi_t$) and macroturbulence ($\zeta_t$) were adjusted simultaneously, giving $\xi_t$~=~0.89~km~s$^{-1}$ and $\zeta_t$~=~3.86~km~s$^{-1}$. Thereafter, log~$gf$ and vdW were adjusted for all lines where the synthetic spectrum did not match the observed solar spectrum. An exception was made for the carbon, magnesium, and potassium lines where the vdW broadening parameters were derived from quantum-mechanical calculations, the so-called ABO theory \citep{Anstee1991, Barklem2000}. For these lines only log~$gf$ was adjusted. The whole procedure was repeated until no significant changes in the fit were observed. We note that changing the vdW parameter had little effect on the synthetic spectra, while significant changes in the oscillator strength were required. A difference compared with the values adopted by On12 was noted, and is believed to arise from a combination of a more recent version of SME and the usage of synthetic flux instead of synthetic disc-centre intensity spectra in the adjustment procedure by On12. The final list of atomic data used in this paper can be seen in Table \ref{gf_values}. (Available in the online version).

\subsubsection{Effective temperature}
\label{teff}

\begin{table*}
\caption{Estimated effective temperatures from different studies. }  
\label{teff_Mdwarfs}
\centering
\begin{tabular}{lccccc}
\hline\hline
Target  & Spectral type & Our work &  On12 & RA12 & Ma15 \\ \hline
HIP~57172 B             & M0-M1                 & $>$ 3900      & 3908          &                         &  \\
GJ~527 B                & M2            & 3425          &                       &                         &  \\
GJ~176                  & M2.2          & 3550                  & 3361          & 3581            & 3680  \\
GJ~250 B                & M2.3          & 3550                  & 3376          & 3569            & 3481 \\
HIP~12048 B             & M2.5          & 3225          &                       &                         & \\
GJ~674                  & M2.5          & 3350                  & 3305          &                         &  \\
GJ~436                  & M2.8          & 3400                  & 3263          & 3469            & 3479  \\
GJ~849                  & M3.1          & 3350                  & 3196          & 3601            & 3530  \\
GJ~581                  & M3.2          & 3350          & 3308          & 3534            & 3395  \\
GJ~317                  & M3.5          & 3375          & 3125          &                         &   \\
GJ~628                  & M3.6          & 3275                  & 3208          & 3380            & 3272  \\
GJ~876                  & M3.7          & 3250                  & 3156          & 3473            & 3247  \\
\hline
\end{tabular}
\begin{flushleft}
\noindent
\newline
{\bf Notes.} Estimated effective temperatures from three different methods. Column 1 contains estimated temperatures from the FeH line strength, Col. 2 contains adopted values by On12 based on the calibration developed by \citet{Casagrande2008} with some adjustments based on FeH line strength, and Col. 3 shows values obtained by RA12 using their $K$ band spectroscopic index. The last column shows the adopted values by Ma15 based on calculated bolometric flux and measured angular diameter.
\end{flushleft}
\end{table*}

Previous studies have shown that the effective temperature is the parameter causing the largest uncertainty when determining the metallicity of M dwarfs. For example, \citet{Bean2006b} found a difference of 0.12~dex in their derived metallicity when changing the effective temperature by 50~K. In order to assess the error due to the adopted effective temperature several methods were compared. In On12 the photometric calibrations by \citet{Casagrande2008}, their Eq. 6, was used together with twelve colour combinations ($V$, $R_c$, $I_c$, $J$, $H$, $K_s$) in order to determine the effective temperature. By comparing the synthetic spectrum with the line strengths of FeH, some of the temperatures were further adjusted if needed with +100 or +200~K. Some of the required photometric data were not available for GJ~527~B and HIP~12048~B; therefore, this method could not be used for these two stars. The binary components of GJ~527 were not resolved in 2MASS, and for HIP~12048~B no $V$-band observation is available. Values from previous studies of the M dwarfs in our sample, or calibrations not based on photometry, were therefore investigated. \citet{Rojas-Ayala2010} and  \citet[hereafter RA12]{Rojas-Ayala2012}, define an index from moderate resolution $K$-band spectra that correlates with effective temperature. The relation was inferred using solar metallicity BT-Settl model atmospheres \citep{Allard2010}. \citet[hereafter Ma15]{Mann2015} derive effective temperatures of M dwarfs from the Stefan-Boltzmann law and calculated angular diameters and bolometric fluxes. The precise angular diameters (uncertainty around 1.5\%) were determined using long-baseline optical interferometry, and the bolometric fluxes were calculated from fits of spectral templates to observed spectral energy distributions. About half of the M dwarfs in our sample overlap with stars in the sample of RA12 and Ma15.

\begin{figure*}
\begin{center}
\includegraphics[width=0.33\textwidth]{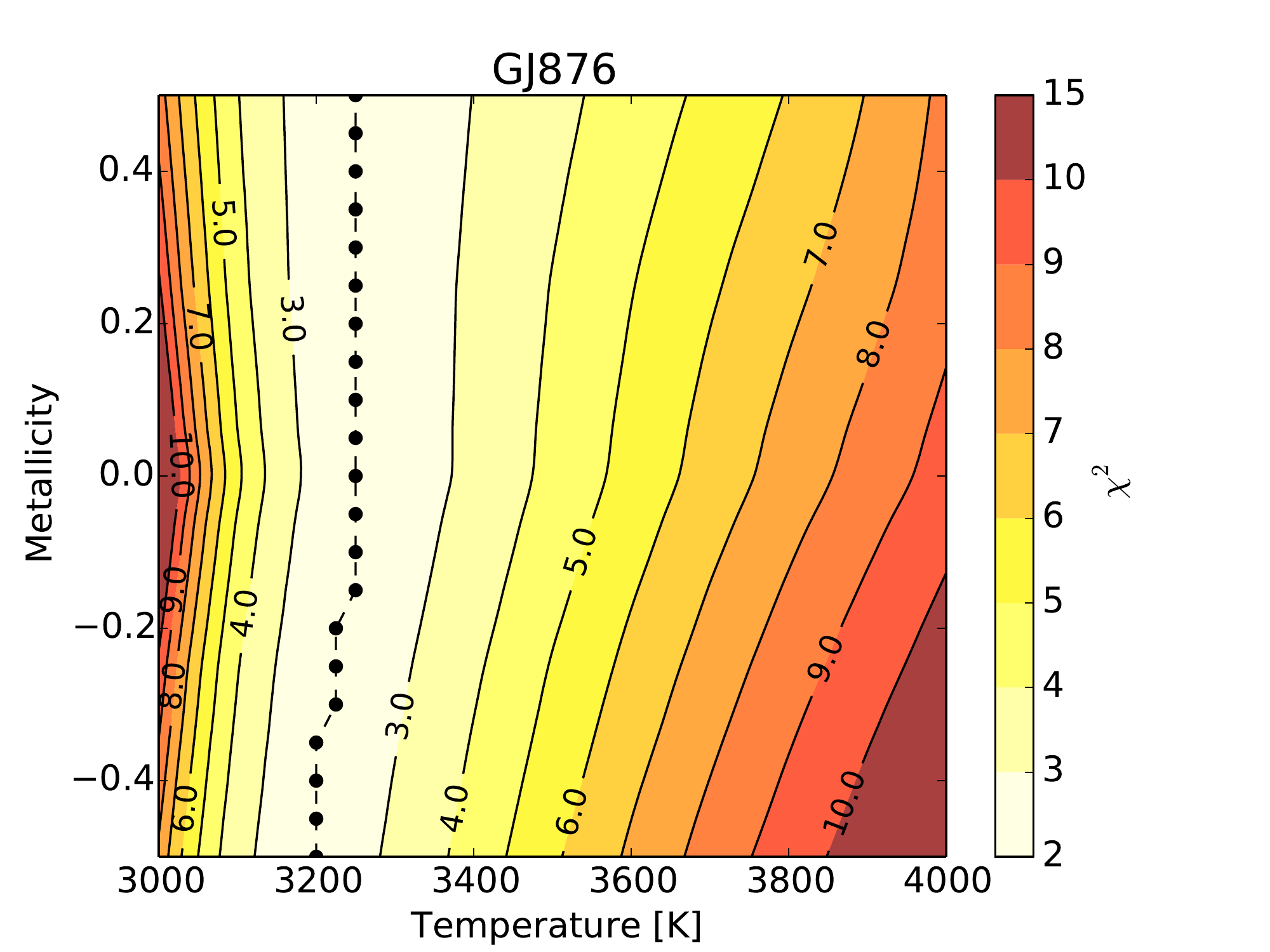}
\includegraphics[width=0.33\textwidth]{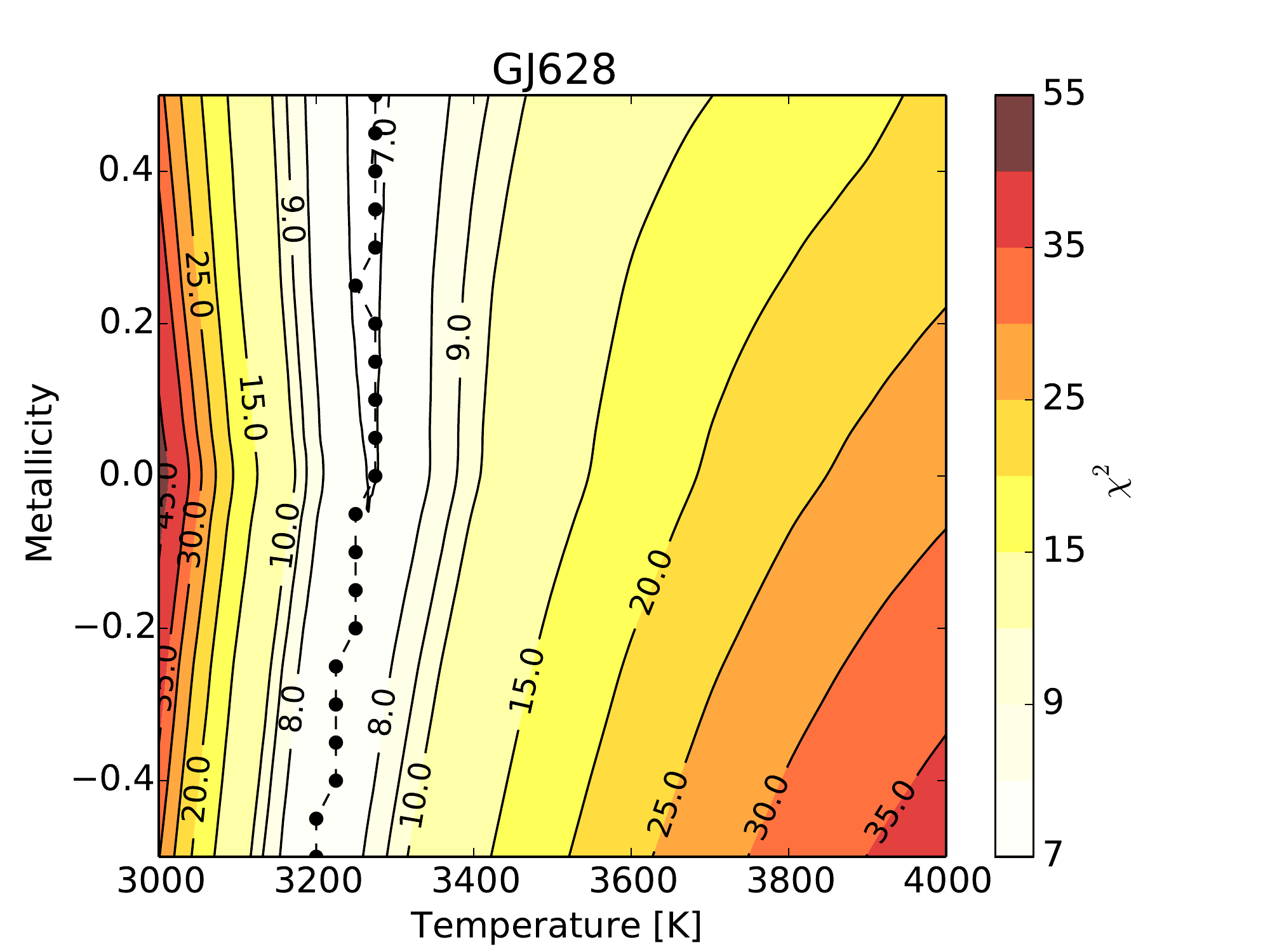}
\includegraphics[width=0.33\textwidth]{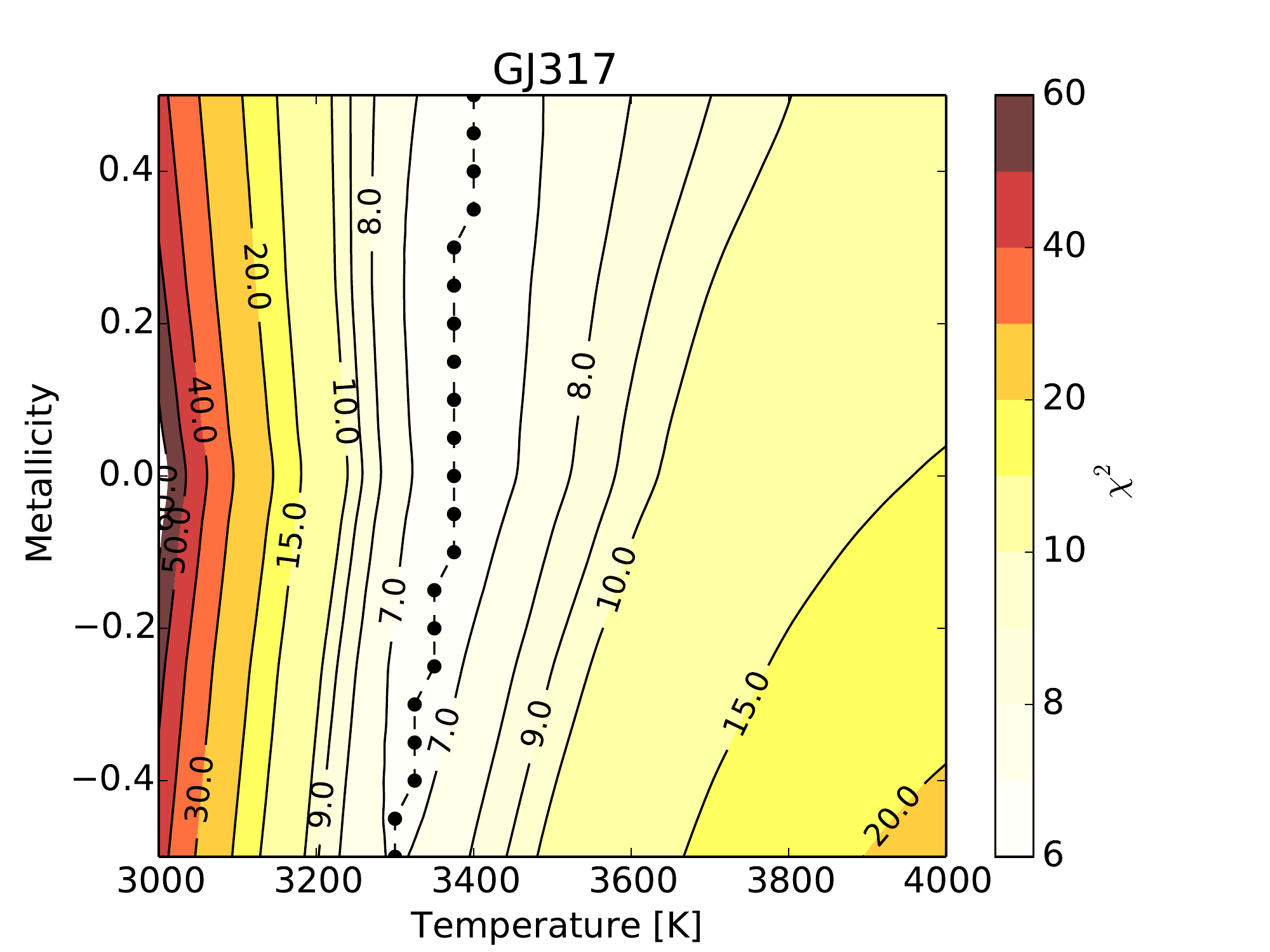}
\includegraphics[width=0.33\textwidth]{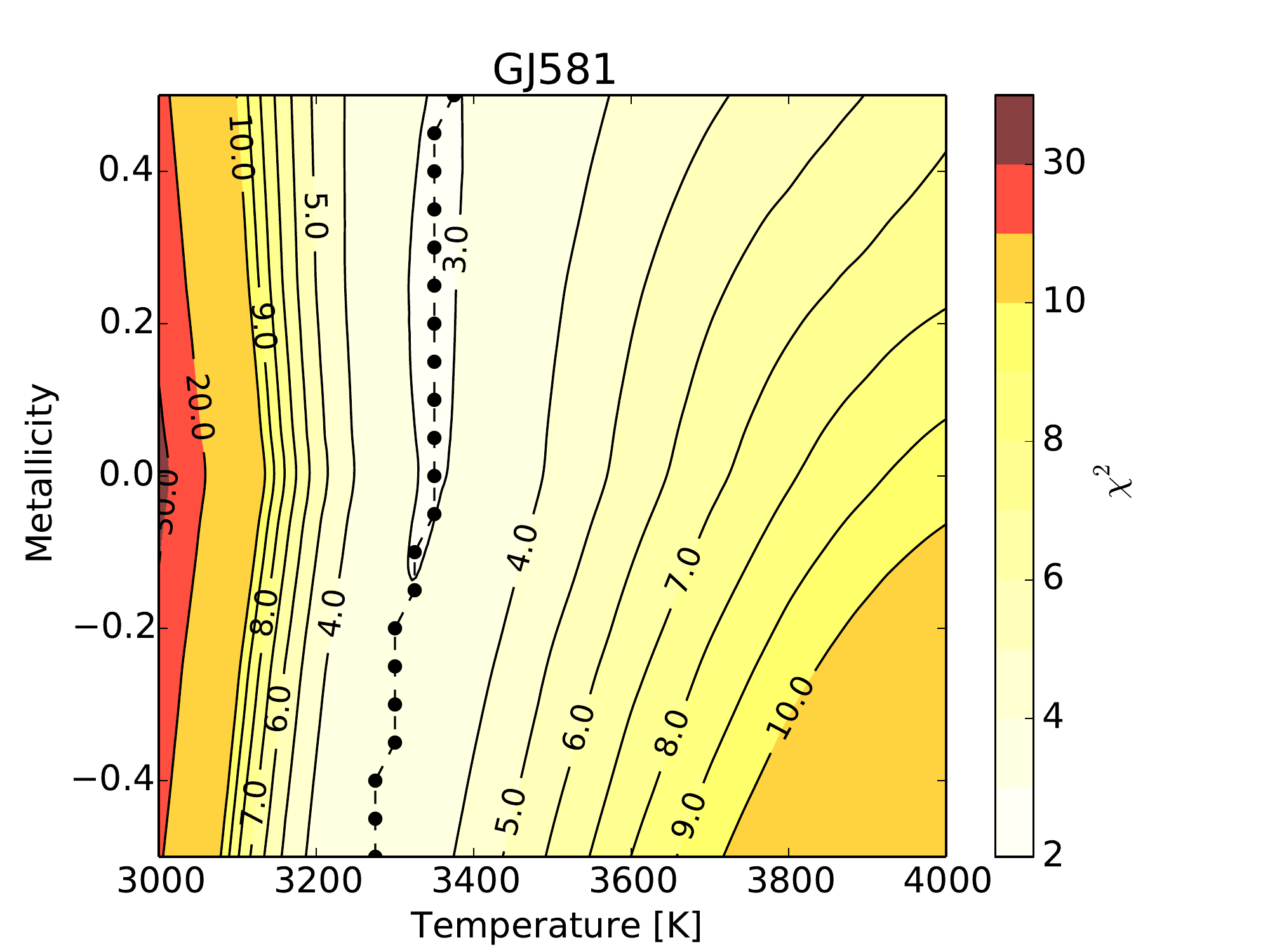}
\includegraphics[width=0.33\textwidth]{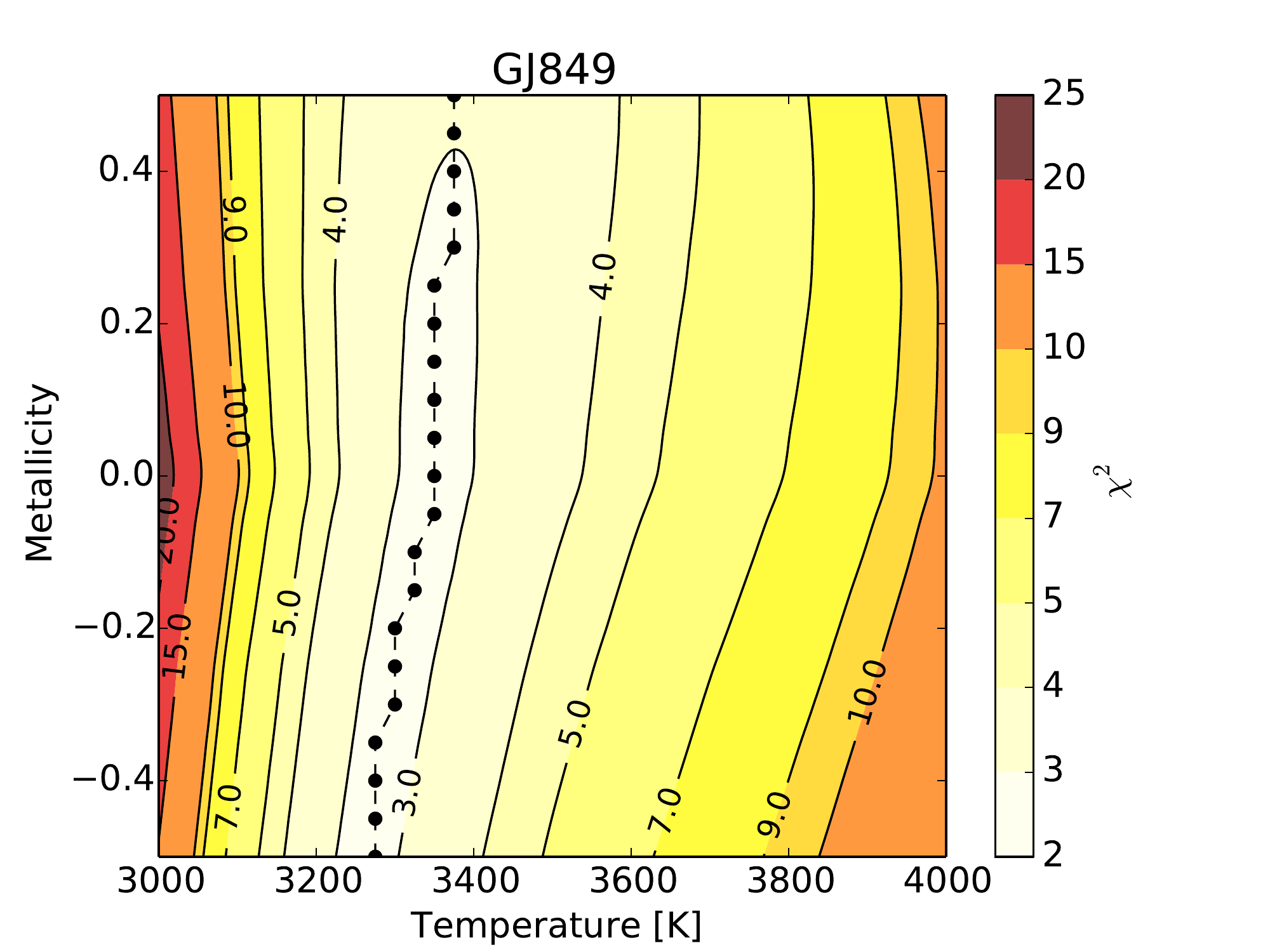}
\includegraphics[width=0.33\textwidth]{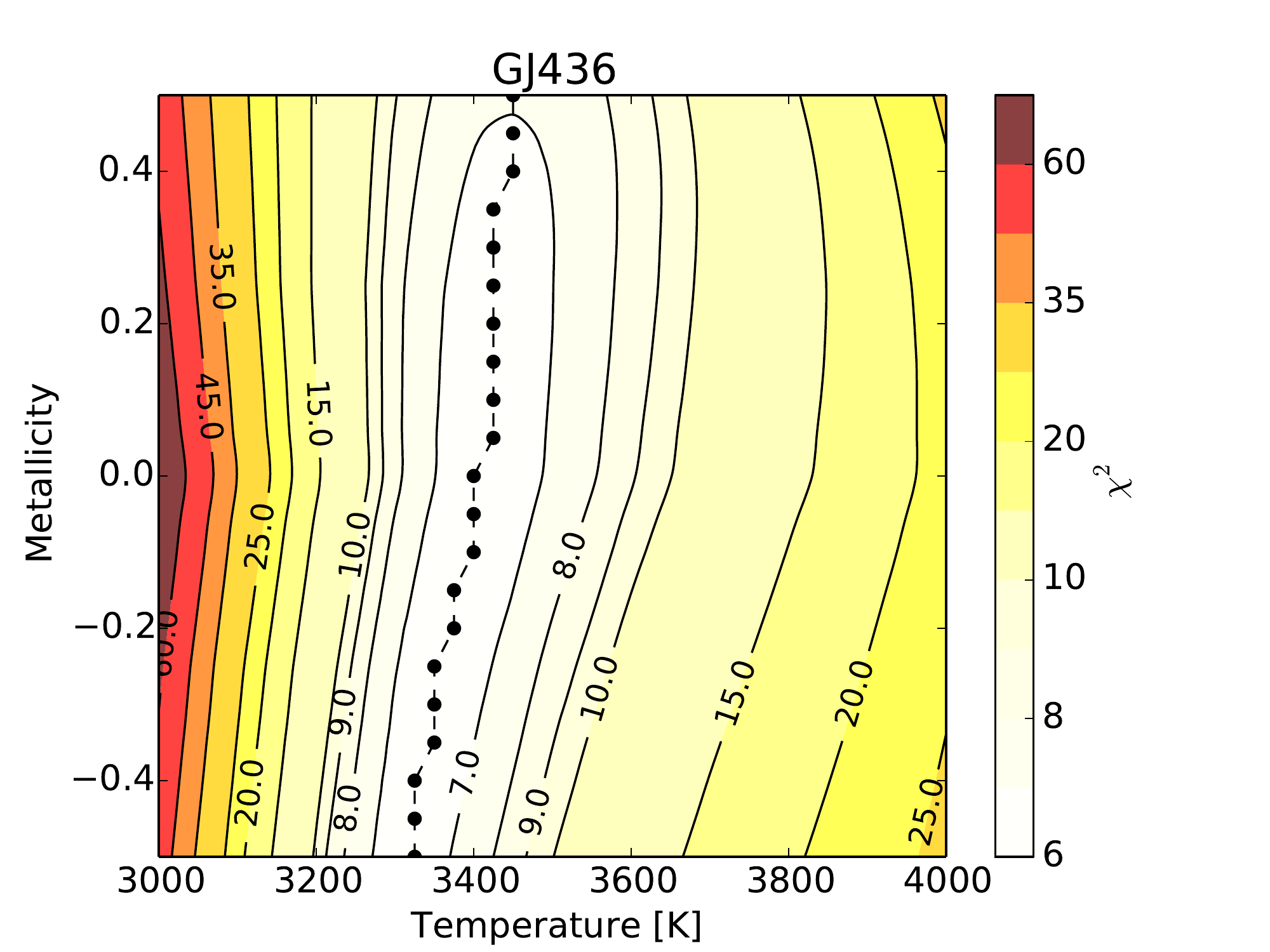}
\includegraphics[width=0.33\textwidth]{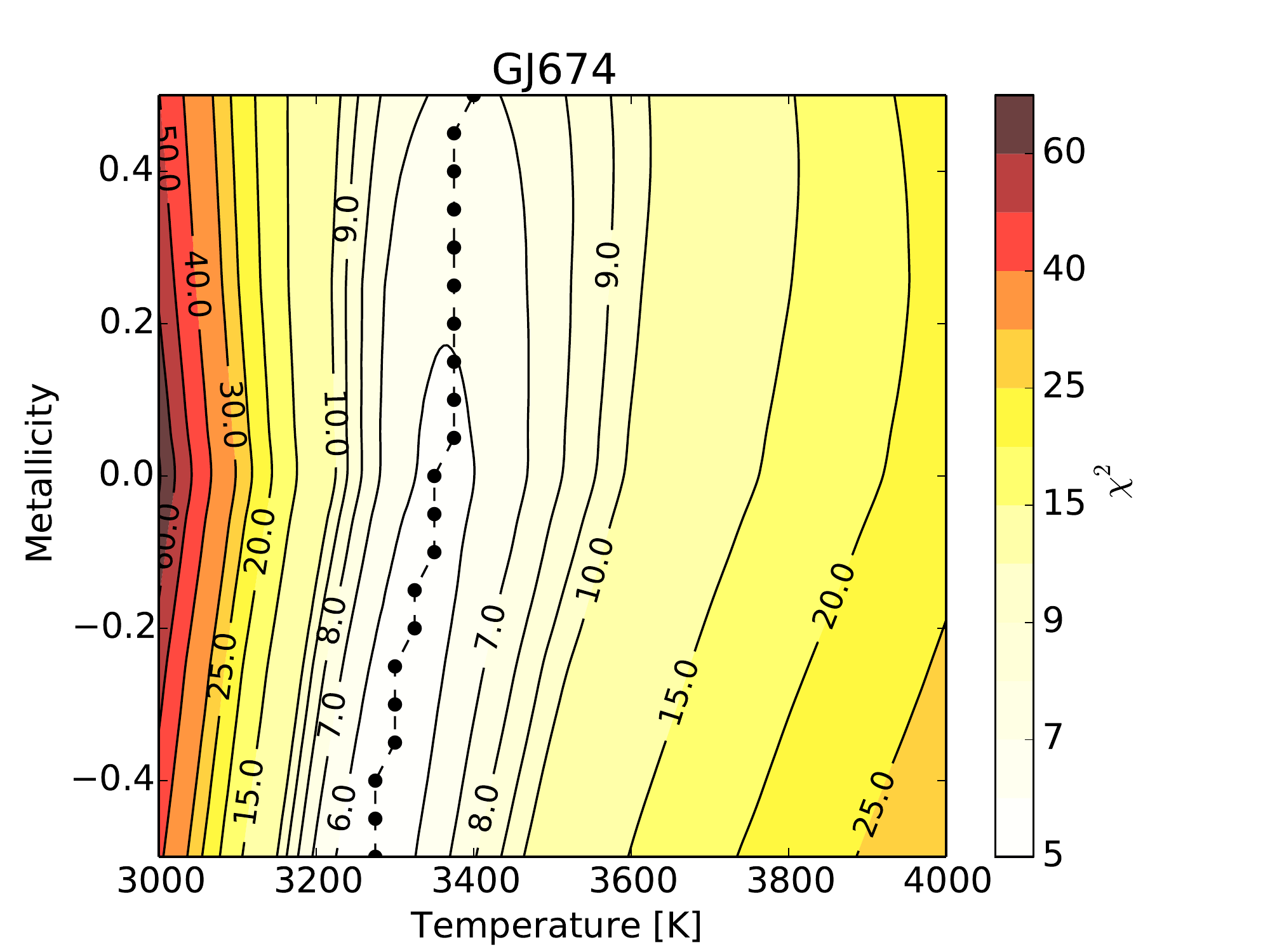}
\includegraphics[width=0.33\textwidth]{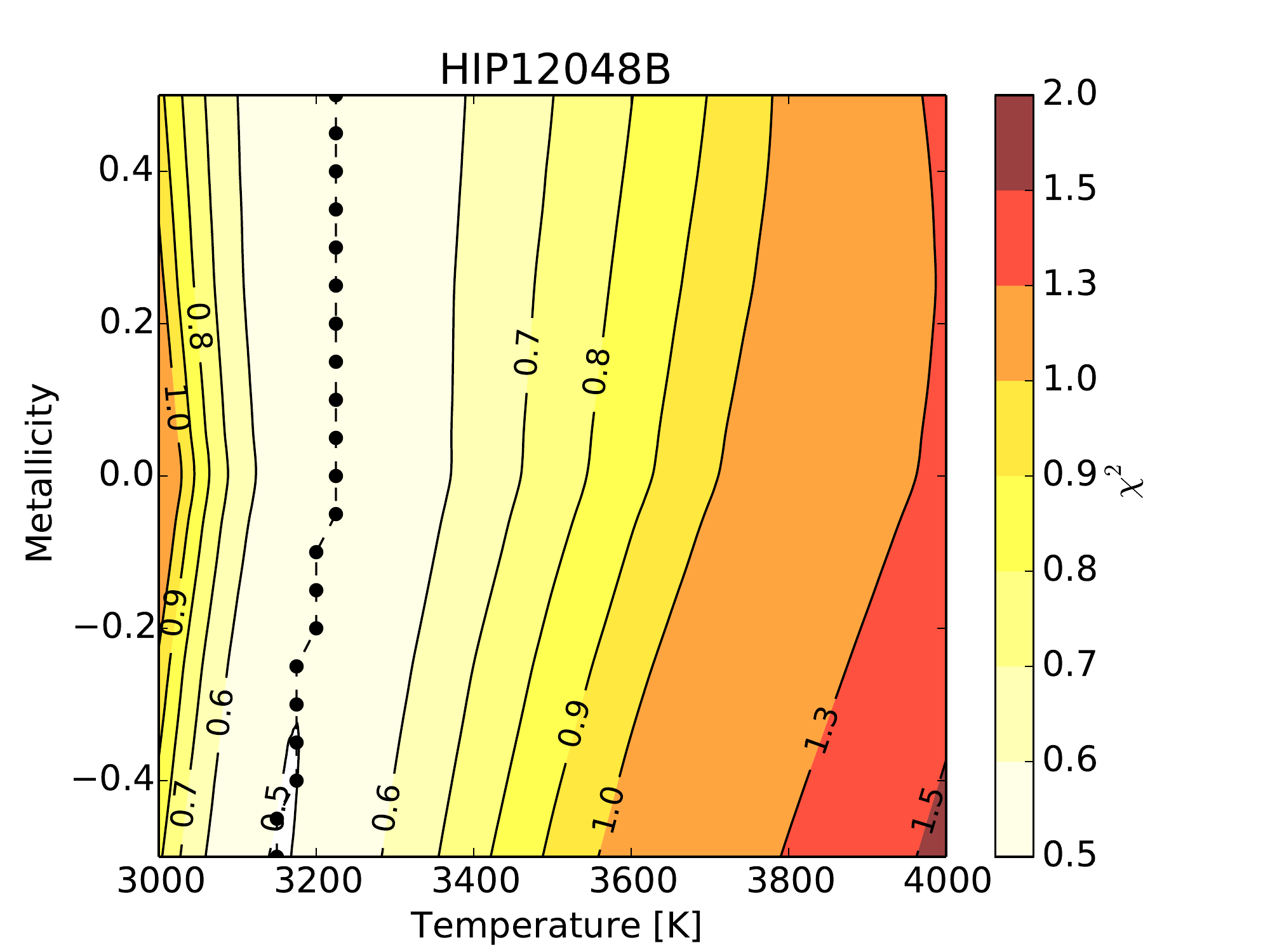}
\includegraphics[width=0.33\textwidth]{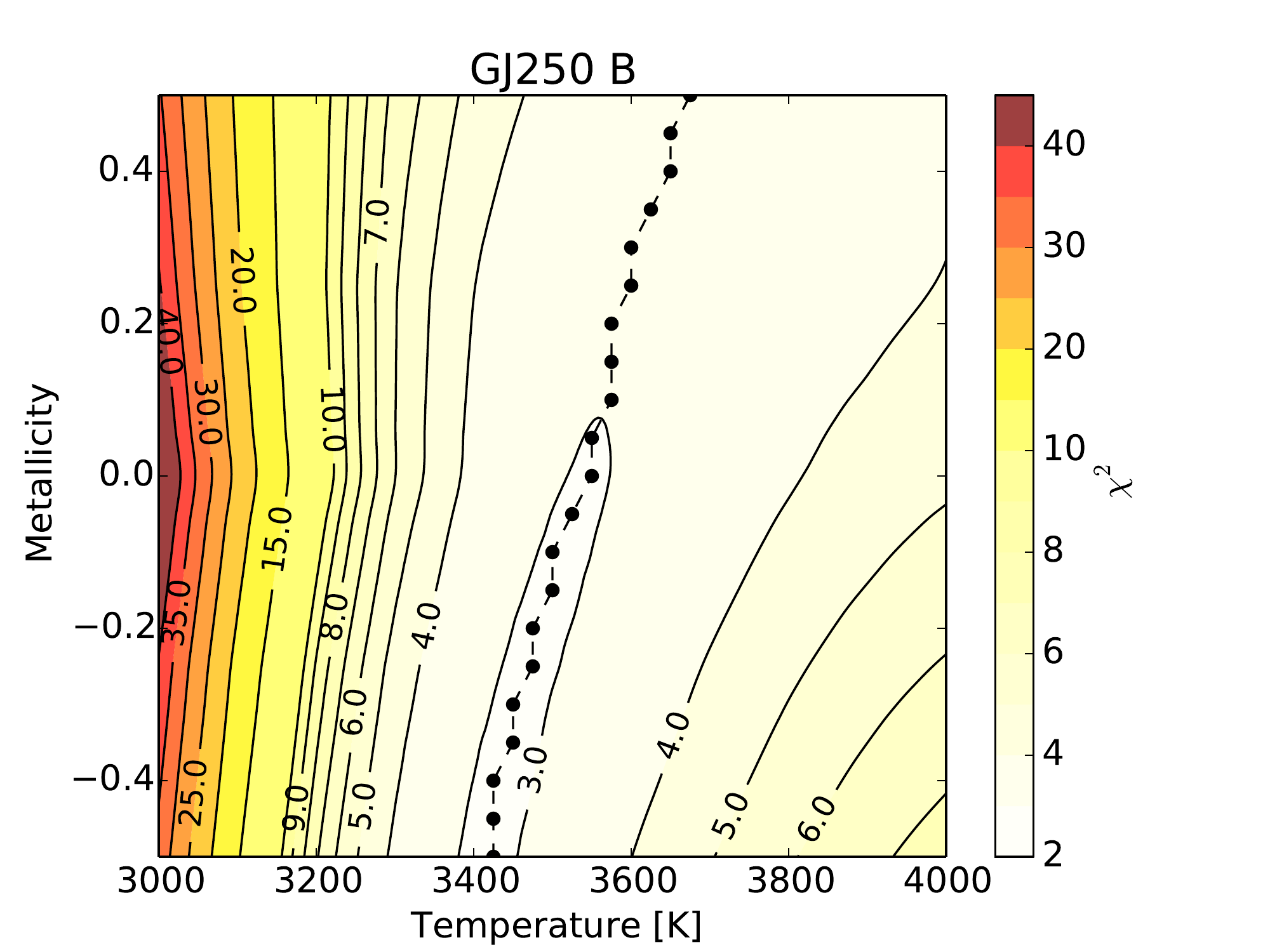}
\includegraphics[width=0.33\textwidth]{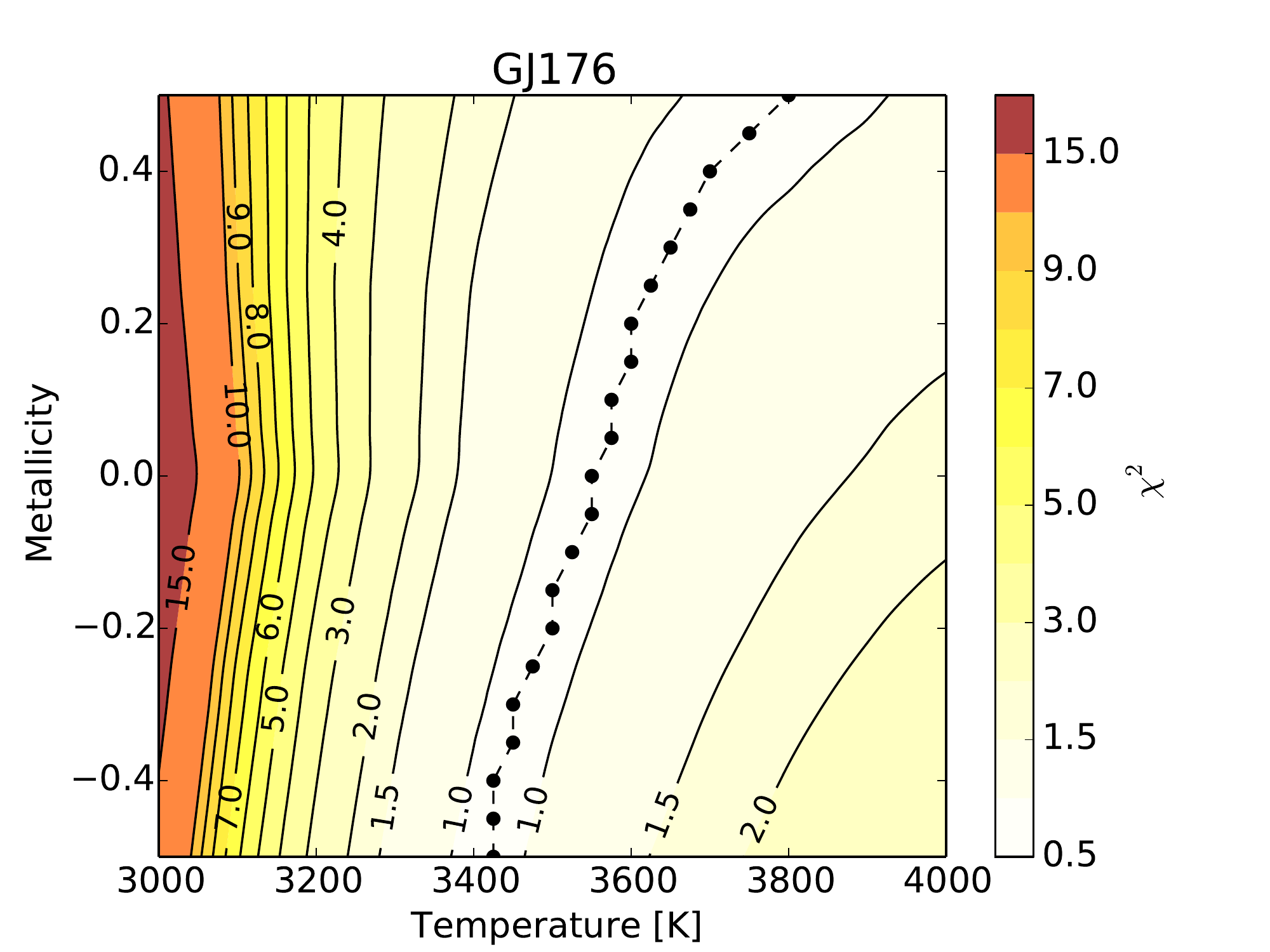}
\includegraphics[width=0.33\textwidth]{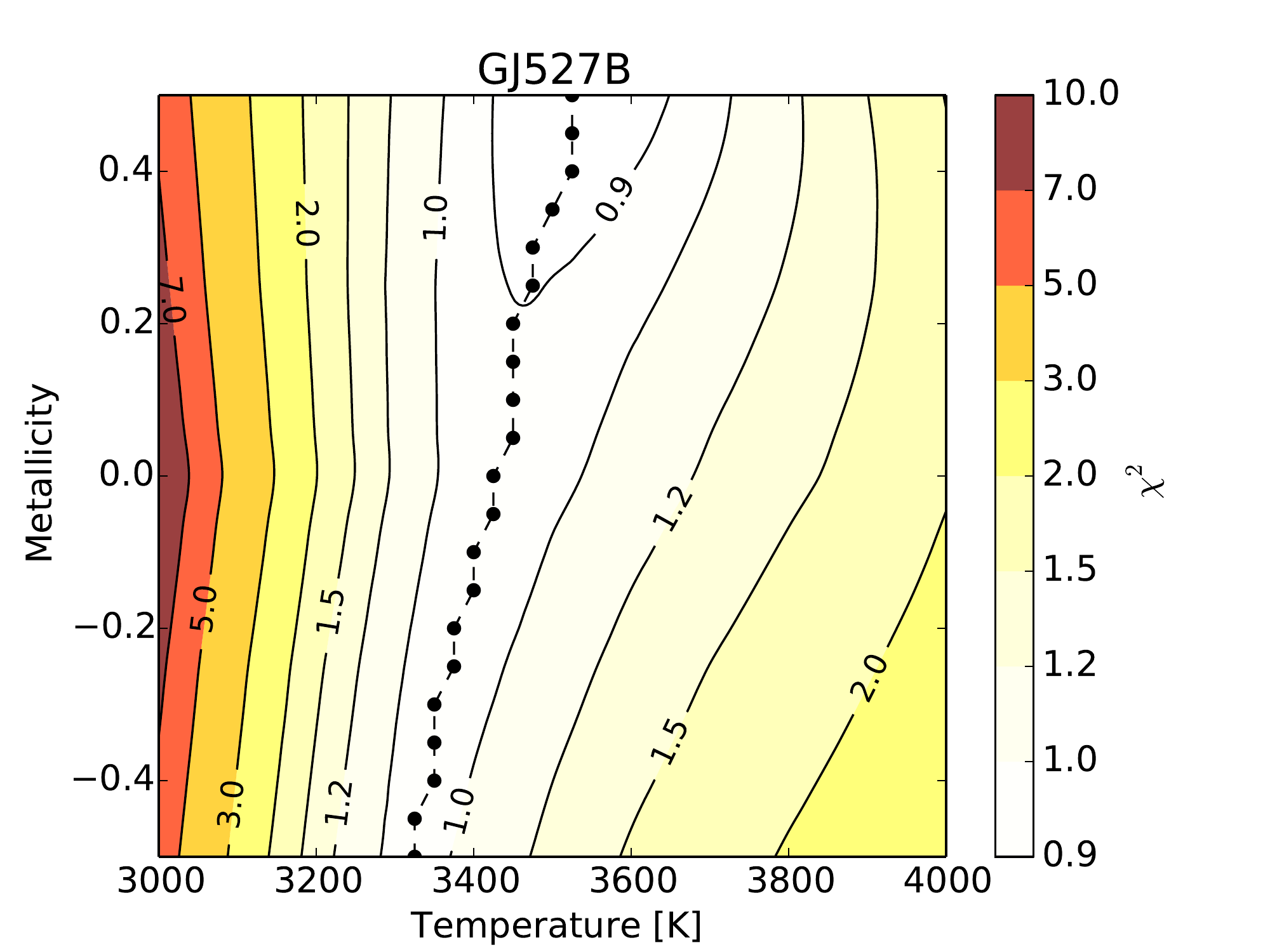}
\includegraphics[width=0.33\textwidth]{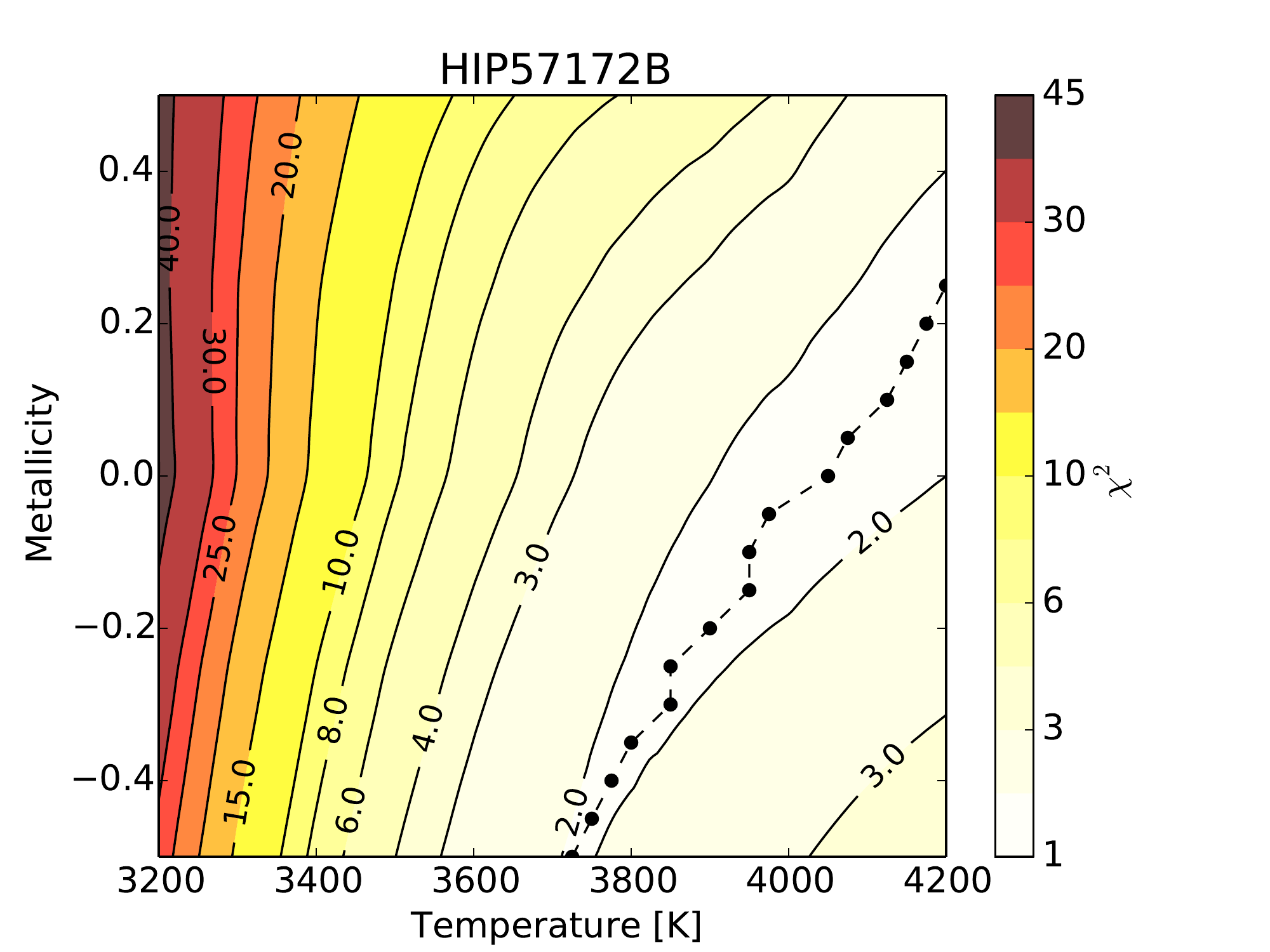}
\end{center}
\caption{Demonstration of FeH line strength dependency on the effective temperature and overall metallicity. The contour plots in this figure show the calculated $\chi^2$ of the fit based on a grid of metallicity and effective temperature of all M dwarfs. The block lines and dots indicate the temperature with the minimum $\chi^2$ for each step in metallicity. The plots are shown in order of spectral type, starting from the coolest going towards the warmest target in our sample. For the eight coolest stars, the $\chi^2$ value strongly depends on the effective temperature, with only a weak dependency on the metallicity. For the warmer stars (above 3400~K) a slight dependency on effective temperature can be seen, and for the warmest M dwarf HIP~57172~B (M0-M1) in our sample, a clear degeneracy between metallicity and effective temperature is found and this method cannot be used to determine the effective temperature for this target. (A colour version of this figure is available in the online journal.) }
\label{contour}
\end{figure*}

Another method explored was to expand the use of the FeH lines in our observed spectra, and to use only these lines to determine the effective temperatures. \citet{Onehag2012} performed several tests varying the effective temperature and surface gravity and visually comparing the obtained synthetic spectra with the observed. They found that within the relevant parameter space and wavelength regions the FeH lines in the synthetic spectra show little difference in strength when changing the surface gravity, but show large differences when changing the effective temperature. These molecular lines should therefore be useful to determine $T_{\rm eff}$. However, there is an expected degeneracy between effective temperature and metallicity since both affect the line strength. The FeH lines become stronger with increasing metallicity (increased Fe abundance), and become weaker with increasing temperature since molecules form at low temperature. Hence, to fit a spectrum with a particular set of line strengths, a higher metallicity might be compensated by a higher temperature. \citet{Onehag2012} therefore used an interactive scheme changing the metallicity and effective temperature in turns. We chose another method, calculating a grid of synthetic spectra covering the relevant parameter space in effective temperature and metallicity used in this study. Using SME, a grid of synthetic spectra was calculated with effective temperatures between 3000 and 4000~K (3200-4200~K for HIP~57172~B) in steps of 25~K, and metallicities between $-$0.5 and +0.5~dex in steps of 0.05~dex. {Since no strong dependence on the surface gravities were found, the values used are the same as for the metallicity determination (see Table \ref{fundamental_parameters}).} Remaining parameters were also kept fixed for all different combinations of $T_{\rm eff}$ and [M/H]; $v$~sin$i$ = 1~km~s$^{-1}$, (except GJ~527~B, where $v$~sin$i$~=~5~km~s$^{-1}$), $\xi_t$ = 1~km~s$^{-1}$, and $\zeta_t$~=~2~km~s$^{-1}$. For each spectrum in the grid the corresponding $\chi^2$ value with respect to the observed spectrum was calculated using SME and a selection of only clear, unblended FeH lines. In total, about 50 FeH lines per star were used. Contour plots of the calculated $\chi^2$ values as a function of $T_{\rm eff}$ and [M/H] were then produced for each star. All plots are shown in Figs. \ref{contour}. The calculated $\chi^2$ values show only a weak dependence on the metallicity, but a clear dependence on the effective temperature. The only exception is HIP~57172~B, where $\chi^2$ depends both on metallicity and temperature. For this star no minimum of $\chi^2$ corresponding to a specific effective temperature was found within our used grid. Previous studies have classified this star as an M0-M1 type dwarf. In conclusion, according to the contour plots there seems to be no strong degeneracy between $T_{\rm eff}$ and [M/H] in the FeH line strength for mid M-type dwarfs. However, a degeneracy is present for the warmer, early M-type dwarfs.

\begin{figure}
\begin{center}
\includegraphics[width=0.45\textwidth]{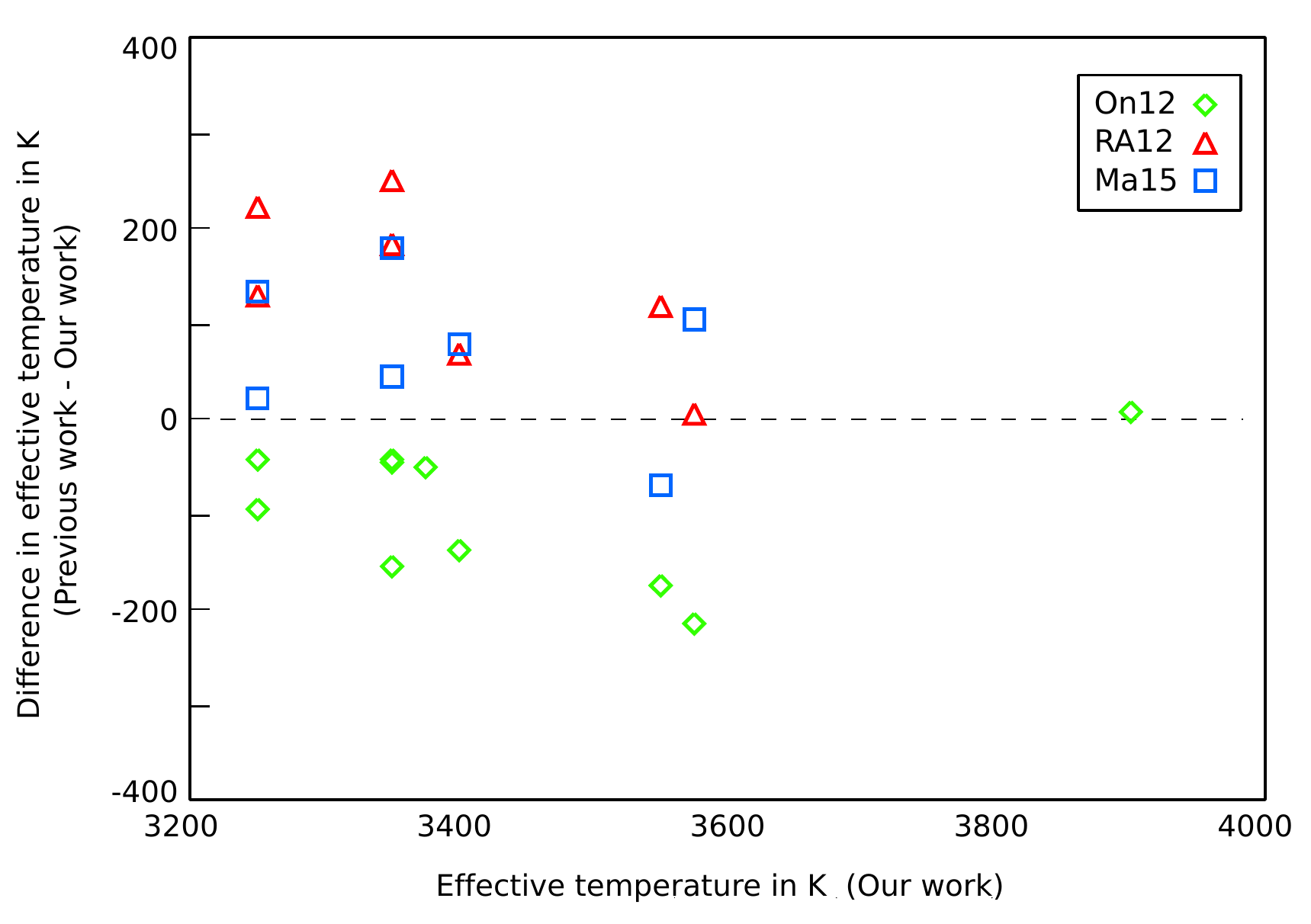}
\end{center}
\caption{Difference in adopted effective temperature between our work and the works by On12 (green diamonds), RA12 (red triangles), and Ma15 (blue squares). A  colour version of this figure is available in the online journal.}
\label{compare_teffs}
\end{figure}

All estimated values of the effective temperature from the FeH lines are shown in Table \ref{teff_Mdwarfs}, together with values adopted by On12, RA12, and Ma15 for comparison. The effective temperatures derived by On12, mainly based on the photometric calibration given by \citet{Casagrande2008}, are systematically lower than our temperatures derived from molecular spectral features. Other work based on molecular features, e.g. \citet{Valenti1998} and \citet{Bean2006b} using TiO bandheads, \citet{Rojas-Ayala2012} using H$_2$O, and \citet{Jones2005} using CO lines, also derived higher effective temperatures than did \citet{Casagrande2008}, who used multiband photometry. This difference between our work and On12 can also be seen graphically in Fig. \ref{compare_teffs}. Furthermore, it can be seen that our derived effective temperatures are lower than the values obtained by RA12 and Ma15 for a majority of the stars. The reason for this difference is unclear, but may be due to different sensitivity to spot coverage of the stellar surface. Although most of the M dwarfs in our sample are expected to show a low level of activity.

For the metallicity determination we adopted the $T_{\rm eff}$ values derived from our FeH plots, except for HIP~57172~B where we adopted the value by On12 because of  the problem with degeneracy between effective temperature and metallicity  described above. For the warmer FGK dwarf companions a value for the effective temperature was calculated from values in the PASTEL catalogue \citep{Soubiran2010} with some supplementary values taken from papers published later. All values published after 1990 were collected and the effective temperature was set to the mean of these values. Adopted values for all stars can be found in Table \ref{fundamental_parameters}. The adopted uncertainties for the primaries are the calculated standard deviations, while for the M dwarfs they were estimated from the widths of the minima in $\chi^2$.

\subsubsection{Surface gravity}
\label{logg}

\citet{Onehag2012} used the empirical relation between log~$g$ and stellar mass $M_\star$ derived by \citet{Bean2006b} to determine the surface gravity for the M dwarfs in their sample. We adopted the same values for the stars in common. To determine this empirical relation, \citet{Bean2006b} used radius measurements from eclipsing binaries or interferometric observations together with the relationship between the absolute K magnitude M$_K$ and M$_\star$ by \citet{Delfosse2000}. The masses used in  On12 were estimated using the M$_K$ - M$_\star$ relationship by \citet{Delfosse2000}. Unfortunately, obtaining masses in the same way is  not possible for HIP~12048~B and GJ~527~B owing to the lack of required photometric data. For HIP~12048~B an estimation of its mass was found in \citet{Mugrauer2007}. In their paper they assumed an age of 5~Gyr and by using the stellar evolutionary models by \citet{Baraffe1998} they derived a probable spectral class of M2.5 and a mass of 0.286 $\pm$ 0.017~M$_{\sun}$. Inserting this into the log~$g$ - M$_\star$ relation by \citet{Bean2006b} we determined a surface gravity of 4.9~log(cm~s$^{-2}$). Unfortunately no estimation of the mass of GJ~527~B could be found in the literature, and is why log~$g$ was set to 4.9 for this star as well. In the same manner as when calculating the effective temperature for the FGK dwarfs, the surface gravity was set to an average of literature values. Again, values were collected from the PASTEL catalogue \citep{Soubiran2010} with some supplementary values taken from papers published later. The associated uncertainties are given as the calculated standard deviation. Adopted values for all stars can be found in Table \ref{fundamental_parameters}.

\subsubsection{Metallicity}

All derived metallicities in this work refer to the overall metallicity determined by the fit of lines from all atomic species available in our spectra. However, all carbon lines were excluded from the analysis owing to possible non-LTE effects in the FGK dwarfs (e.g. \citealt{Asplund2005}, section 3.3 and On12, section 4.3). We also found that two of the four present potassium lines (K~I~11772.830~\AA~and K~I~12522.110~\AA) could not be matched with the same parameters as the remainder of the lines in the spectra for the M dwarfs, which is why we also chose to exclude  potassium from our analysis. This leaves, on average, 15-20 lines from species Fe, Ti, Mg, Ca, Si, Cr, Co, and Mn for the M~dwarfs, and about twice as many lines for the primaries. The spectra from HIP~12048~AB and GJ~527~AB are shown in Fig. \ref{spectra1}-\ref{spectra4} in Appendix A (available in the online journal). The spectra of GJ~527~A contains fewer lines because of its larger $v$~sin$i$, where the increased rotation broadening makes the weaker lines undetectable. A list of all the atomic lines used in the analysis is shown in table B.1 in Appendix B (available in the online journal). We note that not all lines were used in the analysis of all stars. For some stars a few lines were affected by imperfect removal of the telluric lines and the difference in wavelength settings between observational period 82 and 84 also affected which lines were available.

During the analysis of the FGK dwarfs, several tests were performed to check the validity of the fundamental parameters taken from the literature. Different runs allowing combinations of parameters to vary simultaneously were done. Small changes in the effective temperature and surface gravity was found, but most lie within the calculated standard divination, and the derived metallicities only differed  by 0.02-0.04~dex between different settings and the calculated $\chi^2$ remained relatively unchanged. The tabulated values in Table \ref{fundamental_parameters} are our adjusted values, and are the values used for the metallicity determination. For M dwarfs no additional optimisation of the effective temperature or surface gravity was done during the determination of the metallicity. The limited wavelength region of our spectra does not contain the number or types of lines required to make such an analysis with good precision and accuracy. Instead, only [M/H] and $\zeta_t$ were set to vary simultaneously until a minimum of $\chi^2$ was found. To ensure that the optimisation did not depend on the initial value, different starting values were tested: [M/H] = $-$0.1, 0.0 and +0.1~dex, and $\zeta_t$ = 0,1,2, and 3 km s$^{-2}$. For all M dwarfs $\xi_t$ was set to 1~km~s~$^{-2}$ and $v$~sin$i$ = 1~km~s~$^{-2}$. The exception is GJ~527 B, where previous studies and our analysis showed that both binary companions have a faster rotation. From our analysis a value of $v$~sin$i$ = 5~km~s$^{-2}$ was determined for GJ~527~B, and $v$~sin$i$ = 14~km~s$^{-2}$ was determined for its companion, GJ~527~A.

\section{Results and discussion}

\begin{table*}
\caption{Determined stellar parameters of all stars in our sample.} 
\label{fundamental_parameters}
\centering
\begin{tabular}{lccccccccc}
\hline\hline
Target & Metallicity & Metallicity & $T_{\rm eff}$ & log~$g$ & $\xi_t$ & $\zeta_t$  & $v$~sin$i$ & References \\ 
 & Our work & Avg. lit. ($\sigma$)  & [K] & [cm~s$^{-2}$] & [km~s~$^{-2}$] & [km~s~$^{-2}$] & [km~s~$^{-2}$] &  \\ \hline
HIP~12048 A             & +0.13 $\pm$ 0.04              & +0.14 (0.04)          & 5802 $\pm$ 43           & 4.23 $\pm$ 0.09               & 1.35          & 2.76            & 1.93          &  1,4,6,9,10,11,12,13,14,16,   \\
                                &                                       &                               &                                       &                                       &                       &                       &                       &  17,19,20,22,23,24,25,26        \\
HIP~12048 B             & +0.14 $\pm$ 0.15              &                               & 3225 $\pm$ 100          & 4.90 $\pm$ 0.10               & 1.00          & 2.00            & 1.00          &                                               \\
GJ~527 A                & +0.22 $\pm$ 0.04              &  +0.28 (0.07)                 & 6446 $\pm$ 90           & 4.49 $\pm$ 0.21               & 1.49          & 6.67            & 14.0          &   3,5,6,7,8,10,11,15,16,      \\
                                &                                       &                               &                                       &                                       &                       &                       &                       &  20,22,24,26,27                         \\
GJ~527 B                & +0.21 $\pm$ 0.10              &                               & 3325 $\pm$ 100          & 4.90 $\pm$ 0.10               & 1.00          & 2.00            & 5.00          &                                               \\
GJ~250 A                & $-$0.03 $\pm$ 0.07    & $-$0.04 (0.10)        & 4676 $\pm$ 150          & 4.54 $\pm$ 0.26               & 1.07          & 2.91            & 1.00          &   2,11,15,16,18,19,20                 \\
GJ~250 B                & $-$0.07 $\pm$ 0.04    &                               & 3550 $\pm$ 100          & 4.80 $\pm$ 0.08               & 1.00          & 2.91            & 1.00          &                                               \\
HIP~57172 A             & +0.15 $\pm$ 0.06              & +0.17 (0.05)          & 5030 $\pm$ 51           & 4.22 $\pm$ 0.06               & 0.99          & 0.05            & 1.00          &   1,6,21,22,23,25                     \\
HIP~57172 B             & +0.16 $\pm$ 0.13              &                               & 3900 $\pm$ 100          & 4.46 $\pm$ 0.09               &  1.00                 & 0.39            & 1.00          &                                               \\
GJ~176                  & +0.11 $\pm$ 0.09              &                               & 3550 $\pm$ 100          & 4.76 $\pm$ 0.08               &  1.00                 & 2.01            & 1.00          &                                               \\
GJ~317                  & +0.16 $\pm$ 0.08              &                               & 3375 $\pm$ 100          & 4.97 $\pm$ 0.12               & 1.00          & 0.10            & 1.00          &                                               \\
GJ~436                  & +0.03 $\pm$ 0.06              &                               & 3400 $\pm$ 100          & 4.80 $\pm$ 0.08               & 1.00          & 0.08            & 1.00          &                                               \\
GJ~581                  & $-$0.02 $\pm$ 0.13    &                               & 3350 $\pm$ 100          & 4.92 $\pm$ 0.08               & 1.00          & 0.33            & 1.00          &                                               \\
GJ~628                  & +0.12 $\pm$ 0.14              &                               & 3275 $\pm$ 100          & 4.93 $\pm$ 0.08               &  1.00                 & 0.04            & 1.00          &                                               \\
GJ~674                  & $-$0.01 $\pm$ 0.11    &                               & 3350 $\pm$ 100          & 4.88 $\pm$ 0.08               &  1.00                 & 2.43            & 1.00          &                                               \\
GJ~849                  & +0.28 $\pm$ 0.07              &                               & 3350 $\pm$ 100          & 4.76 $\pm$ 0.08               &  1.00                 & 0.05            & 1.00          &                                               \\
GJ~876                  & +0.19 $\pm$ 0.15              &                               & 3250 $\pm$ 100          & 4.89 $\pm$ 0.08               & 1.00          & 0.05            & 1.00          &                                               \\

\hline
\end{tabular}
\begin{flushleft}
\noindent
\newline {\bf Notes.} References used to calculate the average literature metallicities, mean effective temperature, and surface gravity values: 
1. \citet{Adibekyan2012b}, 
2. \citet{Bonfils2005a},
3. \citet{Erspamer2003},
4. \citet{Fuhrmann1998a},
5. \citet{Fuhrmann1998b},  
6. \citet{Ghezzi2010}, 
7. \citet{Gonzalez1997},
8. \citet{Gonzalez2000},
9. \citet{Gonzalez2001},
10. \citet{Gonzalez2010}, 
11. \citet{Heiter2003}, 
12. \citet{Huang2005}, 
13. \citet{Kang2011},
14. \citet{Laws2003},
15. \citet{Lee2011},
16. \citet{Luck2006}, 
17. \citet{Maldonado2013},
18. \citet{Mishenina2004},
19. \citet{Santos2001},
20. \citet{Santos2004}, 
21. \citet{Santos2005} 
22. \citet{Santos2013},
23. \citet{Sousa2008},
24. \citet{Takeda2005}, 
25. \citet{Tsantaki2013},
26. \citet{Valenti2005},
27. \citet{Zhao2002}.
The surface gravity values for the M dwarfs are taken from On12, with the exception of HIP~12048~B and GJ~527~B.  

\end{flushleft}
\end{table*}

As mentioned before, the largest error source to the metallicity is believed to come from the uncertainty of the adopted effective temperature. The values of the surface gravity also have an associated uncertainty adding to the tabulated uncertainty in the metallicity. To estimate the uncertainty, these parameters were varied, $T_{\rm eff}$ with $\pm$ 100~K and log~$g$ with $\pm$ 0.10~dex, and new values for the metallicity were derived for each of these four combinations. Thus the abundance sensitivities to the parameter uncertainties could be computed. The differences found due to the changed input parameters were then co-added in quadrature to estimate the uncertainty in derived metallicity. The same procedure was applied for both the M dwarfs and the FGK dwarfs in our sample.

\subsection{Binaries}
In this work we performed a careful analysis of both components in four FGK+M binaries with the goal of determining the reliability of metallicity determination of M dwarfs using spectral fitting of high-resolution infrared spectra. The results are very promising, with resulting differences of only 0.01-0.04~dex between the metallicities of the primary and secondary components. The derived metallicities of all binary components as well as the eight single M dwarfs are shown in Table \ref{fundamental_parameters}. We also list the adopted micro- and macroturbulence parameters, and the adopted rotation velocities. We note that when deriving the metallicity for all stars, the potassium lines were excluded;  additionally, for the G dwarf and F dwarf all carbon lines were excluded due to possible non-LTE effects. As a test we derived the metallicity including the carbon lines, keeping the remaining settings the same. For the F dwarf GJ~527~A we derived the same metallicity with and without including the carbon lines, while for the G dwarf HIP~12048~A the derived metallicity was lowered by 0.03~dex when including carbon.

To be able to see how our results agree with other studies, our determined metallicity values were compared with values of the warmer component taken from the literature. All  previously determined metallicity values found in the PASTEL catalogue \citep{Soubiran2010}, together with some supplementary values from papers published later, were used to calculate a mean metallicity value and standard deviation. We note that these studies were based on optical spectra. Averaged literature values and corresponding standard deviations are shown in Table \ref{fundamental_parameters}. Our metallicities agree well with the mean reference values from the literature, especially considering the estimated uncertainty of our analysis and the spread between literature values. We note that none of the studies in the literature included all four FGK dwarfs in our sample.

\subsubsection{HIP~12048}

\begin{figure}
\center
\includegraphics[width=0.45\textwidth]{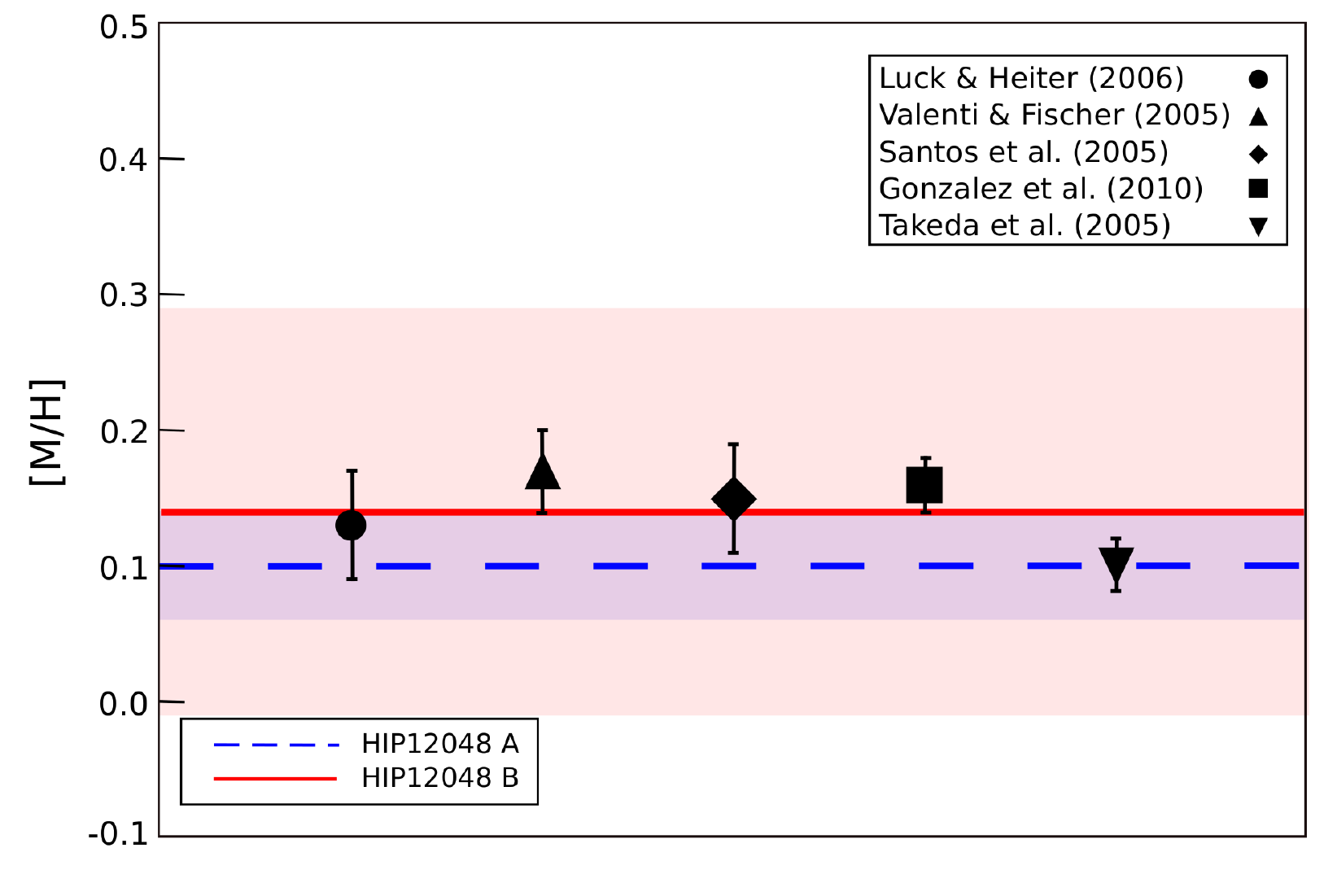}
\caption{Derived metallicity for the G+M binary (HIP~12048), compared with five optical high-resolution studies. The metallicity of the M dwarf is indicated by the  solid red line and the metallicity of the primary is indicated by the dashed blue line. The derived uncertainty for the primary in this work is shown with the semi-transparent blue area, and the corresponding uncertainty for the M dwarf in semi-transparent red area. Values and error bars for the primary from five studies selected from the literature are shown with different symbols. (A colour version of this figure is available in the online journal.) }
\label{comparison_HIP12048}
\end{figure}

As this was the first time the M dwarfs in the binaries HIP~12048 and GJ~527 had been analysed, an additional comparison with several individual high-resolution studies was made. Five studies containing both HIP~12048 and GJ~527, all  based on high-resolution spectra, were chosen \citep{Luck2006, Valenti2005, Santos2005, Gonzalez2010, Takeda2005}. In Fig. \ref{comparison_HIP12048} the result for HIP~12048 is shown. As can be seen, our determined metallicity of both the primary and the M dwarf is in good agreement with all selected studies and values from all six studies lie well within the respective error margins.

\subsubsection{GJ~527}

\begin{figure}
\center
\includegraphics[width=0.45\textwidth]{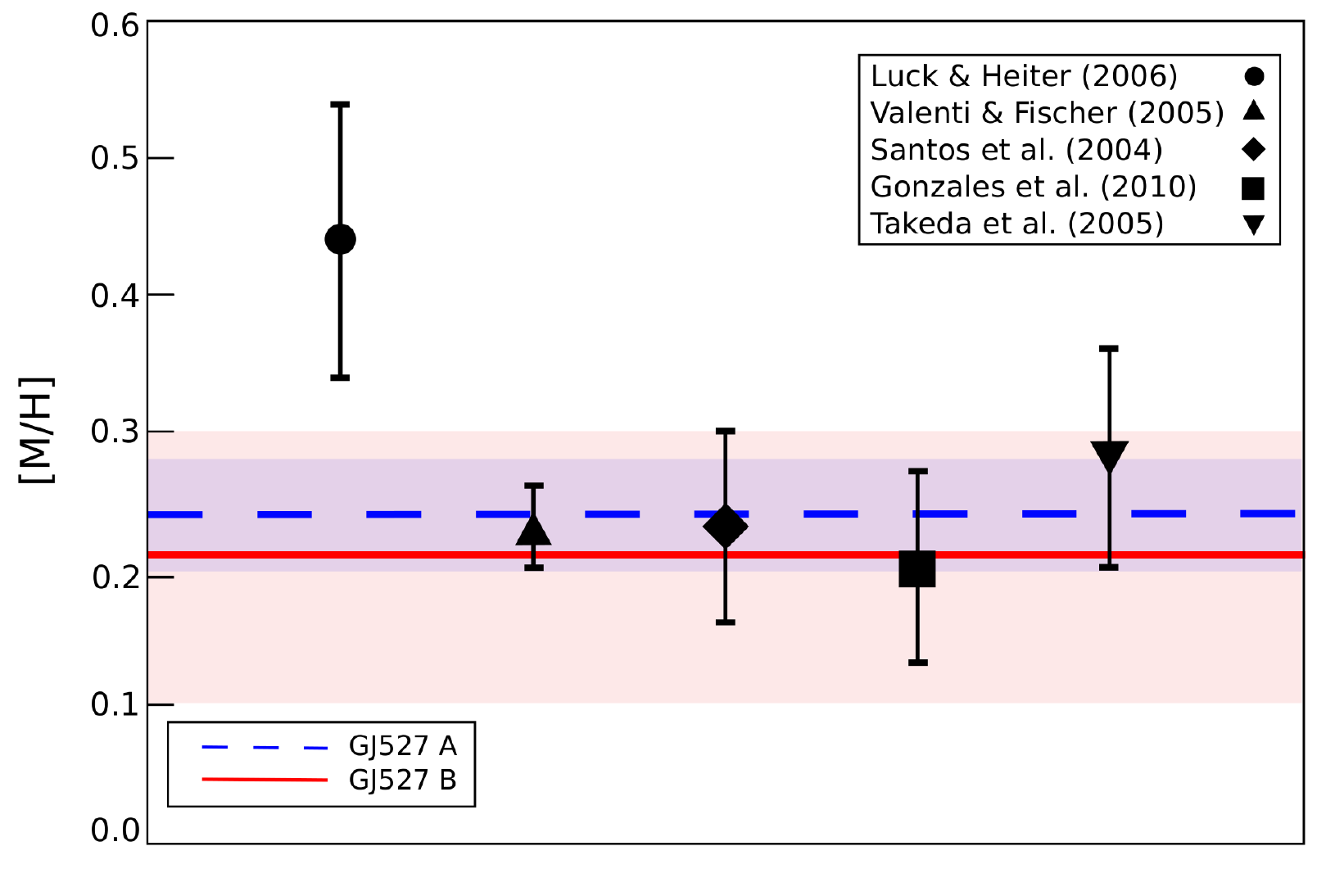}
\caption{Results for the F+M binary (GJ~527) compared with five optical high-resolution studies. The symbols and colours are the same as described in Fig. \ref{comparison_HIP12048}. We note that the minimum and maximum values on the y-axis are different from Fig. \ref{comparison_HIP12048}, but the range is the same for easier comparison. (A  colour version of this figure is available in the online journal.) }
\label{comparison_GJ527}
\end{figure}

As shown in Table \ref{fundamental_parameters}, our derived metallicity for both companions in the binary system GJ~527 is lower than the mean value from the literature. However, compared with the chosen high-resolution studies shown in Fig. \ref{comparison_GJ527}, our determined values of both the primary as well as the M dwarf are in good agreement with four of the shown studies, and lie well within their error margin. The exception is the value obtained by \citet{Luck2006} for GJ~527~A, which is substantially higher than our value and the error bars of the two studies do not overlap. However, this star was among the stars with highest rotational velocity and highest effective temperature in the sample of \citet{Luck2006}, for which their applied equivalent width technique was probably less reliable.

Compared with the remaining M dwarfs the analysis of GJ~527~B was somewhat more complex for two reasons. Both companions are faster rotators than the remaining stars in our sample, and more importantly, this is the most compact binary. As a consequence of the small separation, some of the stellar light from the F dwarf was blended into the spectrum of the M dwarf (see section \ref{reduction}). This can also clearly be seen when visually comparing the spectrum with and without the application of the de-blending routine. Using the spectra where the de-blending routine had not been applied gave a metallicity of 0.44~dex, which is twice the value we derived for the primary. Instead, using the spectra for which the deblending routine had been applied the derived metallicities agree well. To ensure that the difference was due to extra light from the primary and not to the data reduction process, the de-blending routine was applied to the spectra of the primary. Here no visual differences could be observed and the derived metallicity was the same, independent of the version of the spectra.

Previous studies of the primary have determined $v$~sin$i$ = 14-17.86 km~s$^{-1}$ (e.g. \citealt{Takeda2005, Martinez-Arnaiz2010}). For the primary we set $v$~sin$i$ as a free parameter in SME, resulting in an optimum at 14 km~s$^{-1}$. Given the high rotation rate of the primary, it is not surprising that a faster rotation velocity than 1~km~s$^{-1}$ was needed to fit the spectra  for the M dwarf as well. For the M dwarf the higher noise and fewer lines in the spectrum restricted us. Instead of letting SME fit the rotation velocity together with the other parameters, we changed the $v$~sin$i$ parameter in steps, keeping all other parameters fixed. An optimum fit to the line shape was found at $v$~sin$i$~=~5~km~s$^{-1}$, and was used as an input parameter when determining the effective temperature and metallicity.

\subsection{M dwarf metallicities}

For the M dwarfs, the derived metallicities were compared with values calculated from the photometric metallicity calibrations by Bo05, JA09 and Ne12, as well as with adopted values in the studies by RA12 and Ma15. We chose to exclude the calibration by \citet{Schlaufman2010} since it is very similar to the one by Ne12 and including it in our comparison does not add any new information. Unfortunately, the results for HIP~12048~B and GJ~527~B could not be compared with any other study owing to the lack of the required photometric data, and they were not included in the samples of RA12 or Ma15. Furthermore, the M dwarfs GJ~317 and GJ~674 are not included in the sample of RA12 or Ma15, and HIP~52172~B is not included in the sample of Ma15. 

\begin{figure}
\center
\includegraphics[width=0.45\textwidth]{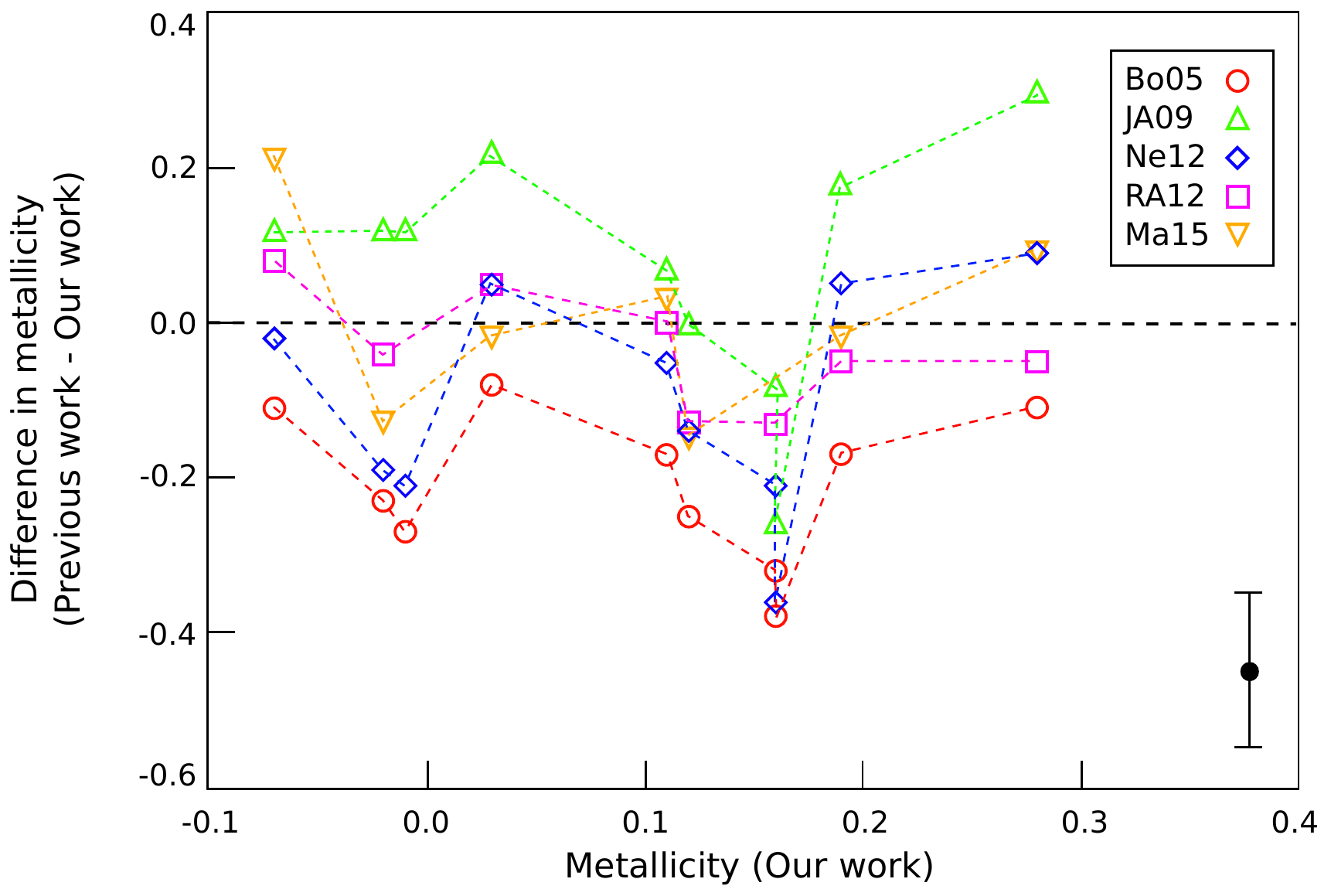}
\includegraphics[width=0.45\textwidth,]{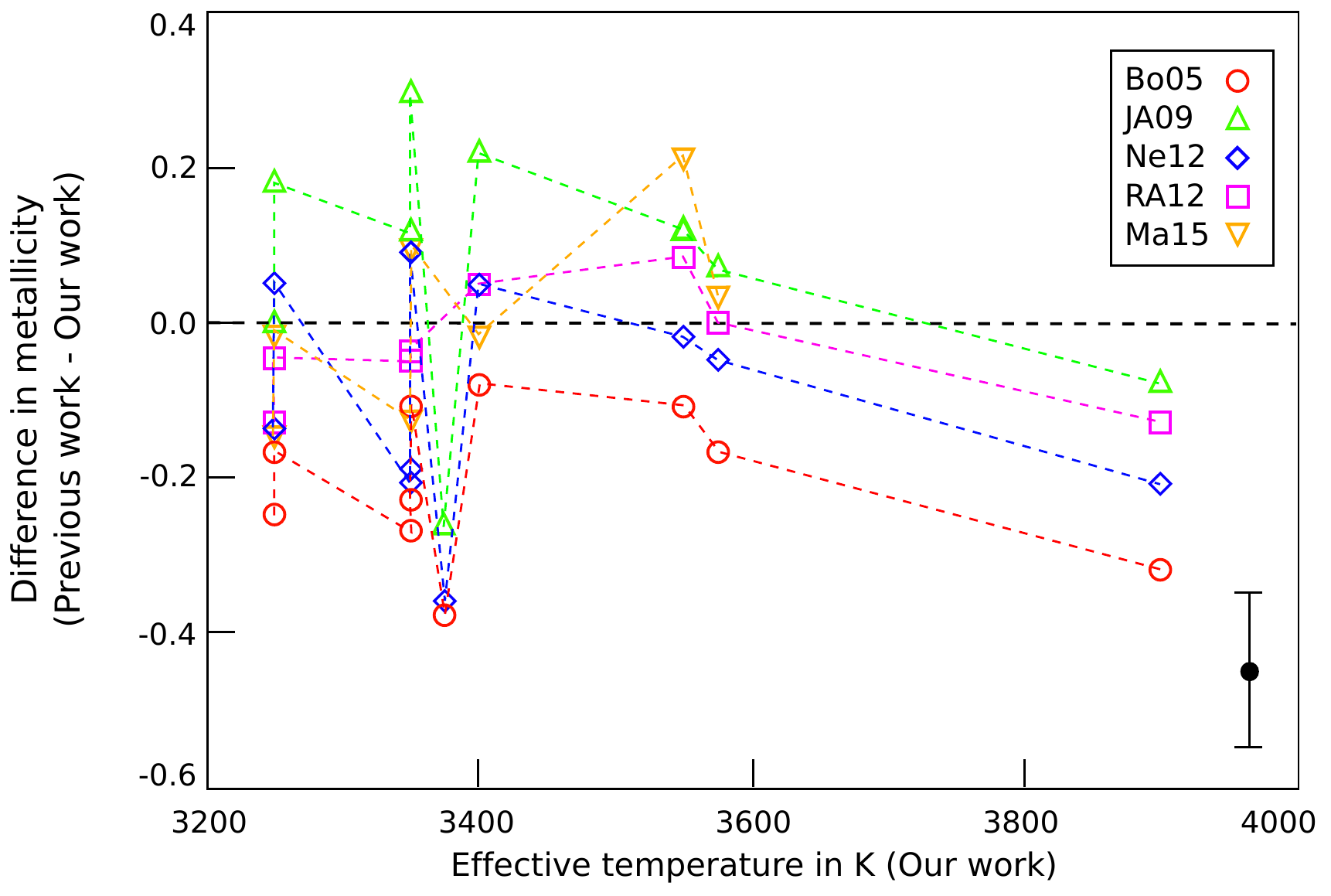}
\caption{Our determined metallicities for ten of the M dwarfs in our sample compared with three photometric calibrations, as well as values adopted by RA12 and Ma15. The differences are plotted  against our derived metallicity and the effective temperature. The size of an uncertainty of $\pm$0.1~dex is indicated by the black symbol in the bottom right. The dashed lines are there to guide the eye which points belong together, not to indicate a trend. (A  colour version of this figure is available in the online journal.)}
\label{comparison_Mdwarfs}
\end{figure}

As shown in Fig. \ref{comparison_Mdwarfs}, the different photometric calibrations give a wide range in metallicity, although they are based on the same photometric data. The range is  of the order 0.2-0.3~dex, which is two or three times larger than our estimated uncertainty. \citet{Neves2012} have shown that compared to their calibration, the  Bo05 calibrations tends to underestimate the metallicity while the  JA09 calibration overestimates it. As shown in Fig. \ref{comparison_Mdwarfs}, our metallicities are systematically higher than the values calculated with the calibration by B05. All values of the  calibration by JA09  are higher than our determination with the exception of GJ~317 and HIP~57172~B, which are lower. This is in line with the conclusion by Ne12. However,  the calibration by Ne12 also seems to underestimate the metallicity compared to our results, and the calculated values for HIP~57172~B, GJ~317, GJ~581, GJ~628, and GJ~674 are substantially lower than our derived metallicities. We note that the Ne12 calibrated metallicities for GJ~317, GJ~581, and GJ~674 lie close to the values using the Bo05 calibration. Since the latter is believed to underestimate the metallicity, this may indicate some problem in the Ne12 calibration for those stars. Furthermore, one of the stars with a large difference between Ne12 and our result is HIP~57172~B for which we found excellent agreement between the binary components and with the metallicities from the literature.

In conclusion, we note that none of the photometric calibrations give results that are consistent with our metallicities for all stars. The calculated mean differences compared to our derived values for the three photometric calibrations are Bo05~=~$-$0.21~($\sigma$~=~0.10), JA09~=~+0.08~($\sigma$~=~0.16), and Ne12~=~$-$0.14~($\sigma$~=~0.15). We also compared our metallicities with the adopted values of RA12 and Ma15, and between our and their works we find a better agreement. This is further strengthened by the calculated mean differences of RA12~=~$-$0.04~($\sigma$~=~0.08) and Ma15~=~0.00~($\sigma$~=~0.13).

\begin{figure}
\center
\includegraphics[width=0.45\textwidth]{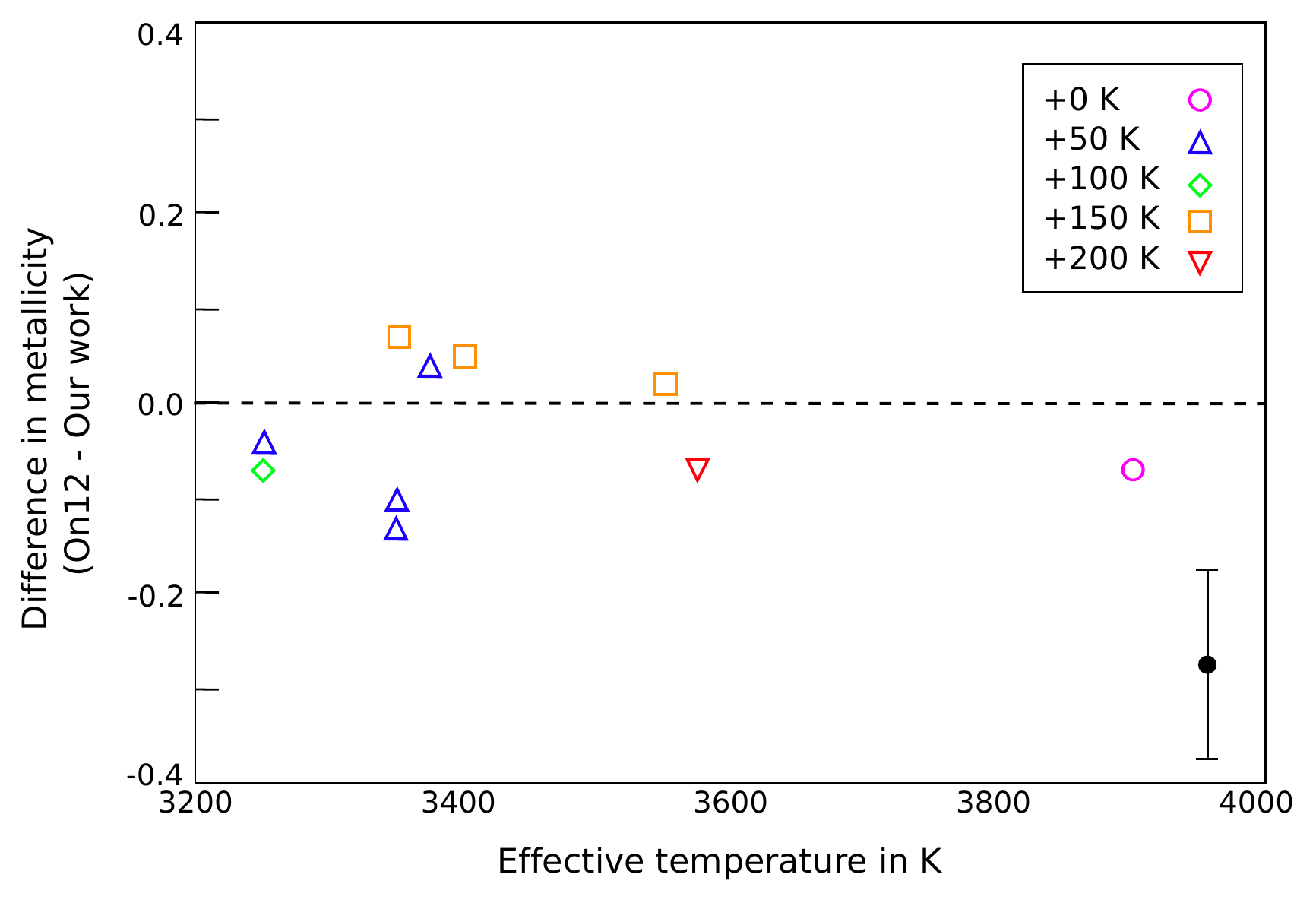}
\caption{Differences between derived values in this study and the values obtained by On12, who used the same observational data. Different symbols indicate the differences in adopted effective temperatures between the two studies. The size of an uncertainty of $\pm$0.1~dex is indicated by the black symbol in the bottom right. (A  colour version of this figure is available in the online journal.)}
\label{comparison_On12}
\end{figure}

Because of our updated oscillator strengths relative to the data used by On12, together with new adopted effective temperatures for all M dwarfs, a comparison was made to see how our derived metallicities correlate with the results by On12, who used the same observed spectra. As shown in Fig. \ref{comparison_On12}, the differences lie within approximately 0.1~dex, which is of the order of the estimated error margins in both studies. The calculated average of the difference between On12 and our work is $-$0.03~dex~($\sigma$~=~0.07).

\section{Conclusions and outlook}

In this paper we derived the metallicity of the M dwarfs in four FGK+M binaries, using high-resolution spectra taken in the $J$ band. Two of those M dwarfs were analysed for the first time. The results from all binaries show good agreement between the derived metallicities of the primary and secondary component ($\Delta$ = 0.01-0.04~dex), and the results are also in good agreement with the literature values. High-resolution infrared spectroscopy can thus provide reliable and accurate metallicity for M dwarfs. To further improve the accuracy and precision, better knowledge of effective temperatures is needed. Also, a sufficiently high signal-to-noise ratio is required ($\gtrsim$ 50) to achieve uncertainties below 0.1~dex.

In the ideal case the effective temperature and surface gravity should both be derived spectroscopically, instead of partly relying on external sources, as is the case for the surface gravity in our analysis. This is currently impossible due to the limited wavelength coverage of CRIRES. The situation will improve in the near future with the on-going update of the CRIRES spectrograph \citep{Follert2014}, the upcoming CARMENES spectrograph \citep{Quirrenbach2014, Quirrenbach2010} and the GIANO spectrograph at TNG \citep{Origlia2014}, all providing a larger wavelength coverage in the infrared.

In addition to confirming the reliability of the method, we re-analysed eight single M dwarfs with the same method and updated line data. Furthermore, we introduced a new method to determine the effective temperature using FeH lines. For the majority of the stars, the effective temperatures were increased by 100-200 K compared with the values adopted by On12, and show better agreement with independent determinations by RA12 and Ma15. Our re-determined metallicities agree within the estimated error margins with the values of On12. Compared to our results of individual M dwarfs the photometric calibrations by Bo05 and Ne12 were found, on average, to underestimate the metallicity while the calibration by JA09 slightly overestimates it. If we instead compare our results to the determination values by RA12 and Ma15, we found a better average agreement.

In this paper the analysis of data observed between  2008 and 2010 have been presented. This sample is too small to draw any conclusions on the metallicity of planet hosts. However, an additional sample of 22 M dwarfs has been observed more recently with the CRIRES spectrograph, which we plan to analyse in a future paper with the same methodology. This will expand the parameter space explored with our analysis technique to cover a metallicity span of approximately 2 dex and cover subtypes M0-M6. With this we hope to provide a good basis for an updated photometric calibration, which can be used to address questions about a metallicity-planet connection for M dwarfs. \\

\begin{acknowledgements}

This work has made use the VALD database (operated at Uppsala University, the Institute of Astronomy RAS in Moscow, and the University of Vienna) and of NSO/Kitt Peak FTS data, produced by NSF/NOAO. We acknowledge Bertrand Plez for providing us with the linelist of FeH. UH acknowledges support from the Swedish National Space Board (Rymdstyrelsen). SL acknowledges Thomas Nordlander and Bengt Edvardsson for many useful discussions and ideas used in this work.

\end{acknowledgements}

\bibliographystyle{aa}
\bibliography{biblio}

\begin{thebibliography}{152}
\expandafter\ifx\csname natexlab\endcsname\relax\def\natexlab#1{#1}\fi

\bibitem[{{Adibekyan} {et~al.}(2012){Adibekyan}, {Sousa}, {Santos}, {Delgado
  Mena}, {Gonz{\'a}lez Hern{\'a}ndez}, {Israelian}, {Mayor}, \&
  {Khachatryan}}]{Adibekyan2012b}
{Adibekyan}, V.~Z., {Sousa}, S.~G., {Santos}, N.~C., {et~al.} 2012, \aap, 545,
  A32

\bibitem[{{Allard} \& {Freytag}(2010)}]{Allard2010}
{Allard}, F. \& {Freytag}, B. 2010, Highlights of Astronomy, 15, 756

\bibitem[{{Allard} {et~al.}(2013){Allard}, {Homeier}, \&
  {Freytag}}]{Allard2013}
{Allard}, F., {Homeier}, D., \& {Freytag}, B. 2013, \memsai, 84, 1053

\bibitem[{{Anglada-Escud{\'e}} {et~al.}(2012){Anglada-Escud{\'e}}, {Boss},
  {Weinberger}, {Thompson}, {Butler}, {Vogt}, \& {Rivera}}]{Anglada-Escude2012}
{Anglada-Escud{\'e}}, G., {Boss}, A.~P., {Weinberger}, A.~J., {et~al.} 2012,
  \apj, 746, 37

\bibitem[{{Anstee} \& {O'Mara}(1991)}]{Anstee1991}
{Anstee}, S.~D. \& {O'Mara}, B.~J. 1991, \mnras, 253, 549

\bibitem[{{Asplund}(2005)}]{Asplund2005}
{Asplund}, M. 2005, \araa, 43, 481

\bibitem[{{Auman}(1969)}]{Auman1969}
{Auman}, Jr., J.~R. 1969, \apj, 157, 799

\bibitem[{{Baraffe} {et~al.}(1998){Baraffe}, {Chabrier}, {Allard}, \&
  {Hauschildt}}]{Baraffe1998}
{Baraffe}, I., {Chabrier}, G., {Allard}, F., \& {Hauschildt}, P.~H. 1998, \aap,
  337, 403

\bibitem[{{Barber} {et~al.}(2006){Barber}, {Tennyson}, {Harris}, \&
  {Tolchenov}}]{Barber2006}
{Barber}, R.~J., {Tennyson}, J., {Harris}, G.~J., \& {Tolchenov}, R.~N. 2006,
  \mnras, 368, 1087

\bibitem[{{Barklem} {et~al.}(2000){Barklem}, {Piskunov}, \&
  {O'Mara}}]{Barklem2000}
{Barklem}, P.~S., {Piskunov}, N., \& {O'Mara}, B.~J. 2000, \aap, 363, 1091

\bibitem[{{Bean} {et~al.}(2006){Bean}, {Benedict}, \& {Endl}}]{Bean2006b}
{Bean}, J.~L., {Benedict}, G.~F., \& {Endl}, M. 2006, \apjl, 653, L65

\bibitem[{{Bean} {et~al.}(2010){Bean}, {Seifahrt}, {Hartman}, {Nilsson},
  {Wiedemann}, {Reiners}, {Dreizler}, \& {Henry}}]{Bean2010}
{Bean}, J.~L., {Seifahrt}, A., {Hartman}, H., {et~al.} 2010, \apj, 713, 410

\bibitem[{{Berger}(2006)}]{Berger2006b}
{Berger}, E. 2006, \apj, 648, 629

\bibitem[{{Bessel}(1990)}]{Bessel1990}
{Bessel}, M.~S. 1990, \aaps, 83, 357

\bibitem[{{Bochanski} {et~al.}(2010){Bochanski}, {Hawley}, {Covey}, {West},
  {Reid}, {Golimowski}, \& {Ivezi{\'c}}}]{Bochanski2010}
{Bochanski}, J.~J., {Hawley}, S.~L., {Covey}, K.~R., {et~al.} 2010, \aj, 139,
  2679

\bibitem[{{Bonfils} {et~al.}(2005{\natexlab{a}}){Bonfils}, {Delfosse}, {Udry},
  {Santos}, {Forveille}, \& {S{\'e}gransan}}]{Bonfils2005a}
{Bonfils}, X., {Delfosse}, X., {Udry}, S., {et~al.} 2005{\natexlab{a}}, \aap,
  442, 635

\bibitem[{{Bonfils} {et~al.}(2005{\natexlab{b}}){Bonfils}, {Forveille},
  {Delfosse}, {Udry}, {Mayor}, {Perrier}, {Bouchy}, {Pepe}, {Queloz}, \&
  {Bertaux}}]{Bonfils2005b}
{Bonfils}, X., {Forveille}, T., {Delfosse}, X., {et~al.} 2005{\natexlab{b}},
  \aap, 443, L15

\bibitem[{{Bonfils} {et~al.}(2007){Bonfils}, {Mayor}, {Delfosse}, {Forveille},
  {Gillon}, {Perrier}, {Udry}, {Bouchy}, {Lovis}, {Pepe}, {Queloz}, {Santos},
  \& {Bertaux}}]{Bonfils2007}
{Bonfils}, X., {Mayor}, M., {Delfosse}, X., {et~al.} 2007, \aap, 474, 293

\bibitem[{{Browning} {et~al.}(2010){Browning}, {Basri}, {Marcy}, {West}, \&
  {Zhang}}]{Browning2010}
{Browning}, M.~K., {Basri}, G., {Marcy}, G.~W., {West}, A.~A., \& {Zhang}, J.
  2010, \aj, 139, 504

\bibitem[{{Burrows} {et~al.}(2002){Burrows}, {Ram}, {Bernath}, {Sharp}, \&
  {Milsom}}]{Burrows2002}
{Burrows}, A., {Ram}, R.~S., {Bernath}, P., {Sharp}, C.~M., \& {Milsom}, J.~A.
  2002, \apj, 577, 986

\bibitem[{{Butler} {et~al.}(2006){Butler}, {Johnson}, {Marcy}, {Wright},
  {Vogt}, \& {Fischer}}]{Butler2006}
{Butler}, R.~P., {Johnson}, J.~A., {Marcy}, G.~W., {et~al.} 2006, \pasp, 118,
  1685

\bibitem[{{Butler} {et~al.}(1997){Butler}, {Marcy}, {Williams}, {Hauser}, \&
  {Shirts}}]{Butler1997}
{Butler}, R.~P., {Marcy}, G.~W., {Williams}, E., {Hauser}, H., \& {Shirts}, P.
  1997, \apjl, 474, L115

\bibitem[{{Butler} {et~al.}(2004){Butler}, {Vogt}, {Marcy}, {Fischer},
  {Wright}, {Henry}, {Laughlin}, \& {Lissauer}}]{Butler2004}
{Butler}, R.~P., {Vogt}, S.~S., {Marcy}, G.~W., {et~al.} 2004, \apj, 617, 580

\bibitem[{{Casagrande} {et~al.}(2008){Casagrande}, {Flynn}, \&
  {Bessell}}]{Casagrande2008}
{Casagrande}, L., {Flynn}, C., \& {Bessell}, M. 2008, \mnras, 389, 585

\bibitem[{{Castelli} \& {Kurucz}(2004)}]{Castelli2004}
{Castelli}, F. \& {Kurucz}, R.~L. 2004, ArXiv Astrophysics e-prints

\bibitem[{{Chabrier}(2003)}]{Chabrier2003}
{Chabrier}, G. 2003, \pasp, 115, 763

\bibitem[{{Correia} {et~al.}(2010){Correia}, {Couetdic}, {Laskar}, {Bonfils},
  {Mayor}, {Bertaux}, {Bouchy}, {Delfosse}, {Forveille}, {Lovis}, {Pepe},
  {Perrier}, {Queloz}, \& {Udry}}]{Correia2010}
{Correia}, A.~C.~M., {Couetdic}, J., {Laskar}, J., {et~al.} 2010, \aap, 511,
  A21

\bibitem[{{Covey} {et~al.}(2008){Covey}, {Hawley}, {Bochanski}, {West}, {Reid},
  {Golimowski}, {Davenport}, {Henry}, {Uomoto}, \& {Holtzman}}]{Covey2008}
{Covey}, K.~R., {Hawley}, S.~L., {Bochanski}, J.~J., {et~al.} 2008, \aj, 136,
  1778

\bibitem[{{Cumming} {et~al.}(1999){Cumming}, {Marcy}, \&
  {Butler}}]{Cumming1999}
{Cumming}, A., {Marcy}, G.~W., \& {Butler}, R.~P. 1999, \apj, 526, 890

\bibitem[{{Cutri} {et~al.}(2003){Cutri}, {Skrutskie}, {van Dyk}, {Beichman},
  {Carpenter}, {Chester}, {Cambresy}, {Evans}, {Fowler}, {Gizis}, {Howard},
  {Huchra}, {Jarrett}, {Kopan}, {Kirkpatrick}, {Light}, {Marsh}, {McCallon},
  {Schneider}, {Stiening}, {Sykes}, {Weinberg}, {Wheaton}, {Wheelock}, \&
  {Zacarias}}]{Cutri2003}
{Cutri}, R.~M., {Skrutskie}, M.~F., {van Dyk}, S., {et~al.} 2003, VizieR Online
  Data Catalog, 2246, 0

\bibitem[{{de Jager} \& {Neven}(1957)}]{deJager1957}
{de Jager}, C. \& {Neven}, L. 1957, {Spectroscopic data for 50 model
  photospherese}

\bibitem[{{Delfosse} {et~al.}(1998){Delfosse}, {Forveille}, {Perrier}, \&
  {Mayor}}]{Delfosse1998}
{Delfosse}, X., {Forveille}, T., {Perrier}, C., \& {Mayor}, M. 1998, \aap, 331,
  581

\bibitem[{{Delfosse} {et~al.}(2000){Delfosse}, {Forveille}, {S{\'e}gransan},
  {Beuzit}, {Udry}, {Perrier}, \& {Mayor}}]{Delfosse2000}
{Delfosse}, X., {Forveille}, T., {S{\'e}gransan}, D., {et~al.} 2000, \aap, 364,
  217

\bibitem[{{Dhital} {et~al.}(2012){Dhital}, {West}, {Stassun}, {Bochanski},
  {Massey}, \& {Bastien}}]{Dhital2012}
{Dhital}, S., {West}, A.~A., {Stassun}, K.~G., {et~al.} 2012, \aj, 143, 67

\bibitem[{{Dommanget} \& {Nys}(2002)}]{Dommanget2002}
{Dommanget}, J. \& {Nys}, O. 2002, VizieR Online Data Catalog, 1274, 0

\bibitem[{{Dressing} \& {Charbonneau}(2013)}]{Dressing2013}
{Dressing}, C.~D. \& {Charbonneau}, D. 2013, \apj, 767, 95

\bibitem[{{Dulick} {et~al.}(2003){Dulick}, {Bauschlicher}, {Burrows}, {Sharp},
  {Ram}, \& {Bernath}}]{Dulick2003}
{Dulick}, M., {Bauschlicher}, Jr., C.~W., {Burrows}, A., {et~al.} 2003, \apj,
  594, 651

\bibitem[{{Eggen}(1960)}]{Eggen1960}
{Eggen}, O.~J. 1960, \mnras, 120, 430

\bibitem[{{Endl} {et~al.}(2008){Endl}, {Cochran}, {Wittenmyer}, \&
  {Boss}}]{Endl2008}
{Endl}, M., {Cochran}, W.~D., {Wittenmyer}, R.~A., \& {Boss}, A.~P. 2008, \apj,
  673, 1165

\bibitem[{{Erspamer} \& {North}(2003)}]{Erspamer2003}
{Erspamer}, D. \& {North}, P. 2003, \aap, 398, 1121

\bibitem[{{Follert} {et~al.}(2014){Follert}, {Dorn}, {Oliva}, {Lizon},
  {Hatzes}, {Piskunov}, {Reiners}, {Seemann}, {Stempels}, {Heiter}, {Marquart},
  {Lockhart}, {Anglada-Escude}, {L{\"o}winger}, {Baade}, {Grunhut}, {Bristow},
  {Klein}, {Jung}, {Ives}, {Kerber}, {Pozna}, {Paufique}, {Kaeufl}, {Origlia},
  {Valenti}, {Gojak}, {Hilker}, {Pasquini}, {Smette}, \&
  {Smoker}}]{Follert2014}
{Follert}, R., {Dorn}, R.~J., {Oliva}, E., {et~al.} 2014, in Society of
  Photo-Optical Instrumentation Engineers (SPIE) Conference Series, Vol. 9147,
  Society of Photo-Optical Instrumentation Engineers (SPIE) Conference Series,
  19

\bibitem[{{Forveille} {et~al.}(2009){Forveille}, {Bonfils}, {Delfosse},
  {Gillon}, {Udry}, {Bouchy}, {Lovis}, {Mayor}, {Pepe}, {Perrier}, {Queloz},
  {Santos}, \& {Bertaux}}]{Forveille2009}
{Forveille}, T., {Bonfils}, X., {Delfosse}, X., {et~al.} 2009, \aap, 493, 645

\bibitem[{{Fuhrmann}(1998)}]{Fuhrmann1998a}
{Fuhrmann}, K. 1998, \aap, 338, 161

\bibitem[{{Fuhrmann} {et~al.}(1998){Fuhrmann}, {Pfeiffer}, \&
  {Bernkopf}}]{Fuhrmann1998b}
{Fuhrmann}, K., {Pfeiffer}, M.~J., \& {Bernkopf}, J. 1998, \aap, 336, 942

\bibitem[{{Ghezzi} {et~al.}(2010){Ghezzi}, {Cunha}, {Smith}, {de Ara{\'u}jo},
  {Schuler}, \& {de la Reza}}]{Ghezzi2010}
{Ghezzi}, L., {Cunha}, K., {Smith}, V.~V., {et~al.} 2010, \apj, 720, 1290

\bibitem[{{Gonzalez}(1997)}]{Gonzalez1997}
{Gonzalez}, G. 1997, \mnras, 285, 403

\bibitem[{{Gonzalez} {et~al.}(2010){Gonzalez}, {Carlson}, \&
  {Tobin}}]{Gonzalez2010}
{Gonzalez}, G., {Carlson}, M.~K., \& {Tobin}, R.~W. 2010, \mnras, 403, 1368

\bibitem[{{Gonzalez} \& {Laws}(2000)}]{Gonzalez2000}
{Gonzalez}, G. \& {Laws}, C. 2000, \aj, 119, 390

\bibitem[{{Gonzalez} \& {Laws}(2007)}]{Gonzalez2007}
{Gonzalez}, G. \& {Laws}, C. 2007, \mnras, 378, 1141

\bibitem[{{Gonzalez} {et~al.}(2001){Gonzalez}, {Laws}, {Tyagi}, \&
  {Reddy}}]{Gonzalez2001}
{Gonzalez}, G., {Laws}, C., {Tyagi}, S., \& {Reddy}, B.~E. 2001, \aj, 121, 432

\bibitem[{{Gray} {et~al.}(2006){Gray}, {Corbally}, {Garrison}, {McFadden},
  {Bubar}, {McGahee}, {O'Donoghue}, \& {Knox}}]{Gray2006}
{Gray}, R.~O., {Corbally}, C.~J., {Garrison}, R.~F., {et~al.} 2006, \aj, 132,
  161

\bibitem[{{Gray} {et~al.}(2003){Gray}, {Corbally}, {Garrison}, {McFadden}, \&
  {Robinson}}]{Gray2003}
{Gray}, R.~O., {Corbally}, C.~J., {Garrison}, R.~F., {McFadden}, M.~T., \&
  {Robinson}, P.~E. 2003, \aj, 126, 2048

\bibitem[{{Grevesse} {et~al.}(2007){Grevesse}, {Asplund}, \&
  {Sauval}}]{Grevesse2007}
{Grevesse}, N., {Asplund}, M., \& {Sauval}, A.~J. 2007, \ssr, 130, 105

\bibitem[{{Gustafsson}(1989)}]{Gustafsson1989}
{Gustafsson}, B. 1989, \araa, 27, 701

\bibitem[{{Gustafsson} {et~al.}(2008){Gustafsson}, {Edvardsson}, {Eriksson},
  {J{\o}rgensen}, {Nordlund}, \& {Plez}}]{Gustafsson2008}
{Gustafsson}, B., {Edvardsson}, B., {Eriksson}, K., {et~al.} 2008, \aap, 486,
  951

\bibitem[{{Hauschildt} {et~al.}(1997){Hauschildt}, {Allard}, {Alexander}, \&
  {Baron}}]{Hauschildt1997}
{Hauschildt}, P.~H., {Allard}, F., {Alexander}, D.~R., \& {Baron}, E. 1997,
  \apj, 488, 428

\bibitem[{{Hauschildt} \& {Baron}(1999)}]{Hauschildt1999b}
{Hauschildt}, P.~H. \& {Baron}, E. 1999, Journal of Computational and Applied
  Mathematics, 109, 41

\bibitem[{{Heiter} {et~al.}(2008){Heiter}, {Barklem}, {Fossati}, {Kildiyarova},
  {Kochukhov}, {Kupka}, {Obbrugger}, {Piskunov}, {Plez}, {Ryabchikova},
  {Stempels}, {St{\"u}tz}, \& {Weiss}}]{Heiter2008}
{Heiter}, U., {Barklem}, P., {Fossati}, L., {et~al.} 2008, Journal of Physics
  Conference Series, 130, 012011

\bibitem[{{Heiter} \& {Luck}(2003)}]{Heiter2003}
{Heiter}, U. \& {Luck}, R.~E. 2003, \aj, 126, 2015

\bibitem[{{Herter} {et~al.}(2008){Herter}, {Henderson}, {Wilson}, {Matthews},
  {Rahmer}, {Bonati}, {Muirhead}, {Adams}, {Lloyd}, {Skrutskie}, {Moon},
  {Parshley}, {Nelson}, {Martinache}, \& {Gull}}]{Herter2008}
{Herter}, T.~L., {Henderson}, C.~P., {Wilson}, J.~C., {et~al.} 2008, in Society
  of Photo-Optical Instrumentation Engineers (SPIE) Conference Series, Vol.
  7014, Society of Photo-Optical Instrumentation Engineers (SPIE) Conference
  Series, 0

\bibitem[{{Horne}(1986)}]{Horn1986}
{Horne}, K. 1986, \pasp, 98, 609

\bibitem[{{Howell} {et~al.}(2014){Howell}, {Sobeck}, {Haas}, {Still},
  {Barclay}, {Mullally}, {Troeltzsch}, {Aigrain}, {Bryson}, {Caldwell},
  {Chaplin}, {Cochran}, {Huber}, {Marcy}, {Miglio}, {Najita}, {Smith},
  {Twicken}, \& {Fortney}}]{Howell2014}
{Howell}, S.~B., {Sobeck}, C., {Haas}, M., {et~al.} 2014, \pasp, 126, 398

\bibitem[{{Hu} {et~al.}(2012){Hu}, {Seager}, \& {Bains}}]{Hu2012}
{Hu}, R., {Seager}, S., \& {Bains}, W. 2012, \apj, 761, 166

\bibitem[{{Huang} {et~al.}(2005){Huang}, {Zhao}, {Zhang}, \&
  {Chen}}]{Huang2005}
{Huang}, C., {Zhao}, G., {Zhang}, H.~W., \& {Chen}, Y.~Q. 2005, \mnras, 363, 71

\bibitem[{{H{\"u}nsch} {et~al.}(1999){H{\"u}nsch}, {Schmitt}, {Sterzik}, \&
  {Voges}}]{Hunsch1999}
{H{\"u}nsch}, M., {Schmitt}, J.~H.~M.~M., {Sterzik}, M.~F., \& {Voges}, W.
  1999, \aaps, 135, 319

\bibitem[{{Ida} \& {Lin}(2004{\natexlab{a}})}]{Ida2004a}
{Ida}, S. \& {Lin}, D.~N.~C. 2004{\natexlab{a}}, \apj, 604, 388

\bibitem[{{Ida} \& {Lin}(2004{\natexlab{b}})}]{Ida2004b}
{Ida}, S. \& {Lin}, D.~N.~C. 2004{\natexlab{b}}, \apj, 616, 567

\bibitem[{{Irwin} {et~al.}(2009){Irwin}, {Charbonneau}, {Nutzman}, \&
  {Falco}}]{Irwin2009}
{Irwin}, J., {Charbonneau}, D., {Nutzman}, P., \& {Falco}, E. 2009, in IAU
  Symposium, Vol. 253, IAU Symposium, ed. F.~{Pont}, D.~{Sasselov}, \& M.~J.
  {Holman}, 37--43

\bibitem[{{Jansen} {et~al.}(2001){Jansen}, {Lumb}, {Altieri}, {Clavel}, {Ehle},
  {Erd}, {Gabriel}, {Guainazzi}, {Gondoin}, {Much}, {Munoz}, {Santos},
  {Schartel}, {Texier}, \& {Vacanti}}]{Jansen2001}
{Jansen}, F., {Lumb}, D., {Altieri}, B., {et~al.} 2001, \aap, 365, L1

\bibitem[{{Johns-Krull} \& {Valenti}(1996)}]{Johns-Krull1996}
{Johns-Krull}, C.~M. \& {Valenti}, J.~A. 1996, \apjl, 459, L95

\bibitem[{{Johnson} \& {Apps}(2009)}]{Johnson2009}
{Johnson}, J.~A. \& {Apps}, K. 2009, \apj, 699, 933

\bibitem[{{Johnson} {et~al.}(2007){Johnson}, {Butler}, {Marcy}, {Fischer},
  {Vogt}, {Wright}, \& {Peek}}]{Johnson2007}
{Johnson}, J.~A., {Butler}, R.~P., {Marcy}, G.~W., {et~al.} 2007, \apj, 670,
  833

\bibitem[{{Jones} {et~al.}(2005){Jones}, {Pavlenko}, {Viti}, {Barber},
  {Yakovina}, {Pinfield}, \& {Tennyson}}]{Jones2005}
{Jones}, H.~R.~A., {Pavlenko}, Y., {Viti}, S., {et~al.} 2005, \mnras, 358, 105

\bibitem[{{Kaeufl} {et~al.}(2004){Kaeufl}, {Ballester}, {Biereichel},
  {Delabre}, {Donaldson}, {Dorn}, {Fedrigo}, {Finger}, {Fischer}, {Franza},
  {Gojak}, {Huster}, {Jung}, {Lizon}, {Mehrgan}, {Meyer}, {Moorwood}, {Pirard},
  {Paufique}, {Pozna}, {Siebenmorgen}, {Silber}, {Stegmeier}, \&
  {Wegerer}}]{Kaeufl2004}
{Kaeufl}, H.-U., {Ballester}, P., {Biereichel}, P., {et~al.} 2004, in Society
  of Photo-Optical Instrumentation Engineers (SPIE) Conference Series, Vol.
  5492, Ground-based Instrumentation for Astronomy, ed. A.~F.~M. {Moorwood} \&
  M.~{Iye}, 1218--1227

\bibitem[{{Kang} {et~al.}(2011){Kang}, {Lee}, \& {Kim}}]{Kang2011}
{Kang}, W., {Lee}, S.-G., \& {Kim}, K.-M. 2011, \apj, 736, 87

\bibitem[{{Kharchenko}(2001)}]{Kharchenko2001}
{Kharchenko}, N.~V. 2001, Kinematika i Fizika Nebesnykh Tel, 17, 409

\bibitem[{{Koen} {et~al.}(2010){Koen}, {Kilkenny}, {van Wyk}, \&
  {Marang}}]{Koen2010}
{Koen}, C., {Kilkenny}, D., {van Wyk}, F., \& {Marang}, F. 2010, \mnras, 403,
  1949

\bibitem[{{Kornet} {et~al.}(2005){Kornet}, {Bodenheimer}, {R{\'o}{\.z}yczka},
  \& {Stepinski}}]{Kornet2005}
{Kornet}, K., {Bodenheimer}, P., {R{\'o}{\.z}yczka}, M., \& {Stepinski}, T.~F.
  2005, \aap, 430, 1133

\bibitem[{{Kupka} {et~al.}(2000){Kupka}, {Ryabchikova}, {Piskunov}, {Stempels},
  \& {Weiss}}]{Kupka2000}
{Kupka}, F.~G., {Ryabchikova}, T.~A., {Piskunov}, N.~E., {Stempels}, H.~C., \&
  {Weiss}, W.~W. 2000, Baltic Astronomy, 9, 590

\bibitem[{{Laws} {et~al.}(2003){Laws}, {Gonzalez}, {Walker}, {Tyagi},
  {Dodsworth}, {Snider}, \& {Suntzeff}}]{Laws2003}
{Laws}, C., {Gonzalez}, G., {Walker}, K.~M., {et~al.} 2003, \aj, 125, 2664

\bibitem[{{Lee} {et~al.}(2011){Lee}, {Beers}, {Allende Prieto}, {Lai},
  {Rockosi}, {Morrison}, {Johnson}, {An}, {Sivarani}, \& {Yanny}}]{Lee2011}
{Lee}, Y.~S., {Beers}, T.~C., {Allende Prieto}, C., {et~al.} 2011, \aj, 141, 90

\bibitem[{{Livingston} \& {Wallace}(1991)}]{Livingston1991}
{Livingston}, W. \& {Wallace}, L. 1991, {An atlas of the solar spectrum in the
  infrared from 1850 to 9000 cm-1 (1.1 to 5.4 micrometer)}

\bibitem[{{Lovis} {et~al.}(2005){Lovis}, {Mayor}, {Bouchy}, {Pepe}, {Queloz},
  {Santos}, {Udry}, {Benz}, {Bertaux}, {Mordasini}, \& {Sivan}}]{Lovis2005}
{Lovis}, C., {Mayor}, M., {Bouchy}, F., {et~al.} 2005, \aap, 437, 1121

\bibitem[{{Luck} \& {Heiter}(2006)}]{Luck2006}
{Luck}, R.~E. \& {Heiter}, U. 2006, \aj, 131, 3069

\bibitem[{{Maldonado} {et~al.}(2013){Maldonado}, {Villaver}, \&
  {Eiroa}}]{Maldonado2013}
{Maldonado}, J., {Villaver}, E., \& {Eiroa}, C. 2013, \aap, 554, A84

\bibitem[{{Mann} {et~al.}(2013){Mann}, {Brewer}, {Gaidos}, {L{\'e}pine}, \&
  {Hilton}}]{Mann2013a}
{Mann}, A.~W., {Brewer}, J.~M., {Gaidos}, E., {L{\'e}pine}, S., \& {Hilton},
  E.~J. 2013, \aj, 145, 52

\bibitem[{{Mann} {et~al.}(2014){Mann}, {Deacon}, {Gaidos}, {Ansdell}, {Brewer},
  {Liu}, {Magnier}, \& {Aller}}]{Mann2014}
{Mann}, A.~W., {Deacon}, N.~R., {Gaidos}, E., {et~al.} 2014, \aj, 147, 160

\bibitem[{{Mann} {et~al.}(2015){Mann}, {Feiden}, {Gaidos}, {Boyajian}, \& {von
  Braun}}]{Mann2015}
{Mann}, A.~W., {Feiden}, G.~A., {Gaidos}, E., {Boyajian}, T., \& {von Braun},
  K. 2015, \apj, 804, 64

\bibitem[{{Marcy} {et~al.}(2000){Marcy}, {Butler}, \& {Vogt}}]{Marcy2000}
{Marcy}, G.~W., {Butler}, R.~P., \& {Vogt}, S.~S. 2000, \apjl, 536, L43

\bibitem[{{Marcy} {et~al.}(1998){Marcy}, {Butler}, {Vogt}, {Fischer}, \&
  {Lissauer}}]{Marcy1998}
{Marcy}, G.~W., {Butler}, R.~P., {Vogt}, S.~S., {Fischer}, D., \& {Lissauer},
  J.~J. 1998, \apjl, 505, L147

\bibitem[{{Marcy} \& {Chen}(1992)}]{Marcy1992}
{Marcy}, G.~W. \& {Chen}, G.~H. 1992, \apj, 390, 550

\bibitem[{{Mart{\'{\i}}nez-Arn{\'a}iz}
  {et~al.}(2010){Mart{\'{\i}}nez-Arn{\'a}iz}, {Maldonado}, {Montes}, {Eiroa},
  \& {Montesinos}}]{Martinez-Arnaiz2010}
{Mart{\'{\i}}nez-Arn{\'a}iz}, R., {Maldonado}, J., {Montes}, D., {Eiroa}, C.,
  \& {Montesinos}, B. 2010, \aap, 520, A79

\bibitem[{{Mason} {et~al.}(2001){Mason}, {Wycoff}, {Hartkopf}, {Douglass}, \&
  {Worley}}]{Mason2001}
{Mason}, B.~D., {Wycoff}, G.~L., {Hartkopf}, W.~I., {Douglass}, G.~G., \&
  {Worley}, C.~E. 2001, \aj, 122, 3466

\bibitem[{{Mayor} \& {Queloz}(1995)}]{Mayor1995}
{Mayor}, M. \& {Queloz}, D. 1995, \nat, 378, 355

\bibitem[{{Mel{\'e}ndez} \& {Barbuy}(1999)}]{Melendez1999}
{Mel{\'e}ndez}, J. \& {Barbuy}, B. 1999, \apjs, 124, 527

\bibitem[{{Miller} {et~al.}(1994){Miller}, {Tennyson}, {Jones}, \&
  {Longmore}}]{Miller1994}
{Miller}, S., {Tennyson}, J., {Jones}, H.~R.~A., \& {Longmore}, A.~J. 1994,
  {IAU Colloq. 146, Molecules in the Stellar Environment}, Vol. 296

\bibitem[{{Mishenina} {et~al.}(2004){Mishenina}, {Soubiran}, {Kovtyukh}, \&
  {Korotin}}]{Mishenina2004}
{Mishenina}, T.~V., {Soubiran}, C., {Kovtyukh}, V.~V., \& {Korotin}, S.~A.
  2004, \aap, 418, 551

\bibitem[{{Mordasini} {et~al.}(2008){Mordasini}, {Alibert}, {Benz}, \&
  {Naef}}]{Mordasini2008}
{Mordasini}, C., {Alibert}, Y., {Benz}, W., \& {Naef}, D. 2008, in Astronomical
  Society of the Pacific Conference Series, Vol. 398, Extreme Solar Systems,
  ed. D.~{Fischer}, F.~A. {Rasio}, S.~E. {Thorsett}, \& A.~{Wolszczan}, 235

\bibitem[{{Mould}(1976{\natexlab{a}})}]{Mould1976a}
{Mould}, J.~R. 1976{\natexlab{a}}, \aap, 48, 443

\bibitem[{{Mould}(1976{\natexlab{b}})}]{Mould1976b}
{Mould}, J.~R. 1976{\natexlab{b}}, \apj, 210, 402

\bibitem[{{Mould}(1978)}]{Mould1978}
{Mould}, J.~R. 1978, \apj, 226, 923

\bibitem[{{Mugrauer} {et~al.}(2004){Mugrauer}, {Neuh{\"a}user}, {Mazeh},
  {Alves}, \& {Guenther}}]{Mugrauer2004}
{Mugrauer}, M., {Neuh{\"a}user}, R., {Mazeh}, T., {Alves}, J., \& {Guenther},
  E. 2004, \aap, 425, 249

\bibitem[{{Mugrauer} {et~al.}(2005){Mugrauer}, {Neuh{\"a}user}, {Seifahrt},
  {Mazeh}, \& {Guenther}}]{Mugrauer2005}
{Mugrauer}, M., {Neuh{\"a}user}, R., {Seifahrt}, A., {Mazeh}, T., \&
  {Guenther}, E. 2005, \aap, 440, 1051

\bibitem[{{Mugrauer} {et~al.}(2007){Mugrauer}, {Seifahrt}, \&
  {Neuh{\"a}user}}]{Mugrauer2007}
{Mugrauer}, M., {Seifahrt}, A., \& {Neuh{\"a}user}, R. 2007, \mnras, 378, 1328

\bibitem[{{Neves} {et~al.}(2012){Neves}, {Bonfils}, {Santos}, {Delfosse},
  {Forveille}, {Allard}, {Nat{\'a}rio}, {Fernandes}, \& {Udry}}]{Neves2012}
{Neves}, V., {Bonfils}, X., {Santos}, N.~C., {et~al.} 2012, \aap, 538, A25

\bibitem[{{Neves} {et~al.}(2013){Neves}, {Bonfils}, {Santos}, {Delfosse},
  {Forveille}, {Allard}, \& {Udry}}]{Neves2013}
{Neves}, V., {Bonfils}, X., {Santos}, N.~C., {et~al.} 2013, \aap, 551, A36

\bibitem[{{Newton} {et~al.}(2014){Newton}, {Charbonneau}, {Irwin},
  {Berta-Thompson}, {Rojas-Ayala}, {Covey}, \& {Lloyd}}]{Newton2014}
{Newton}, E.~R., {Charbonneau}, D., {Irwin}, J., {et~al.} 2014, \aj, 147, 20

\bibitem[{{Noyes} {et~al.}(1984){Noyes}, {Weiss}, \& {Vaughan}}]{Noyes1984}
{Noyes}, R.~W., {Weiss}, N.~O., \& {Vaughan}, A.~H. 1984, \apj, 287, 769

\bibitem[{{{\"O}nehag} {et~al.}(2012){{\"O}nehag}, {Heiter}, {Gustafsson},
  {Piskunov}, {Plez}, \& {Reiners}}]{Onehag2012}
{{\"O}nehag}, A., {Heiter}, U., {Gustafsson}, B., {et~al.} 2012, \aap, 542, A33

\bibitem[{{Origlia} {et~al.}(2014){Origlia}, {Oliva}, {Baffa}, {Falcini},
  {Giani}, {Massi}, {Montegriffo}, {Sanna}, {Scuderi}, {Sozzi}, {Tozzi},
  {Carleo}, {Gratton}, {Ghinassi}, \& {Lodi}}]{Origlia2014}
{Origlia}, L., {Oliva}, E., {Baffa}, C., {et~al.} 2014, in Society of
  Photo-Optical Instrumentation Engineers (SPIE) Conference Series, Vol. 9147,
  Society of Photo-Optical Instrumentation Engineers (SPIE) Conference Series,
  1

\bibitem[{{Patience} {et~al.}(2002){Patience}, {White}, {Ghez}, {McCabe},
  {McLean}, {Larkin}, {Prato}, {Kim}, {Lloyd}, {Liu}, {Graham}, {Macintosh},
  {Gavel}, {Max}, {Bauman}, {Olivier}, {Wizinowich}, \& {Acton}}]{Patience2002}
{Patience}, J., {White}, R.~J., {Ghez}, A.~M., {et~al.} 2002, \apj, 581, 654

\bibitem[{{Plez}(1998)}]{Plez1998}
{Plez}, B. 1998, \aap, 337, 495

\bibitem[{{Poppenhaeger} {et~al.}(2010){Poppenhaeger}, {Robrade}, \&
  {Schmitt}}]{Poppenhaeger2010}
{Poppenhaeger}, K., {Robrade}, J., \& {Schmitt}, J.~H.~M.~M. 2010, \aap, 515,
  A98

\bibitem[{{Poveda} {et~al.}(1994){Poveda}, {Herrera}, {Allen}, {Cordero}, \&
  {Lavalley}}]{Poveda1994}
{Poveda}, A., {Herrera}, M.~A., {Allen}, C., {Cordero}, G., \& {Lavalley}, C.
  1994, \rmxaa, 28, 43

\bibitem[{{Quirrenbach} {et~al.}(2014){Quirrenbach}, {Amado}, {Caballero},
  {Mundt}, {Reiners}, {Ribas}, {Seifert}, {Abril}, {Aceituno},
  {Alonso-Floriano}, {Ammler-von Eiff}, {Antona Jim{\'e}nez},
  {Anwand-Heerwart}, {Azzaro}, {Bauer}, {Barrado}, {Becerril}, {B{\'e}jar},
  {Ben{\'{\i}}tez}, {Berdi{\~n}as}, {C{\'a}rdenas}, {Casal}, {Claret},
  {Colom{\'e}}, {Cort{\'e}s-Contreras}, {Czesla}, {Doellinger}, {Dreizler},
  {Feiz}, {Fern{\'a}ndez}, {Galad{\'{\i}}}, {G{\'a}lvez-Ortiz},
  {Garc{\'{\i}}a-Piquer}, {Garc{\'{\i}}a-Vargas}, {Garrido}, {Gesa}, {G{\'o}mez
  Galera}, {Gonz{\'a}lez {\'A}lvarez}, {Gonz{\'a}lez Hern{\'a}ndez},
  {Gr{\"o}zinger}, {Gu{\`a}rdia}, {Guenther}, {de Guindos},
  {Guti{\'e}rrez-Soto}, {Hagen}, {Hatzes}, {Hauschildt}, {Helmling}, {Henning},
  {Hermann}, {Hern{\'a}ndez Casta{\~n}o}, {Herrero}, {Hidalgo}, {Holgado},
  {Huber}, {Huber}, {Jeffers}, {Joergens}, {de Juan}, {Kehr}, {Klein},
  {K{\"u}rster}, {Lamert}, {Lalitha}, {Laun}, {Lemke}, {Lenzen}, {L{\'o}pez del
  Fresno}, {L{\'o}pez Mart{\'{\i}}}, {L{\'o}pez-Santiago}, {Mall}, {Mandel},
  {Mart{\'{\i}}n}, {Mart{\'{\i}}n-Ruiz}, {Mart{\'{\i}}nez-Rodr{\'{\i}}guez},
  {Marvin}, {Mathar}, {Mirabet}, {Montes}, {Morales Mu{\~n}oz}, {Moya},
  {Naranjo}, {Ofir}, {Oreiro}, {Pall{\'e}}, {Panduro}, {Passegger},
  {P{\'e}rez-Calpena}, {P{\'e}rez Medialdea}, {Perger}, {Pluto}, {Ram{\'o}n},
  {Rebolo}, {Redondo}, {Reffert}, {Reinhardt}, {Rhode}, {Rix}, {Rodler},
  {Rodr{\'{\i}}guez}, {Rodr{\'{\i}}guez-L{\'o}pez},
  {Rodr{\'{\i}}guez-P{\'e}rez}, {Rohloff}, {Rosich}, {S{\'a}nchez-Blanco},
  {S{\'a}nchez Carrasco}, {Sanz-Forcada}, {Sarmiento}, {Sch{\"a}fer},
  {Schiller}, {Schmidt}, {Schmitt}, {Solano}, {Stahl}, {Storz}, {St{\"u}rmer},
  {Su{\'a}rez}, {Ulbrich}, {Veredas}, {Wagner}, {Winkler}, {Zapatero Osorio},
  {Zechmeister}, {Abell{\'a}n de Paco}, {Anglada-Escud{\'e}}, {del Burgo},
  {Klutsch}, {Lizon}, {L{\'o}pez-Morales}, {Morales}, {Perryman}, {Tulloch}, \&
  {Xu}}]{Quirrenbach2014}
{Quirrenbach}, A., {Amado}, P.~J., {Caballero}, J.~A., {et~al.} 2014, in
  Society of Photo-Optical Instrumentation Engineers (SPIE) Conference Series,
  Vol. 9147, Society of Photo-Optical Instrumentation Engineers (SPIE)
  Conference Series, 1

\bibitem[{{Quirrenbach} {et~al.}(2010){Quirrenbach}, {Amado}, {Mandel},
  {Caballero}, {Ribas}, {Reiners}, {Mundt}, \& {CARMENES
  Consortium}}]{Quirrenbach2010}
{Quirrenbach}, A., {Amado}, P.~J., {Mandel}, H., {et~al.} 2010, in Astronomical
  Society of the Pacific Conference Series, Vol. 430, Pathways Towards
  Habitable Planets, ed. V.~{Coud{\'e} du Foresto}, D.~M. {Gelino}, \&
  I.~{Ribas}, 521

\bibitem[{{Rayner} {et~al.}(2003){Rayner}, {Toomey}, {Onaka}, {Denault},
  {Stahlberger}, {Vacca}, {Cushing}, \& {Wang}}]{Rayner2003}
{Rayner}, J.~T., {Toomey}, D.~W., {Onaka}, P.~M., {et~al.} 2003, \pasp, 115,
  362

\bibitem[{{Reid} {et~al.}(1995){Reid}, {Hawley}, \& {Gizis}}]{Reid1995}
{Reid}, I.~N., {Hawley}, S.~L., \& {Gizis}, J.~E. 1995, \aj, 110, 1838

\bibitem[{{Reiners}(2007)}]{Reiners2007b}
{Reiners}, A. 2007, \aap, 467, 259

\bibitem[{{Reiners} {et~al.}(2012){Reiners}, {Joshi}, \&
  {Goldman}}]{Reiners2012}
{Reiners}, A., {Joshi}, N., \& {Goldman}, B. 2012, \aj, 143, 93

\bibitem[{{Rojas-Ayala} {et~al.}(2010){Rojas-Ayala}, {Covey}, {Muirhead}, \&
  {Lloyd}}]{Rojas-Ayala2010}
{Rojas-Ayala}, B., {Covey}, K.~R., {Muirhead}, P.~S., \& {Lloyd}, J.~P. 2010,
  \apjl, 720, L113

\bibitem[{{Rojas-Ayala} {et~al.}(2012){Rojas-Ayala}, {Covey}, {Muirhead}, \&
  {Lloyd}}]{Rojas-Ayala2012}
{Rojas-Ayala}, B., {Covey}, K.~R., {Muirhead}, P.~S., \& {Lloyd}, J.~P. 2012,
  \apj, 748, 93

\bibitem[{{Rosenblatt}(1971)}]{Rosenblatt1971}
{Rosenblatt}, F. 1971, \icarus, 14, 71

\bibitem[{{Russell}(1934)}]{Russell1934}
{Russell}, H.~N. 1934, \apj, 79, 317

\bibitem[{{Rutten} \& {Schrijver}(1987)}]{Rutten1987}
{Rutten}, R.~G.~M. \& {Schrijver}, C.~J. 1987, \aap, 177, 155

\bibitem[{{Saffe} {et~al.}(2005){Saffe}, {G{\'o}mez}, \& {Chavero}}]{Saffe2005}
{Saffe}, C., {G{\'o}mez}, M., \& {Chavero}, C. 2005, \aap, 443, 609

\bibitem[{{Santos} {et~al.}(2001){Santos}, {Israelian}, \&
  {Mayor}}]{Santos2001}
{Santos}, N.~C., {Israelian}, G., \& {Mayor}, M. 2001, \aap, 373, 1019

\bibitem[{{Santos} {et~al.}(2004){Santos}, {Israelian}, \&
  {Mayor}}]{Santos2004}
{Santos}, N.~C., {Israelian}, G., \& {Mayor}, M. 2004, \aap, 415, 1153

\bibitem[{{Santos} {et~al.}(2005){Santos}, {Israelian}, {Mayor}, {Bento},
  {Almeida}, {Sousa}, \& {Ecuvillon}}]{Santos2005}
{Santos}, N.~C., {Israelian}, G., {Mayor}, M., {et~al.} 2005, \aap, 437, 1127

\bibitem[{{Santos} {et~al.}(2013){Santos}, {Sousa}, {Mortier}, {Neves},
  {Adibekyan}, {Tsantaki}, {Delgado Mena}, {Bonfils}, {Israelian}, {Mayor}, \&
  {Udry}}]{Santos2013}
{Santos}, N.~C., {Sousa}, S.~G., {Mortier}, A., {et~al.} 2013, \aap, 556, A150

\bibitem[{{Schlaufman} \& {Laughlin}(2010)}]{Schlaufman2010}
{Schlaufman}, K.~C. \& {Laughlin}, G. 2010, \aap, 519, A105

\bibitem[{{Schneider} {et~al.}(2011){Schneider}, {Dedieu}, {Le Sidaner},
  {Savalle}, \& {Zolotukhin}}]{Schneider2011}
{Schneider}, J., {Dedieu}, C., {Le Sidaner}, P., {Savalle}, R., \&
  {Zolotukhin}, I. 2011, \aap, 532, A79

\bibitem[{{Seager} \& {Mall{\'e}n-Ornelas}(2003)}]{Seager2003}
{Seager}, S. \& {Mall{\'e}n-Ornelas}, G. 2003, \apj, 585, 1038

\bibitem[{{Sneden}(1973)}]{Sneden1973}
{Sneden}, C. 1973, \apj, 184, 839

\bibitem[{{Soubiran} {et~al.}(2010){Soubiran}, {Le Campion}, {Cayrel de
  Strobel}, \& {Caillo}}]{Soubiran2010}
{Soubiran}, C., {Le Campion}, J.-F., {Cayrel de Strobel}, G., \& {Caillo}, A.
  2010, \aap, 515, A111

\bibitem[{{Sousa} {et~al.}(2011){Sousa}, {Santos}, {Israelian}, {Mayor}, \&
  {Udry}}]{Sousa2011}
{Sousa}, S.~G., {Santos}, N.~C., {Israelian}, G., {Mayor}, M., \& {Udry}, S.
  2011, \aap, 533, A141

\bibitem[{{Sousa} {et~al.}(2008){Sousa}, {Santos}, {Mayor}, {Udry},
  {Casagrande}, {Israelian}, {Pepe}, {Queloz}, \& {Monteiro}}]{Sousa2008}
{Sousa}, S.~G., {Santos}, N.~C., {Mayor}, M., {et~al.} 2008, \aap, 487, 373

\bibitem[{{Struve}(1952)}]{Struve1952}
{Struve}, O. 1952, The Observatory, 72, 199

\bibitem[{{Takeda} {et~al.}(2005){Takeda}, {Ohkubo}, {Sato}, {Kambe}, \&
  {Sadakane}}]{Takeda2005}
{Takeda}, Y., {Ohkubo}, M., {Sato}, B., {Kambe}, E., \& {Sadakane}, K. 2005,
  \pasj, 57, 27

\bibitem[{{Terrien} {et~al.}(2012){Terrien}, {Mahadevan}, {Bender},
  {Deshpande}, {Ramsey}, \& {Bochanski}}]{Terrien2012}
{Terrien}, R.~C., {Mahadevan}, S., {Bender}, C.~F., {et~al.} 2012, \apjl, 747,
  L38

\bibitem[{{Tian} {et~al.}(2014){Tian}, {France}, {Linsky}, {Mauas}, \&
  {Vieytes}}]{Tian2014}
{Tian}, F., {France}, K., {Linsky}, J.~L., {Mauas}, P.~J.~D., \& {Vieytes},
  M.~C. 2014, Earth and Planetary Science Letters, 385, 22

\bibitem[{{Torres} {et~al.}(2006){Torres}, {Quast}, {da Silva}, {de La Reza},
  {Melo}, \& {Sterzik}}]{Torres2006}
{Torres}, C.~A.~O., {Quast}, G.~R., {da Silva}, L., {et~al.} 2006, \aap, 460,
  695

\bibitem[{{Tsantaki} {et~al.}(2013){Tsantaki}, {Sousa}, {Adibekyan}, {Santos},
  {Mortier}, \& {Israelian}}]{Tsantaki2013}
{Tsantaki}, M., {Sousa}, S.~G., {Adibekyan}, V.~Z., {et~al.} 2013, \aap, 555,
  A150

\bibitem[{{Tsuji}(1966)}]{Tsuji1966}
{Tsuji}, T. 1966, \pasj, 18, 127

\bibitem[{{Turon} {et~al.}(1993){Turon}, {Creze}, {Egret}, {Gomez}, {Grenon},
  {Jahrei{\ss}}, {Requieme}, {Argue}, {Bec-Borsenberger}, {Dommanget},
  {Mennessier}, {Arenou}, {Chareton}, {Crifo}, {Mermilliod}, {Morin},
  {Nicolet}, {Nys}, {Prevot}, {Rousseau}, {Perryman}, \& {et al.}}]{Turon1993}
{Turon}, C., {Creze}, M., {Egret}, D., {et~al.} 1993, Bulletin d'Information du
  Centre de Donnees Stellaires, 43, 5

\bibitem[{{Valenti} \& {Fischer}(2005)}]{Valenti2005}
{Valenti}, J.~A. \& {Fischer}, D.~A. 2005, \apjs, 159, 141

\bibitem[{{Valenti} \& {Piskunov}(1996)}]{Valenti1996}
{Valenti}, J.~A. \& {Piskunov}, N. 1996, \aaps, 118, 595

\bibitem[{{Valenti} {et~al.}(1998){Valenti}, {Piskunov}, \&
  {Johns-Krull}}]{Valenti1998}
{Valenti}, J.~A., {Piskunov}, N., \& {Johns-Krull}, C.~M. 1998, \apj, 498, 851

\bibitem[{{van Leeuwen}(2007)}]{vanLeeuwe2007}
{van Leeuwen}, F., ed. 2007, Astrophysics and Space Science Library, Vol. 350,
  {Hipparcos, the New Reduction of the Raw Data}

\bibitem[{{Woolf} {et~al.}(2009){Woolf}, {L{\'e}pine}, \&
  {Wallerstein}}]{Woolf2009}
{Woolf}, V.~M., {L{\'e}pine}, S., \& {Wallerstein}, G. 2009, \pasp, 121, 117

\bibitem[{{Woolf} \& {Wallerstein}(2006)}]{Woolf2006}
{Woolf}, V.~M. \& {Wallerstein}, G. 2006, \pasp, 118, 218

\bibitem[{{Zhao} {et~al.}(2002){Zhao}, {Chen}, {Qiu}, \& {Li}}]{Zhao2002}
{Zhao}, G., {Chen}, Y.~Q., {Qiu}, H.~M., \& {Li}, Z.~W. 2002, \aj, 124, 2224

\end{thebibliography}

\onecolumn

\Online

\begin{appendix}

\section{Spectra} \label{app:spectra}

\begin{figure*}[ht!]
\begin{center}
\includegraphics[width=0.92\textwidth]{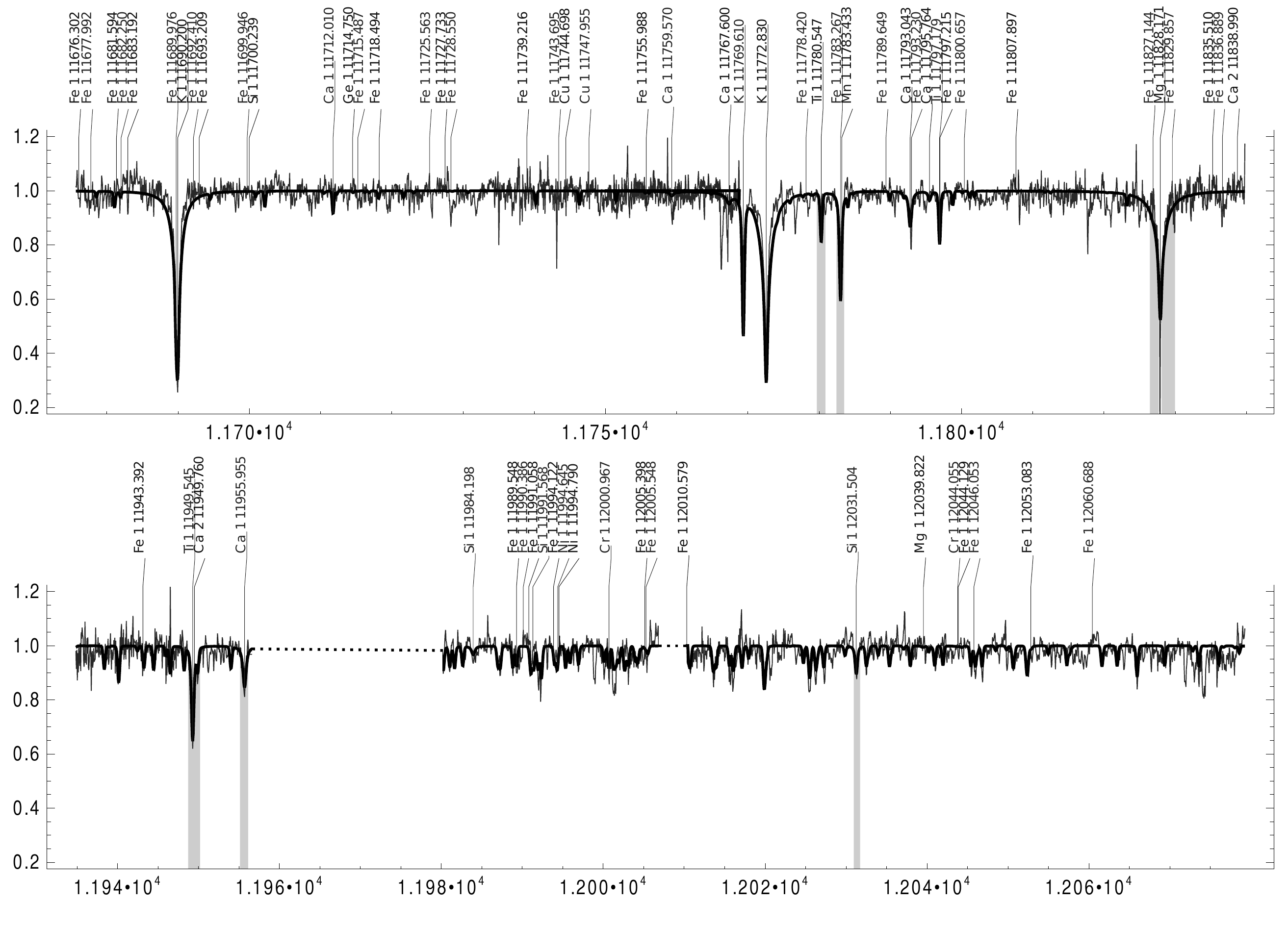}\\
\includegraphics[width=0.92\textwidth]{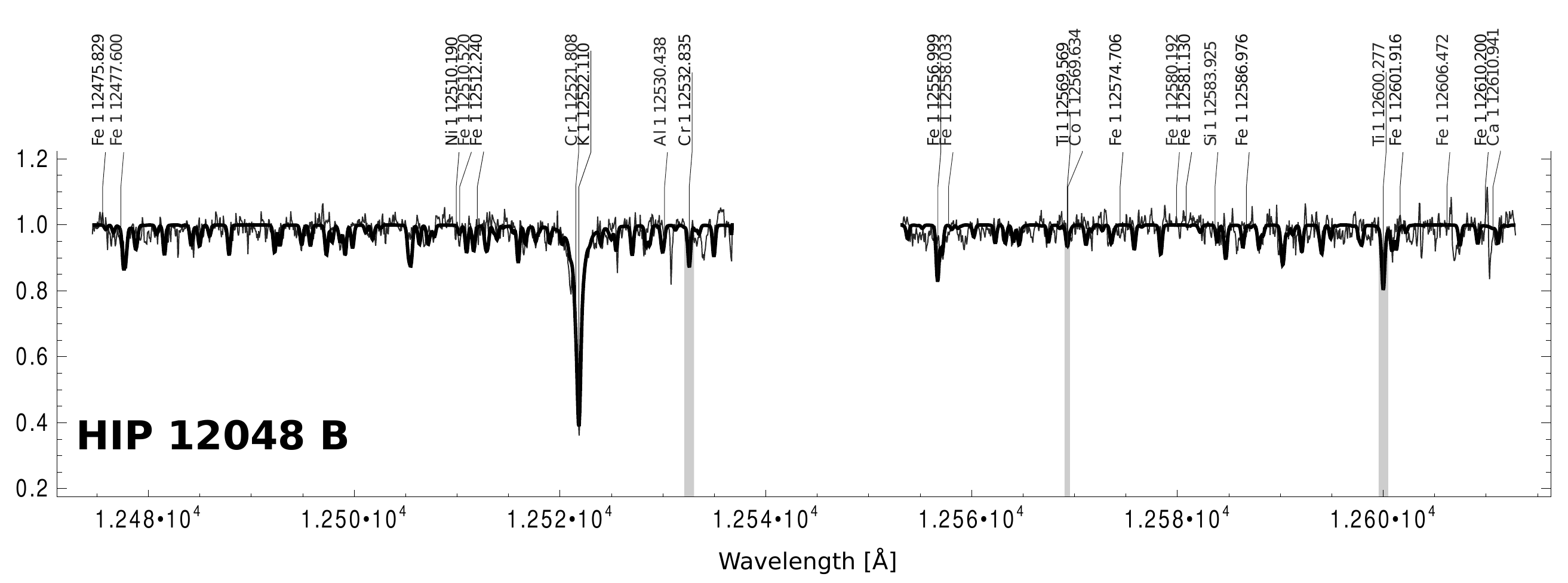}
\end{center}
\caption{Our obtained spectra after continuum rectification, wavelength correction, and removal of telluric features of the M~dwarf HIP~12048~B. The best-fit synthetic spectrum calculated with SME is shown as the overplotted thick black line. The line mask used for the metallicity determination is indicated by grey shading. The dotted lines indicate the two parts that we had to remove from the spectra since a large unphysical absorption feature was present in the observed spectrum of the standard star. Spectra centred at 1177, 1811, and 1204~nm have been co-added where they overlap in this figure. White spacing represents wavelength regions between two chips. We note that the observations centred at 1303~nm have not been used in the analysis owing to problems with the observed spectra in this wavelength for target HIP~12048~B.}
\label{spectra1}
\end{figure*}

\newpage

\begin{figure*}
\begin{center}
\includegraphics[width=0.92\textwidth]{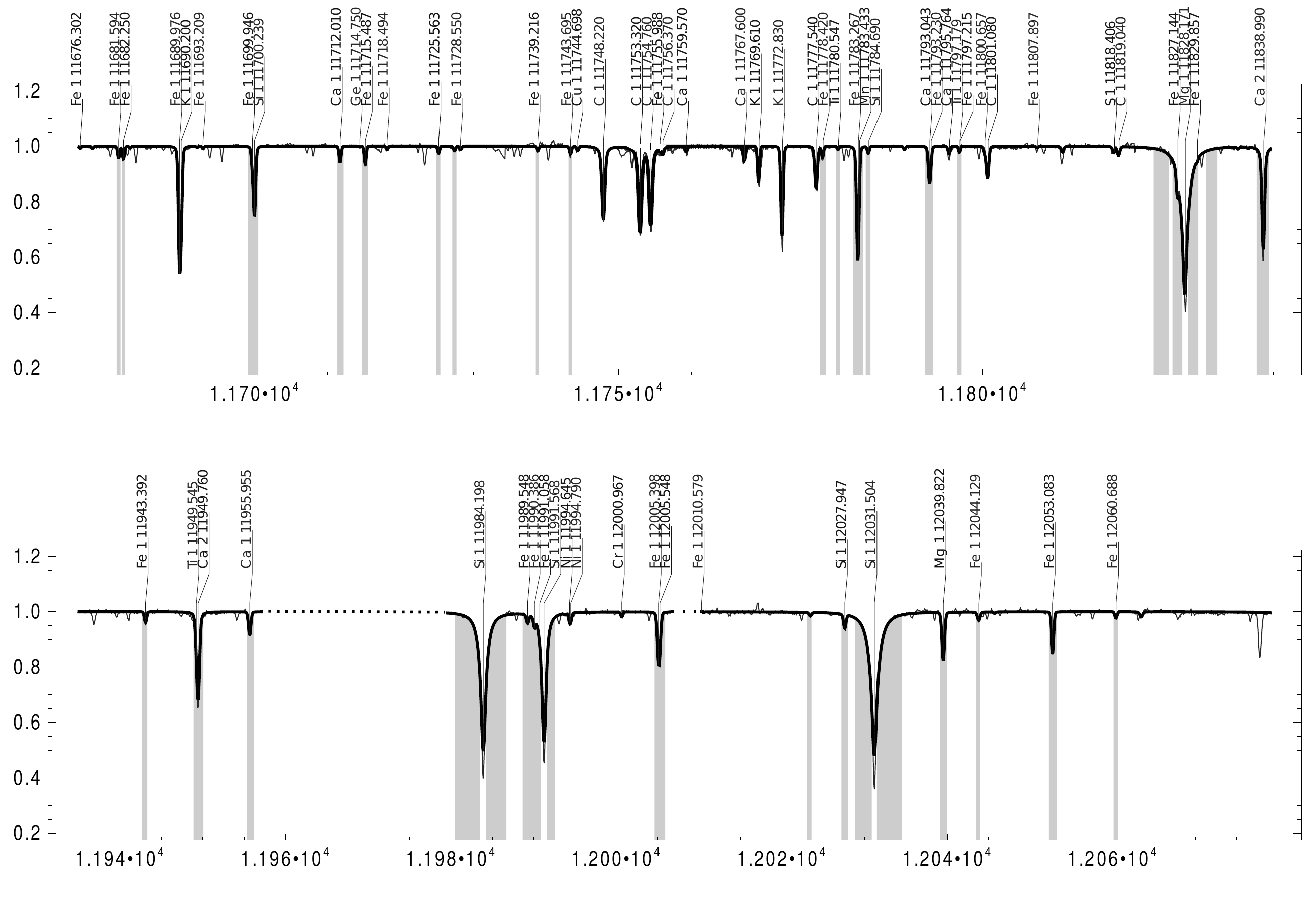}\\
\includegraphics[width=0.92\textwidth]{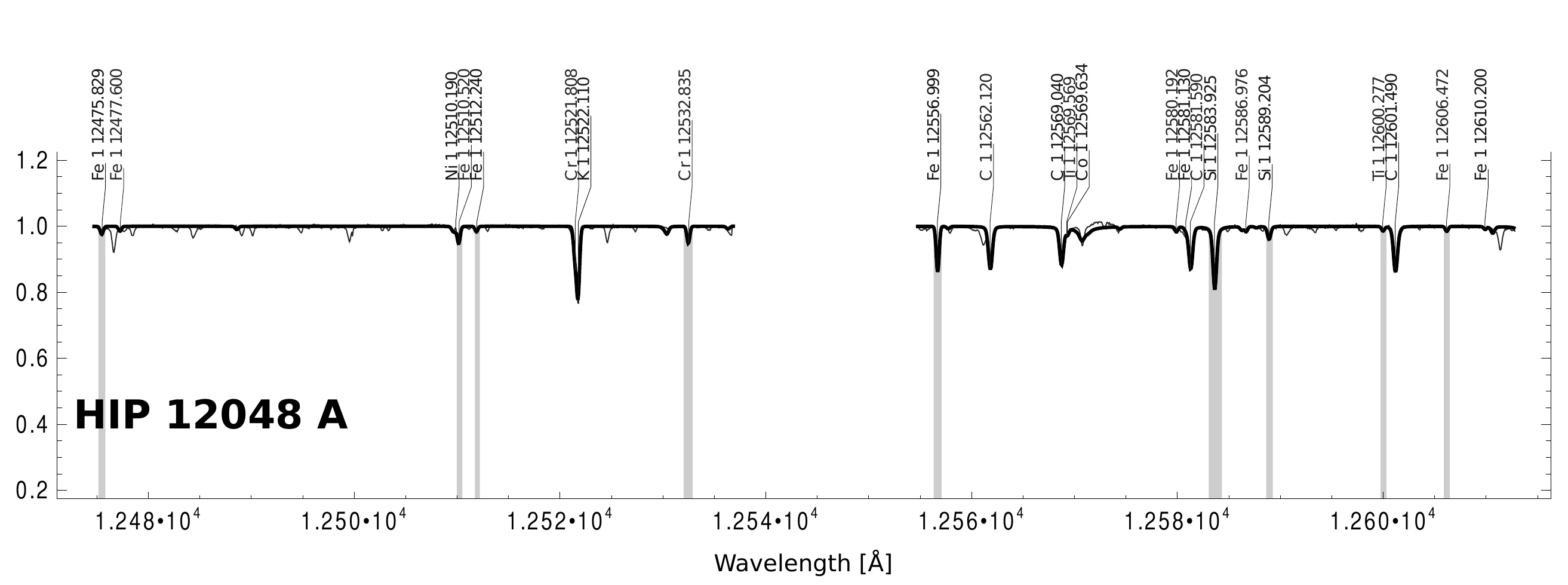}
\end{center}
\caption{Our obtained spectra after continuum rectification, wavelength correction, and removal of telluric features of the G~dwarf HIP~12048~A. The best-fit synthetic spectrum calculated with SME is shown as the overplotted thick black line. The line mask used for the metallicity determination is indicated by  grey shading. The dotted lines indicate two parts that we had to remove from the spectra since a large unphysical absorption feature was present in the observed spectrum of the standard star. Spectra centred at 1177, 1811, and 1204~nm have been co-added where they overlap in this figure. White spacing represents wavelength regions between two chips. We note that the observations centred at 1303~nm have not been used in the analysis owing to problems with the observed spectra in this wavelength for target HIP~12048~A.
}
\label{spectra2}
\end{figure*}

\newpage

\begin{figure*}
\begin{center}
\includegraphics[width=0.92\textwidth]{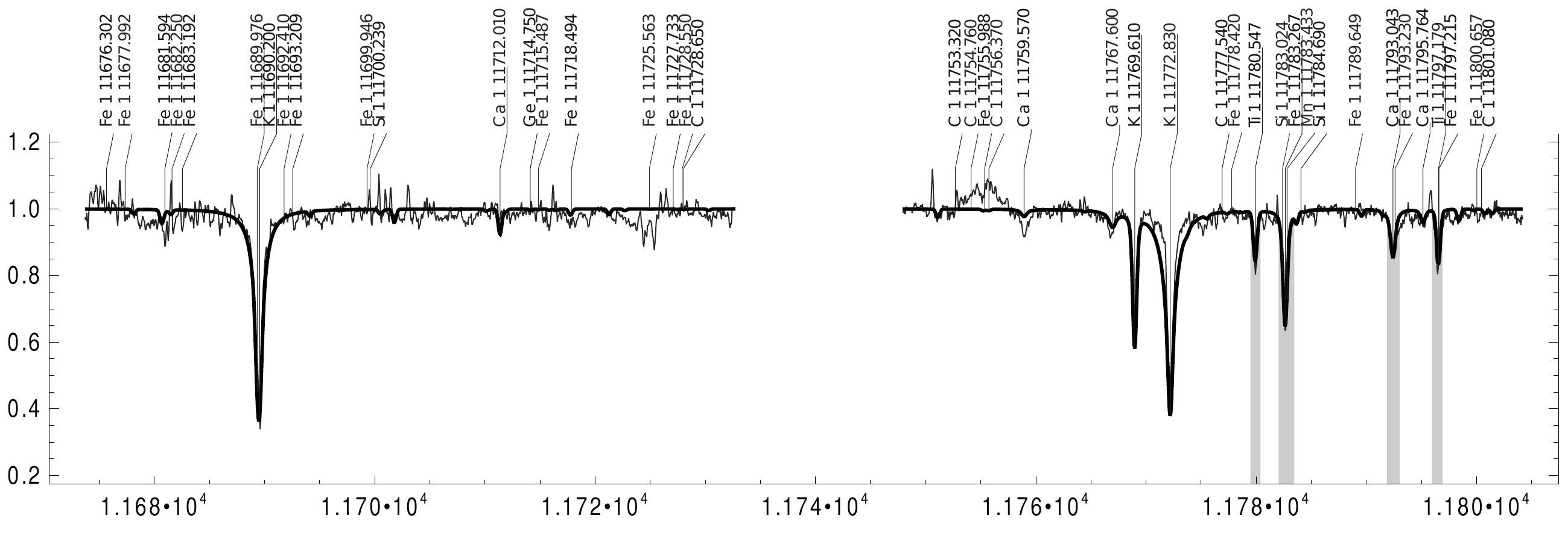}\\
\includegraphics[width=0.92\textwidth]{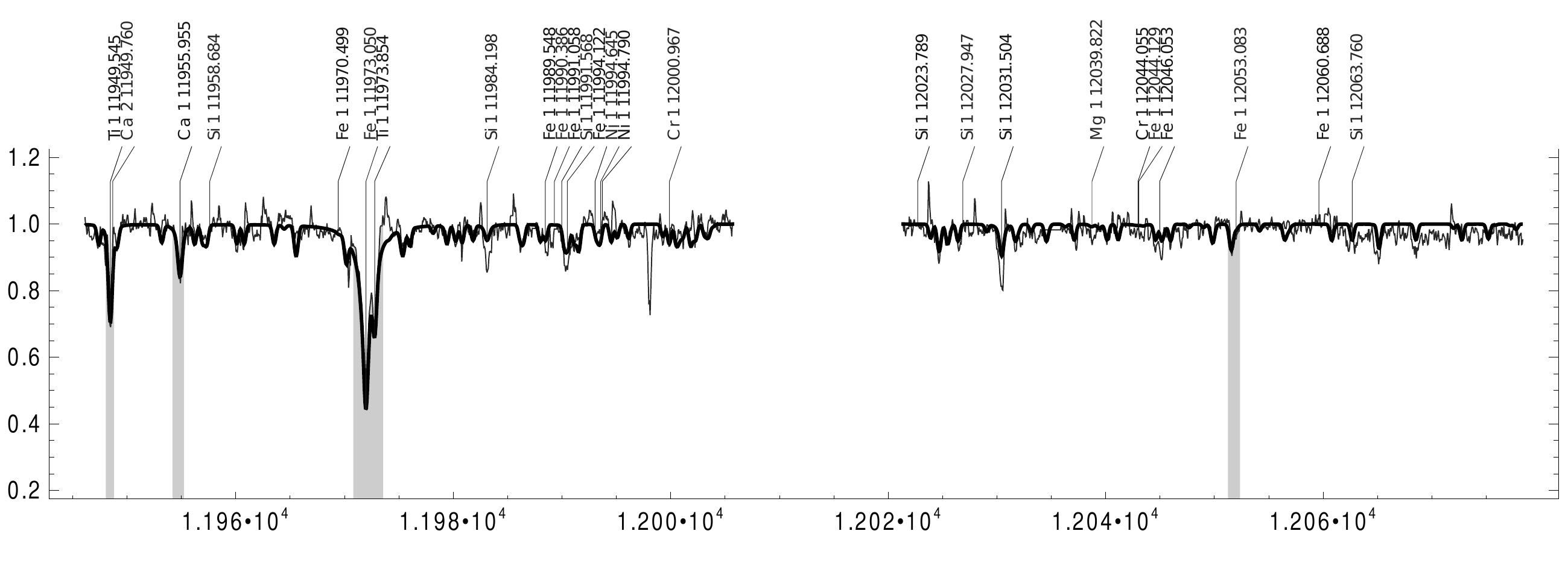}\\
\includegraphics[width=0.92\textwidth]{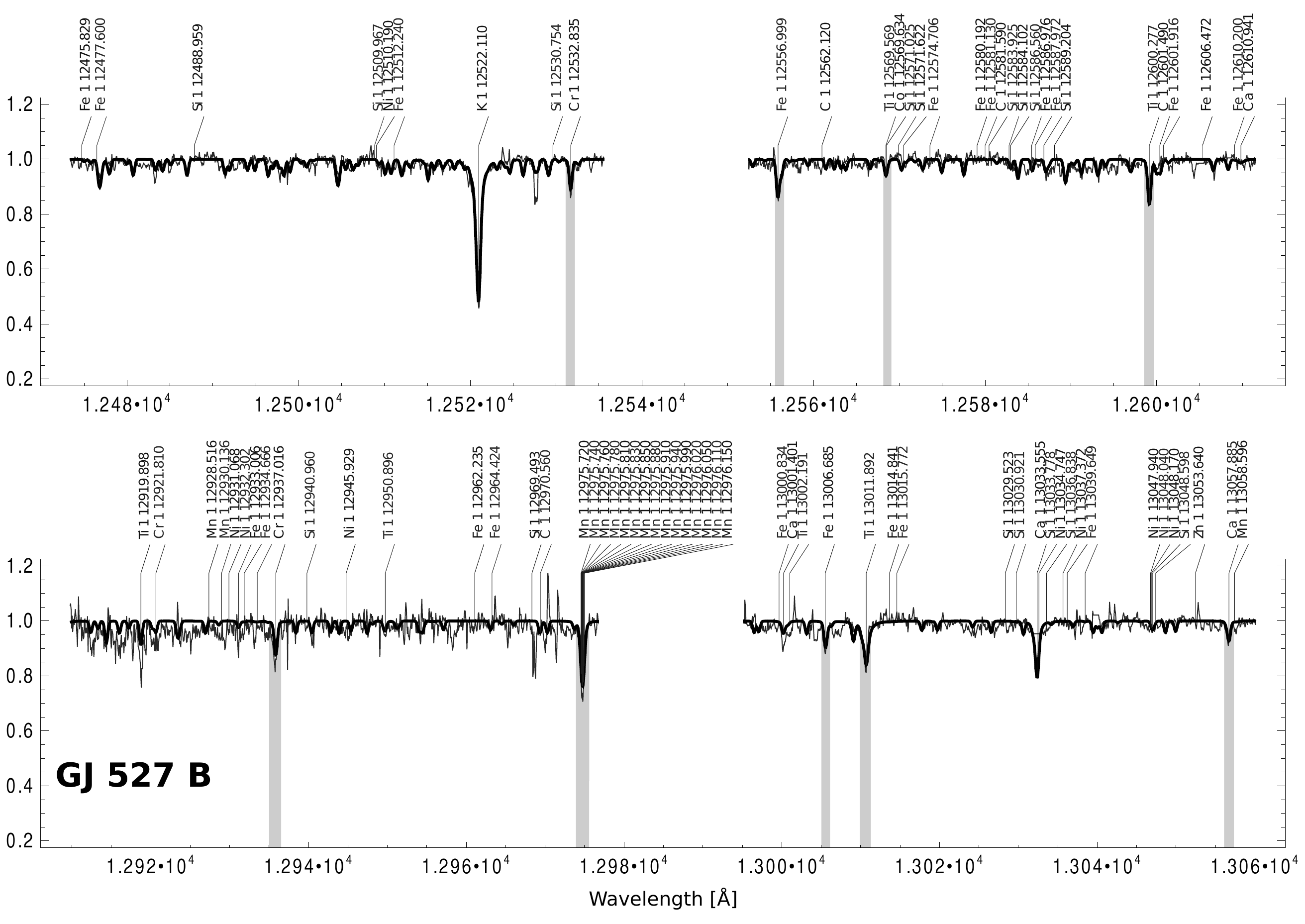}
\end{center}
\caption{Our obtained spectra after continuum rectification, wavelength correction, and removal of telluric features of M~dwarf GJ~527~B. The best-fit synthetic spectrum calculated with SME is shown as the overplotted thick black line. The line mask used for the metallicity determination is indicated by  grey shading. White spacing represents wavelength regions between two chips.}
\label{spectra3}
\end{figure*}

\newpage

\begin{figure*}
\begin{center}
\includegraphics[width=0.92\textwidth]{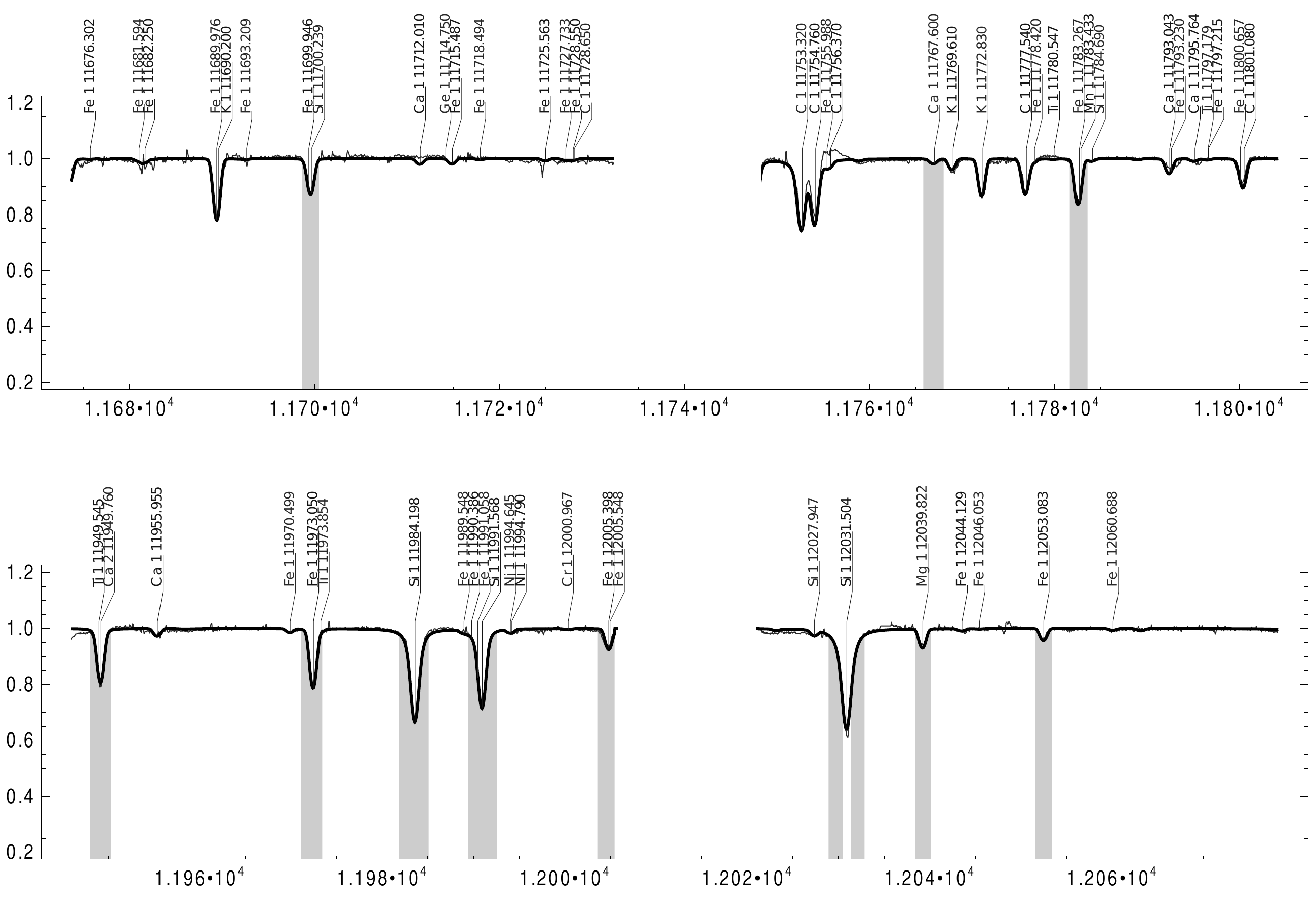}\\
\includegraphics[width=0.92\textwidth]{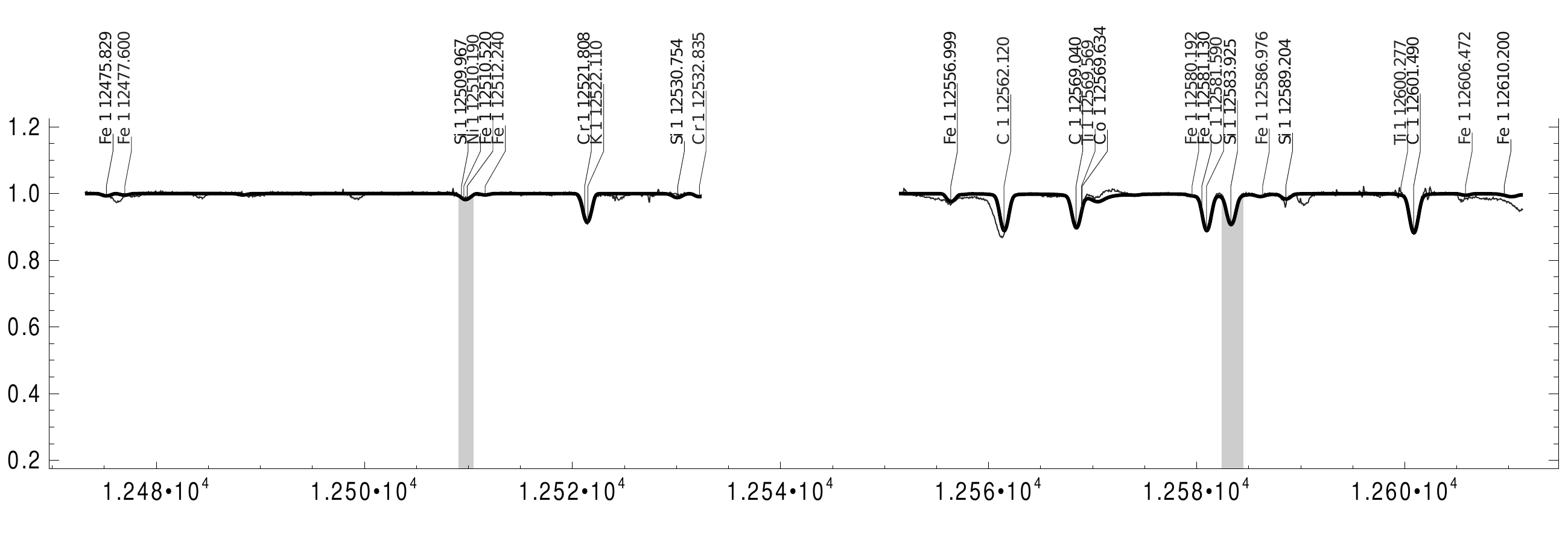}\\
\includegraphics[width=0.92\textwidth]{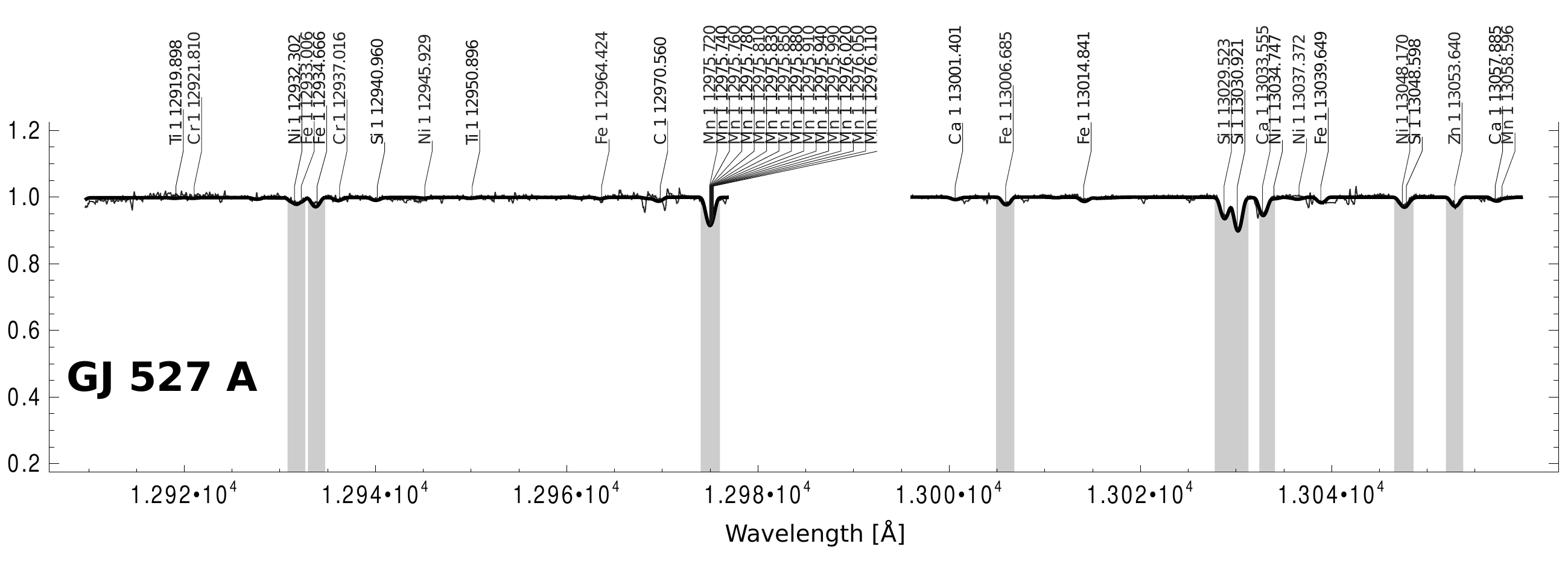}
\end{center}
\caption{Our obtained spectra after continuum rectification, wavelength correction, and removal of telluric features of F~dwarf GJ~527~A. The best-fit synthetic spectrum calculated with SME is shown as the overplotted thick black line. The line mask used for the metallicity determination is indicated by grey shading. White spacing represents wavelength regions between two chips.}
\label{spectra4}
\end{figure*}

\newpage
\noindent
\newline
\newline
\newline
\newline
\newline

\section{Linelist} \label{app:linelist}

\begin{table*}[ht!]
\caption{Some of the data in the used linelist.}
\centering
\begin{tabular}{llccccc}
{\bf Wavelength $[$\AA$]$} & {\bf Species } &  {\bf E$_{low}$ [eV] } & {\bf vdW } (This work)  & {\bf log~$gf$ } (This work) & {\bf log~$gf$ } (On12)  &  \\  \hline \hline
11681.594 &             Fe I &          3.55 &                  $-$6.725 &               $-$3.216 &              $-$3.301 &              FGK \\
11700.239 &             Si I &          6.27 &                  $-$7.294 &               $-$0.513 &              $-$0.526 &              FGK \\
11712.010 &             Ca I &          4.68 &                  $-$8.593 &               $-$0.501 &              $-$1.510 &              FGK \\
11767.600 &             Ca I &          4.532 &                 $-$6.621 &               $-$0.666 &              $-$0.635 &              FGK, M \\
11780.547 &             Ti I &          1.443 &                 $-$7.790 &               $-$2.273 &              $-$2.180 &              FGK, M \\
11783.267 &             Fe I &          2.832 &                 $-$8.012 &               $-$1.615 &              $-$1.520 &              FGK, M \\
11797.179 &             Ti I &          1.430 &                 $-$7.790 &               $-$2.250 &              $-$2.250 &              FGK, M \\
11797.215 &             Fe I &          5.82 &                  $-$7.420 &               $-$1.788 &              $-$1.788 &              FGK, M \\
11800.657 &             Fe I &          6.20 &                  $-$7.160 &               $-$1.090 &              $-$1.090 &              FGK \\
11828.171 &             Mg I &          4.35 &                  862.255 &                       $-$0.220 &               $-$0.046 &              FGK, M \\
11949.545 &             Ti I &          1.443 &                 $-$7.790 &               $-$1.550 &              $-$1.550 &              FGK, M \\
11949.760 &             Ca II   &               6.470 &                 $-$7.743 &               +0.218 &                        $-$0.040 &              FGK \\
11955.955 &             Ca I    &               4.131 &                 $-$7.132 &               $-$0.835 &              $-$0.849 &              FGK, M \\
11973.050 &             Fe I    &               2.176 &                 $-$7.881 &               $-$1.566 &              $-$1.405 &              FGK, M \\
11955.955 &             Ca I &          4.13 &                  $-$7.132 &               $-$0.835 &              $-$0.849 &              FGK \\
11973.050 &             Fe I &          2.18 &                  $-$7.881 &               $-$1.566 &              $-$1.405 &              FGK \\
11973.854 &             Ti I &          1.460 &                 $-$7.380 &               $-$1.558 &              $-$1.591 &              FGK, M \\
11984.198 &             Si I &          4.930 &                 677.228 &                 +0.108 &                        +0.239 &                        FGK, M \\
11991.568 &             Si I &          4.920 &                 674.228 &                       $-$0.257 &               $-$0.109 &              FGK, M \\
12031.504 &             Si I &          4.954 &                 685.229 &                       +0.331 &                       +0.477 &                        FGK, M \\
12039.822 &             Mg I &          5.753 &                 $-$7.239 &               $-$1.525 &              $-$1.530  &             FGK \\
12044.055 &             Cr I &          3.422 &                 $-$6.281 &               $-$1.863 &              $-$1.863 &              FGK \\
12044.129 &             Fe I &          4.988 &                 $-$6.677 &               $-$2.130 &              $-$2.130 &              FGK \\
12053.083 &             Fe I &          4.559 &                 $-$7.439 &               $-$1.535 &              $-$1.543 &              FGK \\
12510.520 &             Fe I &          4.956 &                 $-$7.296 &               $-$1.845 &              $-$1.846 &              FGK \\
12532.835 &             Cr I &          2.709 &                 $-$8.349 &               $-$1.902 &              $-$1.879 &              FGK, M \\
12556.999 &             Fe I &          2.279 &                 $-$8.680 &               $-$3.879 &              $-$3.913 &              FGK, M \\
12569.569 &             Ti I &          2.175 &                 $-$7.810 &               $-$1.867 &              $-$1.867 &              FGK, M \\
12569.634 &             Co I &          3.409 &                 $-$7.730 &               $-$1.605 &              $-$0.992 &              FGK, M \\
12600.277 &             Ti I &          1.443 &                 $-$7.790 &               $-$2.275 &              $-$2.150 &              FGK, M  \\
12909.070 &             Ca I &          4.43 &                  $-$7.471 &               $-$0.461 &              $-$0.426 &              FGK, M \\
12910.087 &             Cr I &          2.708 &                 $-$7.509 &               $-$1.821 &              $-$1.863 &              FGK, M  \\
12919.898 &             Ti I &          2.154 &                 $-$8.970 &               $-$1.611 &              $-$1.553 &              FGK, M \\
12937.016 &             Cr I &          2.710 &                 $-$7.625 &               $-$1.916 &              $-$1.896 &              FGK, M \\
12919.898 &             Ti I &          2.15 &                  $-$8.970 &               $-$1.611 &              $-$1.553 &              FGK, M \\
12975.720-12976.150 $^\dagger$ & Mn I & &                       &                               &                               &                                 FGK, M \\
13001.401 &             Ca I &          4.441 &                 $-$7.396 &               $-$1.273 &              $-$1.139 &              FGK, M  \\
13006.685 &             Fe I &          2.990 &                 $-$7.725 &               $-$3.268 &              $-$3.269  &             FGK, M \\
13011.892 &             Ti I &          1.443 &                 $-$6.792 &               $-$2.242 &              $-$2.180 &              FGK, M  \\
13033.555 &             Ca I &          4.441 &                 $-$7.462 &               $-$0.194 &              $-$0.064  &             FGK, M \\
13057.885 &             Ca I &          4.441 &                 $-$7.716 &               $-$1.054 &              $-$1.092 &              FGK, M  \\ \hline
  \end{tabular}
  \label{gf_values}
\begin{flushleft}
\noindent
\newline
{\bf Notes.} Column descriptions: $E_{\rm low}$: lower level energy. \emph{VdW}: parameter(s) used to calculate line broadening due to van der Waals interaction. Negative numbers: logarithm of the line width per perturber at l0$^4$~K in rad s$^{-1}$ cm$^3$ (log$\gamma_{\rm W}$). Adjusted by fit to a solar spectrum (see section 4.3.1). Positive numbers: integer part: broadening cross-section at a velocity of 10$^4$~m s$^{-1}$ in atomic units, fractional part: velocity parameter (see \citealt{Barklem2000}). \emph{log~$gf$}: oscillator strength, adjusted by fit to a solar spectrum (see section 4.3.1 in this paper, and the values adopted by On12). \emph{FGK/FGK, M}: lines used for the analysis of only the primaries or for both binary components. $\dagger$ Mn I transition with 15 hyperfine structure components. 
\end{flushleft}
\end{table*}

\end{appendix}

\end{document}